\documentclass[12pt]{article}
\pdfoutput=1
\usepackage{amssymb}
\usepackage{amsmath}
\usepackage{latexsym}
\usepackage[usenames]{color}
\usepackage{fancybox}
\usepackage{simplewick}
\usepackage{comment}
\usepackage{cite}
\usepackage{framed}
\definecolor{shadecolor}{rgb}{0.9,0.9,0.95}
\usepackage{setspace}
\usepackage[usenames,dvipsnames]{xcolor}
\definecolor{darkgreen}{rgb}{0,0.5,0}
\definecolor{darkblue}{cmyk}{0.9,0.9,0,0}
\definecolor{darkred}{rgb}{0.6,0,0.3}

\usepackage{graphicx} 
\usepackage[setpagesize=false,pagebackref=false, linktocpage, bookmarksopen=true, colorlinks=true, linkcolor=RoyalBlue,citecolor=Maroon,urlcolor=Maroon]{hyperref}
\usepackage{hyperref}

\renewcommand{\thefootnote}{\arabic{footnote}}

\def\del{\partial}

\def\fn#1{\footnote{#1}}
\def\nn{\nonumber}
\def\eqref#1{(\ref{#1})}
\def\comma{\,,}
\def\period{\,.}

\def\bDelta{\bar{\Delta}}
\def\beq{\begin{equation}}
\def\eeq{\end{equation}}
\def\pmatrix#1#2{\left( 
\begin{array}{#1}
#2\end{array} 
\right)}

\numberwithin{equation}{section}

\begin{document}
\thispagestyle{empty}

\renewcommand{\thefootnote}{\fnsymbol{footnote}}
\setcounter{page}{1}
\setcounter{footnote}{0}
\setcounter{figure}{0}
\begin{flushright}
{\rm PUTP-2614}
\end{flushright}
\vspace{0.7cm}
\begin{center}
{\large \bf
Loop Equation and Exact Soft Anomalous Dimension\\ in $\mathcal{N}=4$ Super Yang-Mills
\par}

\vspace{1.3cm}

\textrm{Simone Giombi$^{\clubsuit}$, Shota Komatsu$^{\diamondsuit}$}
\\ \vspace{1cm}
\footnotesize{\textit{
$^{\clubsuit}$Department of Physics, Princeton University, Princeton, NJ 08544, USA\\
$^{\diamondsuit}$School of Natural Sciences, Institute for Advanced Study, Princeton, NJ 08540, USA
}  
\vspace{1cm}
}

{\tt sgiombi AT princeton.edu, skomatsu AT ias.edu}

\par\vspace{1.5cm}

\textbf{Abstract}\vspace{2mm}
\end{center}
\noindent
BPS Wilson loops in supersymmetric gauge theories have been the subjects of active research since they are often amenable to exact computation. So far most of the studies have focused on loops that do not intersect. In this paper, we derive exact results for intersecting $1/8$~BPS Wilson loops in $\mathcal{N}=4$ supersymmetric Yang-Mills theory, using a combination of supersymmetric localization and the loop equation in 2d gauge theory.  The result is given by a novel matrix-model-like representation which couples multiple contour integrals and a Gaussian matrix model. We evaluate the integral at large $N$, and make contact with the string worldsheet description at strong coupling. As an application of our results, we compute exactly a small-angle limit (and more generally near-BPS limits) of the cross anomalous dimension which governs the UV divergence of intersecting Wilson lines. The same quantity describes the soft anomalous dimension of scattering amplitudes of $W$-bosons in the Coulomb branch.  

\setcounter{page}{1}
\renewcommand{\thefootnote}{\arabic{footnote}}
\setcounter{footnote}{0}
\setcounter{tocdepth}{3}
\newpage
\tableofcontents

\parskip 5pt plus 1pt   \jot = 1.5ex

\newpage
\section{Introduction}
The loop equation was proposed initially in \cite{Makeenko:1979pb,Makeenko:1980vm} as an alternative way to formulate, and possibly solve, the gauge theories (see e.g. \cite{Makeenko:1999hq} for a review). It has the conceptual advantage that it directly constrains the most basic observables, namely the Wilson loops. In lower-dimensional theories such as matrix models \cite{Migdal:1984gj} and two-dimensional Yang-Mills theory \cite{Kazakov:1980zi,Kazakov:1980zj,Boulatov:1993zs,Daul:1993xz,Rusakov:1993gz}, it has proven to be a powerful tool for solving the theories exactly. Unfortunately, solving the loop equation is much harder in higher dimensions and progress remains to be made.

In this paper we demonstrate the power of the loop equation in higher dimensions when used in conjunction with other non-perturbative techniques. Specifically, we consider intersecting $1/8$-BPS Wilson loops in $\mathcal{N}=4$ supersymmetric Yang-Mills ($\mathcal{N}=4$ SYM) and compute their expectation values at finite coupling and finite $N$ using a combination of the loop equation and supersymmetric localization \cite{Pestun:2007rz,Pestun:2009nn}.

The $1/8$-BPS Wilson loop is a supersymmetric Wilson loops in $\mathcal{N}=4$ SYM which can be defined on a arbitrary contour on a two-sphere and preserves four fermionic charges. It was conjectured in \cite{Drukker:2007yx,Drukker:2007qr} and later supported by supersymmetric localization \cite{Pestun:2009nn} that its expectation value (as well as general correlation functions of any number of loops) conicides with that of the standard Wilson loop in two-dimensional Yang-Mills theory (2d YM) in a zero-instanton sector. Based on this, the expectation value of a non-intersecting $1/8$-BPS loop was computed in \cite{Drukker:2007yx,Drukker:2007qr} generalizing the famous result for the $1/2$-BPS loop \cite{Erickson:2000af,Drukker:2000rr,Pestun:2007rz}. It was also used to derive multi-matrix models for correlators of local operators and non-intersecting BPS loops\cite{Giombi:2009ds,Giombi:2012ep}. By now these results have been tested by numerous direct computations \cite{Giombi:2009ms,Giombi:2009ek,Bassetto:2009rt,Bassetto:2009ms,Bonini:2014vta,Bonini:2015fng} and they provide convincing evidence for the equivalence between the BPS sector of $\mathcal{N}=4$ SYM and 2d YM.

The goal of this paper is to generalize them to intersecting loops. The strategy is simple: Using the conjecture above, we relate the intersecting $1/8$-BPS loops in $\mathcal{N}=4$ SYM to intersecting Wilson loops in 2d YM on $S^2$ in the zero-instanton sector. We then solve the loop equation of 2d YM exactly at finite $N$. The loop equation in 2d YM was first solved for loops on $R^2$ at large $N$ \cite{Kazakov:1980zi}. The result was generalized to loops on $R^2$ at finite $N$ in \cite{Kazakov:1980zj} and to loops on $S^2$ at large $N$ in \cite{Boulatov:1993zs,Daul:1993xz,Rusakov:1993gz}. While all these results take simple and compact forms, no such expressions were known for loops on $S^2$ at finite $N$: the only known expression in the literature involves a rather complicated sum over the Young diagrams \cite{Cordes:1994fc}. We show that a simple closed-form expression does exist for loops on $S^2$ at finite $N$ if the theory is restricted to the zero-instanton sector---the sector relevant for the BPS loops in $\mathcal{N}=4$ SYM. 

\begin{figure}[t]
\centering
\begin{minipage}{0.45\hsize}
\centering
\includegraphics[clip,height=2.5cm]{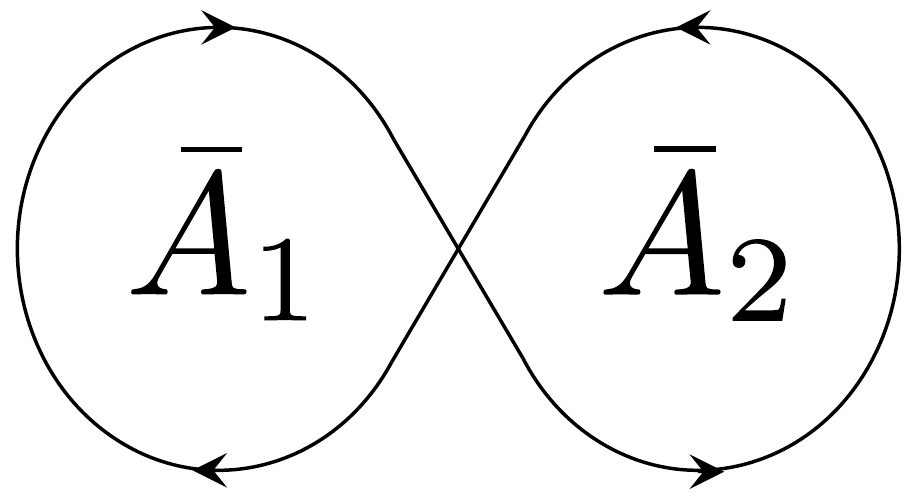} \\
$(a)$
\end{minipage}
\begin{minipage}{0.45\hsize}
\centering
\includegraphics[clip,height=2.5cm]{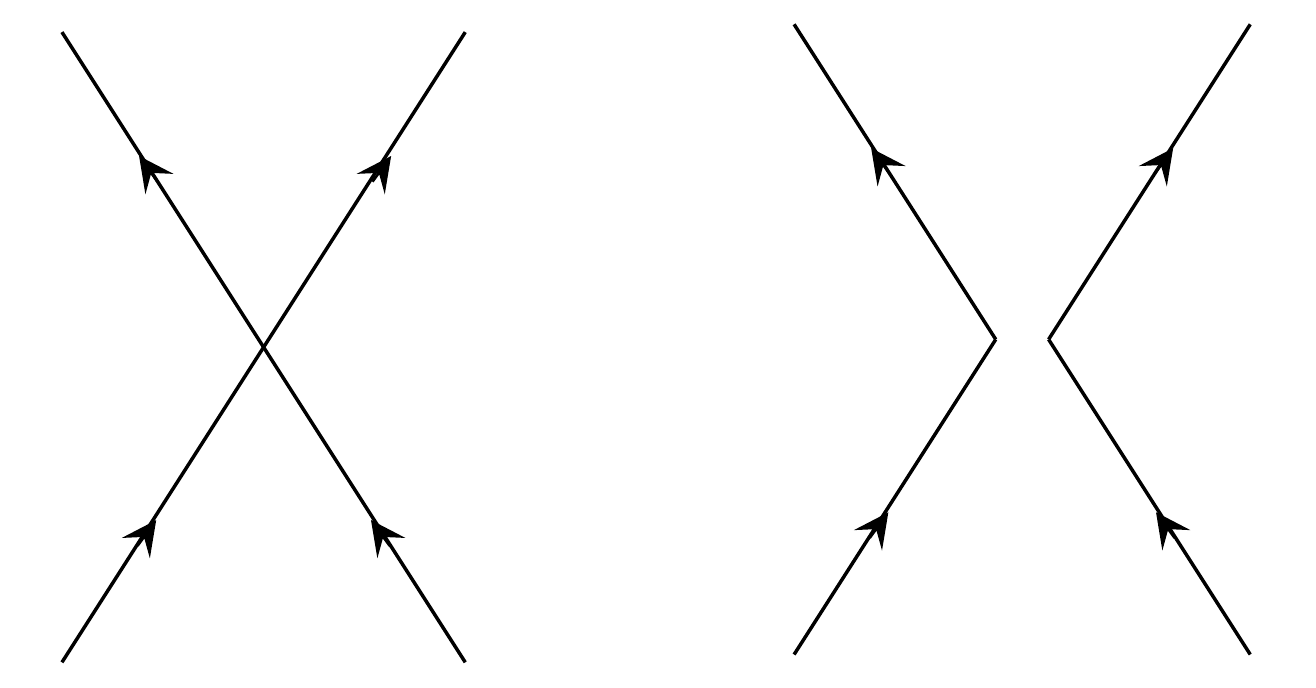} \\
$(b)$
\end{minipage}
\caption{$(a)$ The figure eight Wilson loop with areas $\bar{A}_1$ and $\bar{A}_2$. $(b)$ Two intersecting lines (left) and  two touching lines (right) which mix under the renormalization group flow.} \label{fig:intro}
\end{figure}

The result of our computation is a coupled system of multiple integrals and a Gaussian matrix model. For instance, the expectation value of the {\it figure eight} loop with areas $\bar{A}_1$ and $\bar{A}_2$ reads (see figure \ref{fig:intro})
\beq\label{eq:FigureEight}
\langle \mathcal{W}_{\text{figure-eight}}\rangle=\frac{ i}{\pi \lambda}\oint_{\mathcal{C}_1\prec \mathcal{C}_2} du_1du_2\frac{u_1-u_2}{(u_1-u_2)^2+\left(\frac{\lambda}{4\pi  N}\right)^2} \left< f_{4\pi -\bar{A}_1}(u_1)f_{\bar{A}_2}(u_2)\right>_M\period
\eeq
Here $\lambda:= g_{\rm YM}^2 N$ is the 't Hooft coupling constant and the function $f_A$ is defined by
\beq
f_{A}(u):=e^{i A \left(u-\frac{i \lambda}{8\pi N}\right)}\frac{\det \left(u-M-\frac{i \lambda}{4\pi N}\right)}{\det (u-M)} \comma
\eeq
while $\langle \bullet \rangle_M$ denotes the expectation value of $\bullet$ in the Gaussian matrix model of size $N$,
\beq\label{eq:introGaussian}
\langle \bullet\rangle_M :=\frac{\int [d M] \,\bullet \,\exp\left[-\frac{8\pi^2}{g_{\rm YM}^2}{\rm tr}(M^2)\right]}{\int [d M] \,\,\exp \left[-\frac{8\pi^2}{g_{\rm YM}^2}{\rm tr}(M^2)\right]}\period
\eeq
The integration contours $\mathcal{C}_{1,2}$ encircle the eigenvalues of the Gaussian matrix model \eqref{eq:introGaussian} and they are placed far apart from each other (see sections \ref{sec:setup} and \ref{sec:loopeq} for more details). The integral can be evaluated explicitly at large $N$ and it gives
\beq\nn
\begin{aligned}
&\langle \mathcal{W}_{\text{figure-eight}}\rangle\overset{N\to\infty}{=}\frac{\mathcal{I}_{0}^{\bar{a}_1}\mathcal{I}_{1}^{\bar{a}_2}}{2\pi g_{\bar{a}_2}}+\sum_{k=1}^{\infty}\frac{\rho_{\bar{a}_1}^{k}\mathcal{I}_{k}^{\bar{a}_1}}{4\pi g}\left[\left(\rho_{\bar{a}_2}^{k+1}+\frac{(-1)^{k}}{\rho_{\bar{a}_2}^{k+1}}\right)\mathcal{I}_{k+1}^{\bar{a}_2}+\left(\rho_{\bar{a}_2}^{k-1}+\frac{(-1)^{k}}{\rho_{\bar{a}_2}^{k-1}}\right)\mathcal{I}_{k-1}^{\bar{a}_2}\right]\comma
\end{aligned}
\eeq
where
\beq
g:=\frac{\sqrt{\lambda}}{4\pi}\comma\quad\bar{a}_i:=\frac{\bar{A}_i-2\pi}{2}\comma\quad g_{\bar{a}} :=g\sqrt{1-\frac{\bar{a}^2}{\pi^2}}\comma\quad  \rho_{\bar{a}} :=\sqrt{\frac{\pi+\bar{a}}{\pi-\bar{a}}}\comma
\eeq
and $\mathcal{I}_k^{\bar{a}}:=I_{k}(4\pi g_{\bar{a}})$ is the modified Bessel function. The result at strong coupling reproduces the area of the minimal surface as we show in section \ref{subsubsec:figureeight}.

As an important application, we compute a small angle limit of the {\it cross anomalous dimension} of intersecting Wilson lines. The cross anomalous dimension controls the mixing of two different configurations of Wilson lines, two intersecting lines and  two touching lines (see figure \ref{fig:intro}), under the renormalization group. It also describes the {\it soft anomalous dimension}, which controls how the soft gluons transfer the color degrees of freedom of partons in a scattering process.  
We generalize the analysis of the Bremsstrahlung function in \cite{Correa:2012at,Giombi:2018qox} and relate the small angle limit---and more generally the near-BPS limits---of these anomalous dimensions to our localization computations. The results are exact at finite $\lambda$ and $N$ and reproduce the answers at weak coupling in the literature \cite{Munkler:2018cvu,Korchemsky:1993hr,Korchemskaya:1994qp}.

The formalism in this paper will be used in our upcoming paper \cite{Toappear} on the defect CFT correlators on the higher-rank Wilson loops. There, we consider a configuration in which multiple fundamental Wilson loops with different areas are joined together by a projector to a higher-rank representation. We compute its expectation value by taking an appropriate linear combination of multiply intersecting loops.

The rest of the paper is organized as follows. In section \ref{sec:setup}, we review the $1/8$ BPS Wilson loops and their relation to 2d YM. We then present a new integral representation for correlators of non-intersecting Wilson loops which simplifies the analysis of the loop equation in the subsequent sections.  
 In section \ref{sec:loopeq}, we review the loop equation in 2d YM and solve it in the zero-instanton sector on $S^2$. As a result, we obtain a closed-form integral representation for intersecting loops. We then demonstrate how to evaluate the integral at large $N$ using examples of a figure eight loop  and a two-intersection loop and study the weak- and strong-coupling limits. In section \ref{sec:cross}, we use these results to compute the small-angle and near-BPS limits of the cross anomalous dimension at finite $\lambda$ and $N$.  We then discuss future directions in section \ref{sec:conclusion}. A few appendices are provided to explain technical details.

\section{1/8 BPS Wilson Loop, 2d YM and Matrix Model \label{sec:setup}}
\subsection{1/8 BPS Wilson loops and matrix model} 
\begin{figure}[t]
\centering
\includegraphics[clip,height=3.5cm]{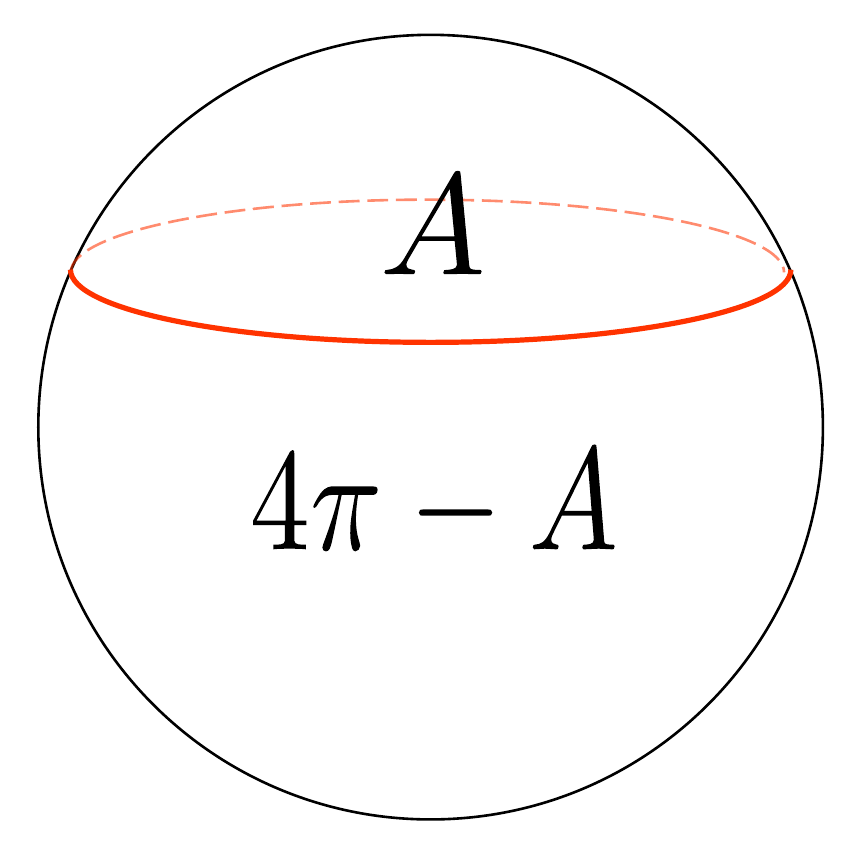}
\caption{The $1/8$-BPS Wilson loop on $S^2$. The $1/8$-BPS loop can be defined on an arbitrary contour on $S^2$ and its coupling to scalars is given by \eqref{eq:eightBPScond}. A single loop divides $S^2$ into two regions, one with area $A$ and the other with area $4\pi-A$. Its expectation value depends only on the area $A$ owing to the relation to 2d YM.} \label{fig:singleloop}
\end{figure}

In supersymmetric gauge theories, one can often define a supersymmetric generalization of the Wilson loop by coupling the loop to scalar fields.
The $1/8$ BPS Wilson loop in $\mathcal{N}=4$ SYM is a special kind of such operators which can be defined for an arbitrary contour $C$ on $S^2$ inside $R^4$ or $S^4$ (see figure \ref{fig:singleloop}):
\beq\label{eq:eightBPScond}
\mathcal{W}_{1/8}:=\frac{1}{N}{\rm tr}\,{\rm Pexp} \left(\oint_{C}\left(iA_j+\epsilon_{kjl}x^{k}\Phi^{l}\right)dx^{j} \right)\period
\eeq
Here all the indices run from $1$ to $3$ and $x^{j}$'s are the coordinates of $S^2$, $\sum_{j=1}^{3}(x^{j})^2=1$. Throughout this paper, we consider the U$(N)$ gauge group unless otherwise stated.

It was conjectured through the perturbative computations \cite{Drukker:2007yx,Drukker:2007qr} and later supported by the localization \cite{Pestun:2009nn} that the computation of the $1/8$ BPS Wilson loop reduces to that of the standard Wilson loop in 2d YM on $S^2$ in the zero-instanton sector,
\beq\label{eq:relation}
\begin{aligned}
\mathcal{W}_{1/8} \quad& \longleftrightarrow \quad \mathcal{W}_{\text{2dYM}}:=\frac{1}{N}{\rm tr}\,{\rm Pexp} \left(\oint_{C}\,iA_j dx^{j} \right)\comma
\end{aligned}
\eeq 
where the coupling constants of 2d YM ($g_{\rm 2d}$) and $\mathcal{N}=4$ SYM ($g_{\rm YM}$) are related by
\beq\label{eq:2dto4d}
g_{\rm 2d}^2 =  -\frac{g_{\rm YM}^2}{2\pi}\comma
\eeq
and the action of 2d YM is
\beq\label{eq:2dYMaction}
S_{2d}=\frac{1}{g_{\rm 2d}^2}\int d^{2}\sigma \sqrt{g}{\rm tr}\left(F_{\mu\nu}F^{\mu\nu}\right)\period
\eeq

The expectation value of a single $1/8$-BPS loop was computed by resumming the perturbative series in 2d YM and re-expressing it as a matrix integral\fn{$Z$ is a partition function of the matrix model, namely a matrix integral without the insertion of ${\rm tr}\left(e^{\Phi}\right)/N$.},
\beq
\begin{aligned}
\langle \mathcal{W}_{1/8}\rangle=\frac{1}{Z}\int [d\Phi] \frac{1}{N}{\rm tr}\left(e^{\Phi}\right) e^{-\frac{(4\pi)^2}{2A(4\pi -A)g_{\rm YM}^2}{\rm tr}\left(\Phi^2\right)}\period
\label{GaussMat}
\end{aligned}
\eeq
By evaluating this matrix integral, we get
\beq
\langle \mathcal{W}_{1/8}\rangle=\frac{1}{N}L_{N-1}^{1}\left(-\frac{\lambda^{\prime}}{4N}\right)e^{\frac{\lambda^{\prime}}{8N}}\comma\qquad \lambda^{\prime}:=\lambda \left(1-\frac{(A-2\pi)^2}{4\pi^2}\right)\comma
\eeq
where $\lambda:=g_{\rm YM}^2N $ is the 't Hooft coupling constant, $L_{N-1}^{1}$ is an associated Laguerre polynomial and $A$ is the area of the region encircled by the Wilson loop. The large $N$ limit of this result gives a generalization of the famous result for the $1/2$-BPS Wilson loop,
\beq
\langle \mathcal{W}_{1/8}\rangle\overset{N\to\infty}{=}\frac{2}{\sqrt{\lambda^{\prime}}}I_{1}(\sqrt{\lambda^{\prime}})\comma
\eeq
with $I_n$ being the modified Bessel function.
 
As another application of the relation \eqref{eq:relation}, a multi-matrix model for the correlation functions of $1/8$-BPS Wilson loops was derived in \cite{Giombi:2012ep}. A heuristic way\fn{See \cite{Giombi:2012ep} for more rigorous discussions.} to derive the matrix model is as follows: First we rewrite the action of 2d YM as a deformed BF theory,
\beq\label{eq:BFaction}
S=\int {\rm tr}\left(B\wedge F\right) -\frac{g_{\rm 2d}^2}{4} \int d^2 \sigma \sqrt{g}\, {\rm tr}\left(B^2\right)\period
\eeq
If one integrates out $B$, one recovers the standard 2d YM action.
Second we use the fact that the theory reduces essentially to the abelian theory and localizes to configurations for which $B$ is piecewise constant with discontinuities across the Wilson loops. In such configurations, the first term of the action \eqref{eq:BFaction} becomes a boundary term
\beq
\int_{\Sigma} {\rm tr}(B\wedge F)= {\rm tr}\left[B\int_{\partial \Sigma}A\right]=-i{\rm tr}\left[B X\right] \comma
\eeq
where $\Sigma$ is a subregion of $S^2$ whose boundaries are given by the Wilson loops and $X:=i\int_{\del \Sigma}A$ is a boundary holonomy.  On the other hand the second term simply gives
\beq
-\frac{g_{\rm 2d}^2}{4} \int_{\Sigma} d^2 \sigma \sqrt{g}\, {\rm tr}\left(B^2\right)=-\frac{g_{\rm 2d}^2A_{\Sigma}}{4}{\rm tr}(B^2)=\frac{g_{\rm YM}^2 A_{\Sigma}}{8\pi}{\rm tr}(B^2)\comma
\eeq
where $A_{\Sigma}$ is the area of the region $\Sigma$.

Performing such rewritings to each region $\Sigma_m$, we obtain the following multi-matrix model action:
\beq\label{eq:multiaction}
\begin{aligned}
\hat{S}&=\sum_{\{\Sigma_m\}}\left[-i{\rm tr}\left(\sum_{j\in \del \Sigma_m} s_j^{(m)}\hat{B}_{\Sigma_m}\hat{X}_j\right)+\frac{g_{\rm YM}^2 A_{\Sigma_m}}{8\pi}{\rm tr}\left(\hat{B}_{\Sigma_m}^2\right)\right]\period
\end{aligned}
\eeq
Here $s_j^{(m)}$ is an orientation factor which takes value $+1$ when the holonomy $\hat{X}_j$ is oriented in the same direction as the boundary $\del\Sigma_m$, and $-1$ otherwise. From this action, the correlator of the Wilson loops can be computed as follows,
\beq\label{eq:correlatoratio}
\begin{aligned}
\left<\prod_{k}\mathcal{W}_k \right> =\frac{\int [d\hat{B}_{\Sigma_m}][d\hat{X}_j]\prod_{k}{\rm tr}_{r_k}(e^{\hat{X}_k})e^{-\hat{S}}}{\int [d\hat{B}_{\Sigma_m}][d\hat{X}_j]e^{-\hat{S}}}\comma
\end{aligned}
\eeq
where $r_k$ is the representation of the $k$-th Wilson loop $\mathcal{W}_k$. As was done in \cite{Giombi:2012ep}, one can integrate out $\hat{B}_{\Sigma_m}$'s and obtain an action purely in terms of $\hat{X}_j$'s. This however is not convenient for analyzing the loop equation since the resulting action depends nonlinearly on the areas $A_{\Sigma_m}$. In this paper, we instead integrate out $\hat{X}_j$ and derive a new integral representation which is more convenient for the application of the loop equation.

\subsection{A new integral representation for Wilson loop correlators\label{subsec:newint}}
To derive an integral representation, it is convenient to first rescale the matrices as 
\beq
X_j:= \frac{4\pi\hat{X}_j}{g_{\rm YM}^2}\comma \qquad\qquad  B_{\Sigma_m}:= \frac{g_{\rm YM}^2\hat{B}_{\Sigma_m}}{4\pi}\period
\eeq 
Then, the action and the correlator read
\beq\label{eq:rescaled}
\begin{aligned}
&S=\sum_{\{\Sigma_m\}}\left[-i{\rm tr}\left(\sum_{j\in \del \Sigma_m} s_j^{(m)}B_{\Sigma_m}X_j\right)+\frac{2\pi A_{\Sigma_m}}{g_{\rm YM}^2}{\rm tr}\left(B_{\Sigma_m}^2\right)\right]\comma\\
&\left<\prod_{k}\mathcal{W}_k \right> =\frac{\int [dB_{\Sigma_m}][dX_j]\prod_{k}{\rm tr}_{r_k}(e^{\epsilon X_k})e^{-S}}{\int [dB_{\Sigma_m}][dX_j]e^{-S}}\comma
\end{aligned}
\eeq
with
\beq
\epsilon:= \frac{g_{\rm YM}^2}{4\pi}=\frac{\lambda}{4\pi N}=\frac{4\pi g^2}{N}\period
\eeq
In the last equality, we used the  notation for the 't Hooft coupling constant commonly used in the integrability literature,
\beq\label{eq:integrabilitycoupling}
g^2 := \frac{\lambda}{16\pi^2}\period
\eeq
\subsubsection{A single fundamental loop\label{subsubsec:integralsinglefund}} 
Let us first discuss the expectation value of a single fundamental Wilson loop with no self-intersections. In the presence of a Wilson loop, $S^2$ is divided into two regions, one with the area $4\pi -A(=:A_0)$ and the other with the area $A (=:A_1)$ (see figure \ref{fig:singleloop}), and the action of the matrix model reads
\beq\label{eq:actionfund}
S=\frac{2\pi A_0}{g_{\rm YM}^2} {\rm tr}(B_0^2)+\frac{2\pi A_1}{g_{\rm YM}^2} {\rm tr}(B_1^2)-i{\rm tr}\left(X(B_0-B_1)\right)\period
\eeq
To compute the expectation value of the Wilson loop in this matrix model, we use the Harish-Chandra-Itzykson-Zuber identity
\beq\label{eq:HCIZ}
\int d\Omega\, e^{i{\rm tr}\left(\Omega^{\dagger}A\Omega B\right)}=\frac{\det e^{ia_ib_j}}{\Delta (a)\Delta(b)}\comma
\eeq
where $A$ and $B$ are diagonal matrices with entries $a_j$'s and $b_j$'s respectively, $\int d\Omega$ is an integral over the unitary matrices and $\Delta$ is the Vandermonde determinant $\Delta(a):=\prod_{i<j}(a_i-a_j)$. 

Using \eqref{eq:HCIZ}, we can reduce the partition function of the matrix model \eqref{eq:actionfund} to integrals of eigenvalues,
\beq
Z=\int [dX][dB_0][dB_1] \,\, e^{-S}=\int d^{N}x \, d^{N}b^{(0)}\, d^{N}b^{(1)}\,\, \mathcal{I}  \comma
\eeq
where the integrand $\mathcal{I} $ is given by
\beq
\begin{aligned}
\label{calI}
\mathcal{I} &=\Delta^2 (x) \Delta^2 (b^{(0)})\Delta^2 (b^{(1)})\frac{\det e^{i x_i b^{(0)}_j}\det e^{-i x_k b^{(1)}_l}}{\Delta^2 (x)\Delta (b^{(0)})\Delta (b^{(1)})} \,\,e^{-\frac{2\pi}{g_{\rm YM}^2}\sum_{j} (A_0(b_j^{(0)})^2+A_1(b_j^{(1)})^2)}\\
&= \Delta (b^{(0)})\Delta (b^{(1)})\det e^{i x_i b^{(0)}_j}\det e^{-i x_k b^{(1)}_l} \,\,e^{-\frac{2\pi}{g_{\rm YM}^2}\sum_{j} (A_0(b_j^{(0)})^2+A_1(b_j^{(1)})^2)}\period
\end{aligned}
\eeq
We next expand the determinants into a sum over permutations:
\beq
\det e^{i x_i b^{(0)}_j} =\sum_{\sigma \in S_N}(-1)^{\sigma}\prod_j e^{i x_i b^{(0)}_{\sigma_j}}\comma\qquad \det e^{-i x_k b^{(1)}_l} =\sum_{\sigma^{\prime} \in S_N}(-1)^{\sigma^{\prime}}\prod_j e^{-i x_i b^{(1)}_{\sigma_j}}\period
\eeq
Owing to the symmetry of the rest of the integrand, all these permutations give the same answer . We thus pick the simplest one ($\sigma_j=j$ and $\sigma^{\prime}_j=j$) and multiply a factor $(N!)^2$. After integrating out $x$'s, we get
\beq
\begin{aligned}
Z&=(2\pi)^{N}(N!)^2\int \left(\prod_{s=0}^{1}d^{N}b^{(s)}\Delta (b^{(s)})\right)\left(\prod_k\delta (b^{(0)}_{k}-b^{(1)}_{k})\right)e^{-\frac{2\pi}{g_{\rm YM}^2}\sum_{j} (A_0(b_j^{(0)})^2+A_1(b_j^{(1)})^2)}\\
&=(2\pi)^{N}(N!)^2 \int d^{N} b \,\Delta^2 (b)\, e^{-\frac{8\pi^2}{g_{\rm YM}^2}\sum_j (b_j)^2} \period
\end{aligned}
\eeq
In the second line, we integrated out $b^{(1)}$'s and denoted the remaining variables $b_j^{(0)}$ as $b_j$. We also used $A_0+A_1=4\pi$ to simplify the exponent.

Similarly, the expectation value of the fundamental Wilson loop can be reduced to the following eigenvalue integral:
\beq\label{eq:singlefirst}
\begin{aligned}
\langle \mathcal{W}\rangle&=\frac{1}{Z}\int [dX][dB_0][dB_1] \frac{{\rm tr}\left(e^{\epsilon X}\right)}{N}e^{-S}=\frac{1}{Z}\int d^{N}x\, d^{N}b^{(0)}\,d^{N}b^{(1)}\,\,\mathcal{I}\,\,\frac{\sum_ke^{\epsilon x_k}}{N}\,,
\end{aligned}
\eeq
with $\mathcal{I}$ defined in (\ref{calI}). 
To evaluate this integral, we again expand the determinants and integrate out $x$'s. The only modification is that one of the delta functions gets shifted by $-i\epsilon$ owing to the factor $e^{\epsilon x_k}$. We then get
\beq\label{eq:singlefundsum}
\begin{aligned}
\langle \mathcal{W}\rangle=&\frac{(2\pi)^{N}(N!)^2}{Z }\int \left(\prod_{s=0}^{1}d^{N}b^{(s)}\Delta (b^{(s)})\right)\sum_k\frac{\delta (b_{k}^{(0)}-b_{k}^{(1)}-i\epsilon)}{N}\\
&\times \left(\prod_{j\neq k}\delta (b^{(0)}_{j}-b^{(1)}_{j})\right)e^{-\frac{2\pi}{g_{\rm YM}^2}\sum_{j} (A_0(b_j^{(0)})^2+A_1(b_j^{(1)})^2)}\\
=&\frac{(2\pi)^{N}(N!)^2}{Z}\int d^{N}b\,\, \Delta^2 (b)e^{-\frac{8\pi^2}{g_{\rm YM}^2}\sum_j (b_j)^2}\frac{\sum_k e^{i A_1 (b_k-\frac{i\epsilon}{2})}\prod_{j\neq k}\frac{b_k-b_j-i\epsilon}{b_k-b_j}}{N}\period
\end{aligned}
\eeq
Now the crucial observation is that the sum $\sum_k$ can be recast into a contour integral,
\beq
\begin{aligned}
\langle \mathcal{W}\rangle&=\frac{(2\pi)^{N}(N!)^2}{Z}\int d^{N}b\,\, \Delta^2 (b)e^{-\frac{8\pi^2}{g_{\rm YM}^2}\sum_j (b_j)^2}\left[\oint_{\mathcal{C}}\frac{du}{8\pi^2 g^2}e^{i A_1(u-\frac{i\epsilon}{2})}\prod_{j}\frac{u-b_j-i\epsilon}{u-b_j}\right]\period
\end{aligned}
\eeq
We can then interpret this as an expectation value of a Gaussian matrix model and get
\beq\label{eq:singleresult}
\langle \mathcal{W}\rangle=\left<\oint_{\mathcal{C}} \frac{du}{8\pi^2g^2} f_{A_1}(u)\right>_M\period
\eeq
Here the integration contour $\mathcal{C}$ encircles all the eigenvalues $b_k$'s and $f_{A}$ is given by
\beq\label{eq:deffA}
f_{A}(u):=e^{i A(u-\frac{i\epsilon}{2})}\det \left[\frac{u-M-i\epsilon}{u-M}\right]\period
\eeq
The symbol $\left<\bullet\right>_M$ denotes the expectation value of $\bullet$ in a Gaussian matrix model with the action $S_M:=8\pi^2{\rm tr}\left(M^2\right)/g_{\rm YM}^2$:
\beq
\left<\bullet \right>_M:=\frac{\int [dM]\,\bullet\, e^{-S_M}}{\int [dM]\,e^{-S_M}}\period
\eeq
The result \eqref{eq:singleresult} may appear more complicated than the expressions known in the literature \cite{Drukker:2007yx,Drukker:2007qr}. However, for the analysis of the loop equation, \eqref{eq:singleresult} is more convenient since the area-dependence is simple. See section \ref{sec:loopeq} for more details.
\subsubsection{Multiple fundamental loops\label{subsubsec:multiple}}
We now generalize the result \eqref{eq:singleresult} to correlators of multiple Wilson loops. 

\begin{figure}[t]
\centering
\includegraphics[clip,height=3.5cm]{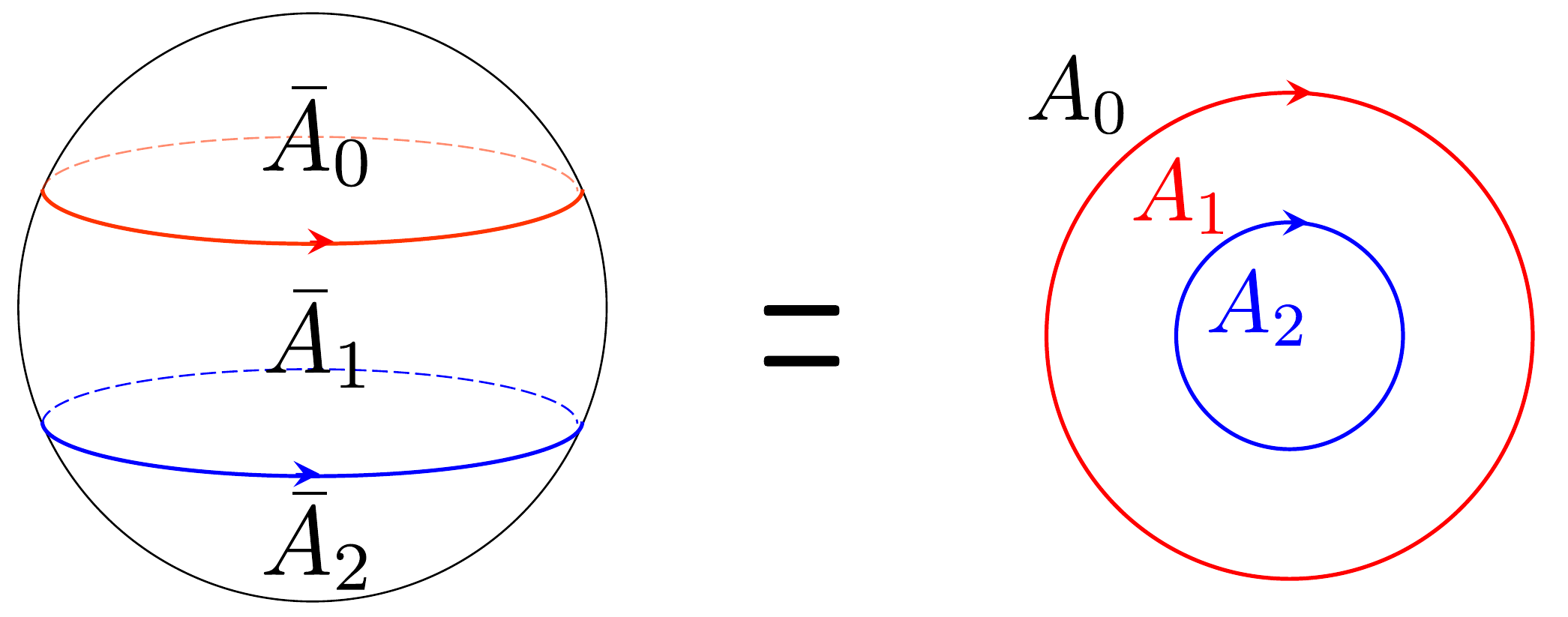}
\caption{Two $1/8$-BPS loops with the same orientations. Left: The loops divide $S^2$ into three disconnected regions with areas $\bar{A}_{0}$, $\bar{A}_1$ and $\bar{A}_2$. Right: The same configuration viewed from the south pole. We denote the areas inside outer (red) and inner (blue) circles by $A_1$ and $A_2$ respectively and the area of the complement by $A_0$. The two sets of areas are related by $\bar{A}_{0}=A_0$, $\bar{A}_{1}=A_1-A_2$ and $\bar{A}_2=A_2$.} \label{fig:twoloop}
\end{figure}

To see how it works, let us first consider the correlator of two fundamental Wilson loops with the same orientation. As depicted in figure \ref{fig:twoloop}, we denote the areas inside outer and inner loops by $A_1$ and $A_2$ respectively and the area of the complement by $A_0$.  Then the matrix model action is given by
\beq
S=\sum_{s=0}^{2}\frac{2\pi \bar{A}_s}{g_{\rm YM}^2}{\rm tr}(B_s^2)-i\sum_{s=1,2}{\rm tr}\left(X_s(B_{s-1}-B_s)\right)\comma
\eeq
with $\bar{A}_{0}:=A_0$, $\bar{A}_{1}:=A_1-A_2$ and $\bar{A}_2:=A_2$.
By reducing the matrix integral to the eigenvalues using the identity \eqref{eq:HCIZ}, we get
\beq\label{eq:partition2}
Z=\int \left(\prod_{s=0}^{2}[dB_s]\right)\left(\prod_{s=1,2}[dX_s]\right)e^{-S}=\int\left(\prod_{s=0}^{2}d^{N}b^{(s)}\right)\left(\prod_{s=1,2}d^{N}x^{(s)}\right)\mathcal{I}_2\comma
\eeq 
with
\beq
\begin{aligned}
\mathcal{I}_2=\Delta (b^{(0)})\Delta (b^{(2)}) \left(\prod_{s=1}^{2}\det e^{i x_i^{(s)}b_j^{(s-1)}}\det e^{-i x_i^{(s)}b_j^{(s)}}\right)e^{-\frac{2\pi}{g_{\rm YM}^2}\sum_{s=0}^{2}\sum_j\bar{A}_s (b_j^{(s)})^2}
\end{aligned}
\eeq
Expanding a product of determinants, we obtain a sum of permutations
\beq
\begin{aligned}
&\prod_{s=1}^{2}\det e^{i x_i^{(s)}b_j^{(s-1)}}\det e^{-i x_i^{(s)}b_j^{(s)}}=\sum_{\sigma^0,\sigma^1,\sigma^{2},\sigma^{3}}(-1)^{\sum_{s}\sigma^{s}}\prod_j e^{i (b^{(0)}_{\sigma^{0}_j}-b^{(1)}_{\sigma^{1}_j})x_j^{(1)}}e^{i (b^{(1)}_{\sigma^{2}_j}-b^{(2)}_{\sigma^{3}_j})x_j^{(2)}}\period
\end{aligned}
\eeq
Again all these permutations turn out to give the same result owing to the symmetry of the integrand. We can therefore replace the integrand $\mathcal{I}_2$ with
\beq
\begin{aligned}
\mathcal{I}_2=(N!)^4\Delta (b^{(0)})\Delta (b^{(2)})\left(\prod_{j}e^{i (b_j^{(0)}-b_j^{(1)})x_j^{(1)}}e^{i (b_j^{(1)}-b_j^{(2)})x_j^{(2)}}\right)e^{-\frac{2\pi}{g_{\rm YM}^2}\sum_{s=0}^{2}\sum_j\bar{A}_s (b_j^{(s)})^2}\period
\end{aligned}
\eeq
Plugging this expression into \eqref{eq:partition2}, we find that the integrals of $x_j^{(k)}$ give the delta functions $\delta (b^{(0)}_j-b^{(1)}_j)$ and $\delta (b^{(1)}_j-b^{(2)}_j)$. Performing the integrals of $b^{(1)}$ and $b^{(2)}$, we get
\beq
Z=(2\pi)^{2N}(N!)^4 \int d^{N}b \,\,\Delta^{2}(b)\,\,e^{-\frac{8\pi^2}{g_{\rm YM}^2}\sum_j(b_j)^2}\period
\eeq
Here we again denoted $b^{(0)}_j$ by $b_j$ and used $\sum_s A_s=4\pi$.

To compute the correlator of the Wilson loops, we insert
\beq
\frac{1}{N^2}{\rm tr}\left(e^{\epsilon X_{1}}\right){\rm tr}\left(e^{\epsilon X_{2}}\right)\quad \mapsto \quad\frac{1}{N^2}\sum_{n,m}e^{\epsilon x_n^{(1)}+\epsilon x_m^{(2)}}\comma
\eeq
to the partition function \eqref{eq:partition2}. We then obtain
\beq
\begin{aligned}
\langle \mathcal{W}_1\mathcal{W}_2\rangle =&\frac{(2\pi)^{2N}(N!)^4}{Z}\int \left(\prod_{s=0}^{2}d^{N}b^{(s)}\right)\Delta (b^{(0)})\Delta (b^{(2)})e^{-\frac{2\pi}{g_{\rm YM}^2}\sum_{s=0}^{2}\sum_j \bar{A}_s (b_j^{(s)})^2}\\
&\times\sum_{n,m}\frac{\delta (b_n^{(0)}-b_n^{(1)}-i\epsilon)\delta (b_m^{(1)}-b_m^{(2)}-i\epsilon)}{N^2}\prod_{\substack{j\neq n\\k\neq m}}\delta (b_j^{(0)}-b_j^{(1)})\delta (b_k^{(1)}-b_k^{(2)})\period
\end{aligned}
\eeq
To proceed, we split the sum into two parts,
\beq\label{eq:splitsum}
\sum_{n,m}=\sum_{n\neq m}+\sum_{n=m}\comma
\eeq
integrate out $b_j^{(1,2)}$'s and recast the sums into integrals. The first term in \eqref{eq:splitsum} gives
\beq\label{eq:integraldouble}
\begin{aligned}
\langle\mathcal{W}_1\mathcal{W}_2\rangle_{n\neq m}=&\frac{(2\pi)^{2N}(N!)^4}{Z}\int d^{N} b \,\Delta^2 (b)\, e^{-\frac{8\pi^2}{g_{\rm YM}^2}\sum_j (b_j)^2}\sum_{n\neq m}\frac{e^{i A_1(b_n-\frac{i\epsilon}{2})}e^{i A_2(b_m-\frac{i\epsilon}{2})}}{N^2}\\
 &\times\frac{(b_n-b_m)^2}{(b_n-b_m)^2+\epsilon^2} \prod_{j\neq n}\frac{b_n-b_j-i\epsilon}{b_n-b_j}\prod_{k\neq m}\frac{b_m-b_k-i\epsilon}{b_m-b_k} \comma\\
 =&\frac{(2\pi)^{2N}(N!)^4}{Z}\int d^{N} b \,\Delta^2 (b) e^{-\frac{8\pi^2}{g_{\rm YM}^2}\sum_j (b_j)^2}\oint_{\mathcal{C}}\frac{du_1\,e^{i A_1(u_1-\frac{i\epsilon}{2})}}{8\pi^2g^2}\frac{du_2\,e^{i A_2(u_2-\frac{i\epsilon}{2})}}{8\pi^2g^2}\\
 &\times  \bar{\Delta} (u_1,u_2) \prod_{j}\frac{u_1-b_j-i\epsilon}{u_1-b_j}\prod_{k}\frac{u_2-b_k-i\epsilon}{u_2-b_k}\comma
\end{aligned}
\eeq
where the ``interaction term'' $\bDelta (u,v)$ is given by
\beq
\bDelta (u,v):=\frac{(u-v)^2}{(u-v)^2+\epsilon^2}\period
\eeq
Again \eqref{eq:integraldouble} can be interpreted as an expectation value in the Gaussian matrix model
\beq\label{eq:gaussdisc}
\langle\mathcal{W}_1\mathcal{W}_2\rangle_{n\neq m}=\left<\oint_{\mathcal{C}} \frac{du_1}{8\pi^2g^2}\frac{du_2}{8\pi^2g^2}\bDelta (u_1,u_2) f_{A_1}(u_1)f_{A_2}(u_2)\right>_M\period
\eeq 
Similarly, the second term in \eqref{eq:splitsum} gives\fn{The term \eqref{eq:gaussconn} resembles the ``bound-state'' contribution in the hexagon approach to the correlation functions \cite{Basso:2015zoa, Jiang:2016ulr}. See also the comments below \eqref{eq:genmultiply}.}
\beq\label{eq:gaussconn}
\begin{aligned}
\langle \mathcal{W}_1\mathcal{W}_2\rangle_{n=m}=&\frac{(2\pi)^{2N}(N!)^4}{Z}\int d^{N} b \,\Delta^2 (b)e^{-\frac{8\pi^2}{g_{\rm YM}^2}\sum_j (b_j)^2}\sum_n \frac{e^{i A_1(b_n- i\epsilon/2)}e^{i A_2(b_n-3i\epsilon/2)}}{N^2}\\
&\times \prod_{j\neq n}\frac{b_n-b_j-2i\epsilon}{b_n-b_j}\\
=&\frac{1}{2N}\left<\oint_{\mathcal{C}}\frac{du}{8\pi^2 g^2}f_{A_1}(u)f_{A_2}(u-i\epsilon)\right>_M\period
\end{aligned}
\eeq
Thus the sum of the two contributions \eqref{eq:gaussdisc}  and \eqref{eq:gaussconn} gives
\begin{align}
&\langle \mathcal{W}_1\mathcal{W}_2\rangle=\label{eq:twocortwo}\\
&\left<\oint_{\mathcal{C}} \frac{du_1}{8\pi^2g^2}\frac{du_2}{8\pi^2g^2}\,\,\bDelta (u_1,u_2) \,\,f_{A_1}(u_1)f_{A_2}(u_2)\right>_M+\frac{1}{2N}\left<\oint_{\mathcal{C}}\frac{du}{8\pi^2 g^2}f_{A_1}(u)f_{A_2}(u-i\epsilon)\right>_M\period\nn
\end{align}

\begin{figure}[t]
\centering
\includegraphics[clip,height=4.5cm]{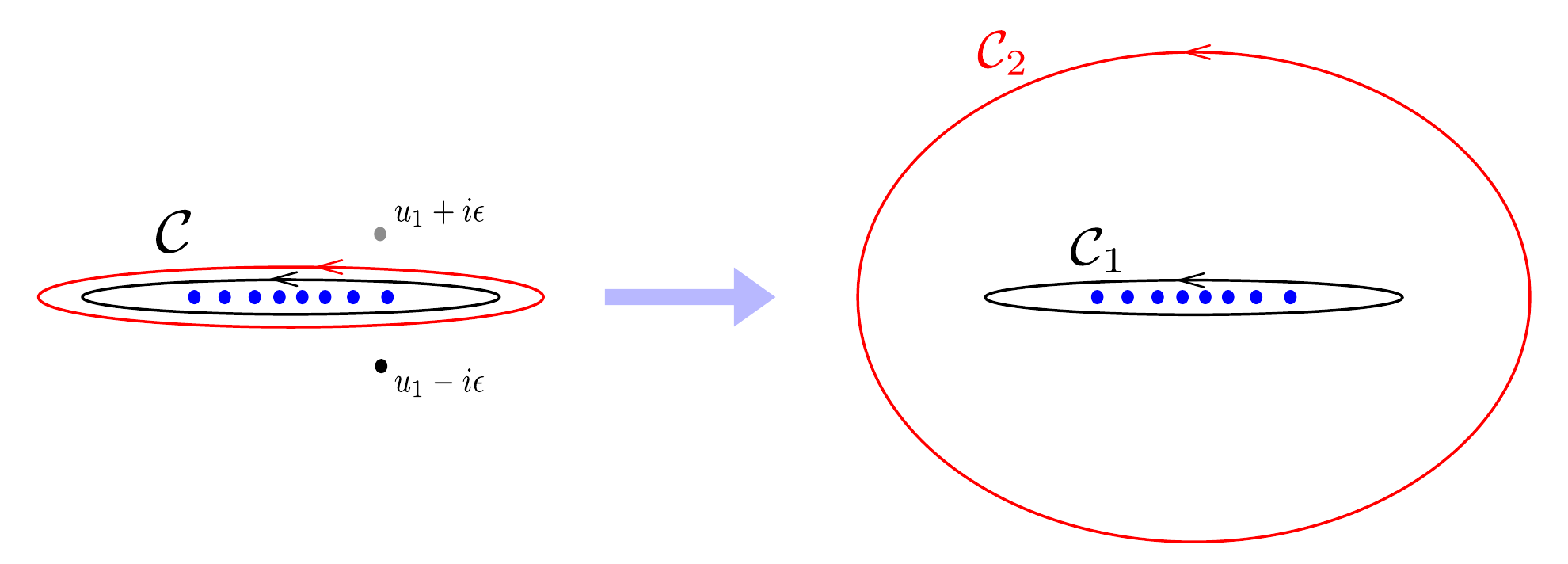}
\caption{Deformation of the integration contours. Left: The integration contours for $u_1$ (black) and $u_2$ (red) in \eqref{eq:twocortwo}. The two contours are on top of each other and encircle the eigenvalues of $M$ (the blue dots in the figure). Right: The integration contours in \eqref{eq:twocorrfinal}. When we deform the contour of $u_2$, the integral picks up a contribution from poles at $u_2=u_1\pm i\epsilon$ (the gray and black dots). The residue at $u_1+i\epsilon$ turns out to vanish while the residue at $u_1-i\epsilon$ cancels the second term in \eqref{eq:twocortwo}. } \label{fig:clustering}
\end{figure}

One can further simplify the expression by combining the two terms in \eqref{eq:twocortwo}: In the first term of \eqref{eq:twocortwo}, the integration contours of $u_{1,2}$ are on top of each other and encircle all the eigenvalues of the matrix model (see figure \ref{fig:clustering}). If one deforms the contour of $u_2$  so that the two contours are far separated, the integral picks up a contribution from poles in the interaction term 
\beq
\bar{\Delta}(u_1,u_2)=\frac{(u_1-u_2)^2}{(u_1-u_2)^2+\epsilon^2}\comma
\eeq
whose residues are
\beq
\begin{aligned}
(\text{Residue at $u_2=u_1\pm i\epsilon$})=\pm\frac{1}{2N}\left<\oint_{\mathcal{C}}\frac{du}{8\pi^2 g^2}f_{A_1}(u)f_{A_2}(u\pm i\epsilon)\right>_M
\end{aligned}
\eeq 
The residue at $u_2=u_1+i\epsilon$ turns out to vanish since the product $f_{A_1}(u)f_{A_2}(u+i\epsilon)$ is
\beq
f_{A_1}(u)f_{A_2}(u+i\epsilon)\propto \det \frac{u-M-i\epsilon}{u-M} \det \frac{u-M}{u-M+i\epsilon}=\det \frac{u-M-i\epsilon}{u-M+i\epsilon}\comma
\eeq
which is nonsingular inside the integration contour $\mathcal{C}$. On the other hand, the residue at $u_2=u_1-i\epsilon$ precisely cancels the second term in \eqref{eq:twocortwo}. We therefore arrive at the following simple expression for the correlator of two fundamental loops:
\beq\label{eq:twocorrfinal}
\langle \mathcal{W}_1\mathcal{W}_2\rangle =\left<\oint_{\mathcal{C}_1\prec \mathcal{C}_2}\frac{du_1}{8\pi^2 g^2}\frac{du_2}{8\pi^2 g^2}\bar{\Delta}(u_1,u_2)f_{A_1}(u_1)f_{A_2}(u_2)\right>_M\period
\eeq
The notation $\mathcal{C}_1\prec\mathcal{C}_2$ means that the contour $\mathcal{C}_1$ is inside the contour $\mathcal{C}_2$ and they are far separated from each other.

\begin{figure}[t]
\centering
\includegraphics[clip,height=3.7cm]{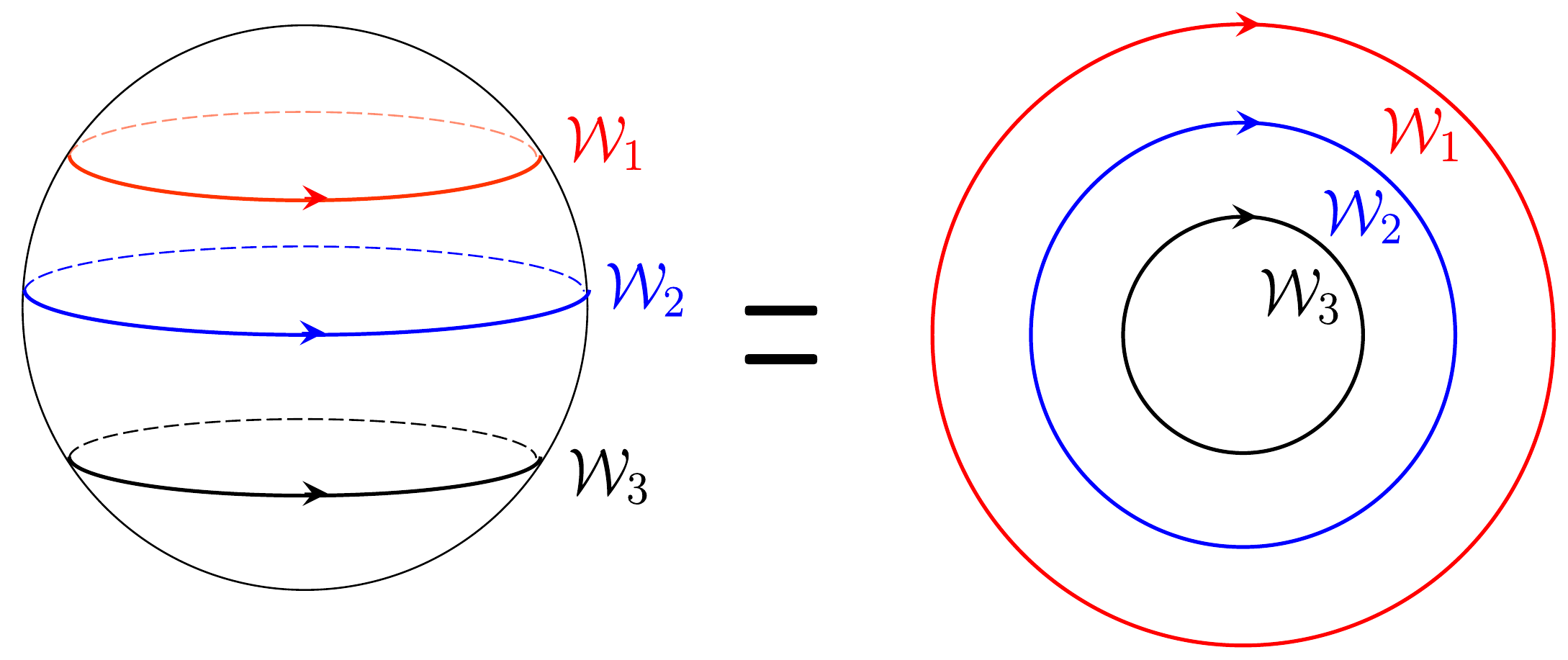}
\caption{Multiple $1/8$-BPS loops with the same orientations. The left figure is a configuration drawn on $S^2$ while the right figure is the same configuration viewed from the south pole and projected to a plane. We denote the area inside $\mathcal{W}_j$ (in the right figure) by $A_j$.} \label{fig:threeloop}
\end{figure}

Carrying out the same analysis for correlators of more than two Wilson loops of the same orientations (see figure \ref{fig:threeloop} for details of the setup), we obtain a simple generalization of \eqref{eq:twocorrfinal},
\beq\label{eq:genmultiply}
\left< \prod_{j=1}^{n}\mathcal{W}_j\right>=\left<\oint_{\mathcal{C}_1\prec\cdots \prec\mathcal{C}_n}\prod_{j=1}^{n}\frac{du_j f_{A_j}(u_j)}{8\pi^2 g^2}  \prod_{j<k}\bDelta (u_j,u_k)\right>_{M}\period
\eeq
Interestingly this integral resembles multiparticle integrals \cite{Basso:2015zoa,Jiang:2016ulr,Fleury:2016ykk,Eden:2018vug,Coronado:2018ypq,Bargheer:2019exp,Kostov:2019stn,Kostov:2019auq,Belitsky:2019fan,deLeeuw:2019qvz,Belitsky:2020qrm} in the hexagon approach to the correlation functions and we will use this connection in our upcoming paper \cite{Toappear} to analyze the defect CFT correlators on the higher-rank Wilson loop.
\section{Loop Equation and Intersecting Wilson Loops\label{sec:loopeq}}
We now discuss intersecting $1/8$ BPS Wilson loops. In section \ref{subsec:loop2dYM}, we review the derivation of the loop equation in 2d YM in \cite{Kazakov:1980zi}. We then apply it to the results in the previous section and derive integral representations for the intersecting BPS Wilson loops in section \ref{subsec:integralintersect}. 

Let us make a remark before we proceed: Below we consider the Wilson loops without the normalization factor $1/N$ since they are often more convenient for analyzing the loop equation. To distinguish them from a more standard definition \eqref{eq:relation}, we will put a tilde as
\beq
\tilde{\mathcal{W}}_{\mathcal{C}}:={\rm tr}{\rm P}e^{i \oint_{\mathcal{C}}A_{\mu}dy^{\mu}}\period
\eeq 

\subsection{Review of loop equation in 2d YM \label{subsec:loop2dYM}}
\begin{figure}[t]
\centering
\begin{minipage}{0.45\hsize}
\centering
\includegraphics[clip,height=3cm]{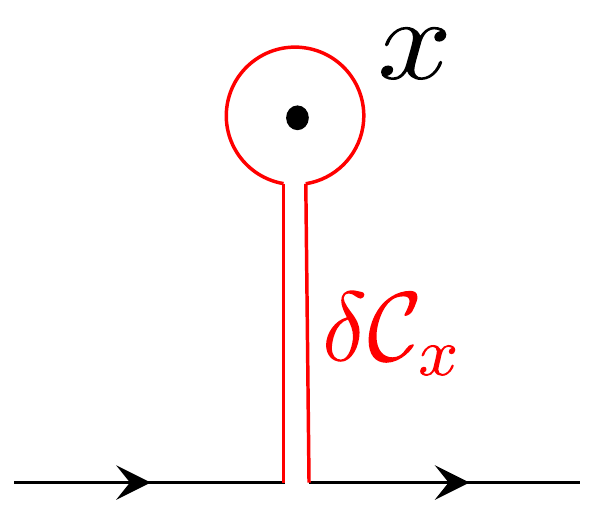}\\
$(a)$
\end{minipage}
\begin{minipage}{0.45\hsize}
\centering
\includegraphics[clip,height=3cm]{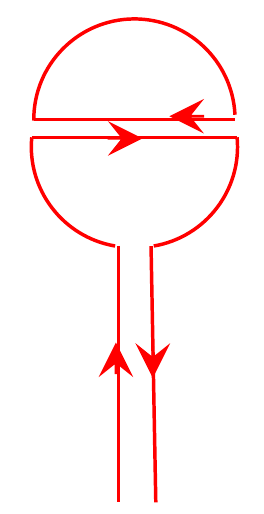}\\
$(b)$
\end{minipage}
\caption{$(a)$ Deformation of the contour $\delta\mathcal{C}_x$ denoted in red. We pick a point $x$ close to the Wilson loop, draw a small circle and connect it to the original contour. $(b)$ One-loop correction to $\langle {\tt Sym}\rangle$ coming from the gluon exchange. Here we used the double-line notation to make manifest the contraction of indices.} \label{fig:loopeqderivation}
\end{figure}

We consider an infinitesimal deformation of the contour $\mathcal{C}\to \mathcal{C}+\delta \mathcal{C}_x$. Specifically, we pick a point $x$ close to but {\it not} on top of the Wilson loop. We then add a small circle  around $x$ and connect it the original contour as depicted in figure \ref{fig:loopeqderivation}. This defines the {\it area derivative} 
\beq
\frac{\delta \langle \tilde{\mathcal{W}}_{\mathcal{C}}\rangle}{\delta \sigma^{\mu\nu}(x)}:=\lim_{\delta \sigma^{\mu\nu}\to 0}\frac{\langle \tilde{\mathcal{W}}_{\mathcal{C}+\delta\mathcal{C}_x}\rangle-\langle \tilde{\mathcal{W}}_{\mathcal{C}}\rangle}{\delta \sigma^{\mu\nu}}\period
\eeq
Here $\delta \sigma^{\mu\nu}$ is an {\it area tensor} of $\delta \mathcal{C}_x$, which is a product of the area $|\delta\sigma|$ and the {\it orientation tensor} $n^{\mu\nu}$:
\beq
\delta \sigma^{\mu\nu}:=|\delta \sigma| n^{\mu\nu}\period
\eeq
The orientation tensor is unit-normalized and parameterizes the orientation of $\delta \mathcal{C}_x$,
\beq
\frac{1}{2}n_{\mu\nu}n^{\mu\nu}=1\comma\qquad |\delta \sigma|=\frac{1}{2}n_{\mu\nu}\delta \sigma^{\mu\nu}\period
\eeq

We then expand the path-ordered exponential to get
\beq
\begin{aligned}
&\langle \tilde{\mathcal{W}}_{\mathcal{C}+\delta\mathcal{C}_x}\rangle-\langle \tilde{\mathcal{W}}_{\mathcal{C}}\rangle=\left< {\rm tr}{\rm P}\left[\left(e^{\oint_{\delta\mathcal{C}_x}iA_{\mu}dy^{\mu}}-1\right)e^{\oint_{\mathcal{C}}iA_{\mu}dy^{\mu}}\right]\right>\\
&=\left<{\rm tr}{\rm P}\left[\left(\oint_{\delta \mathcal{C}_x}iA_{\mu}dy^{\mu}-\frac{1}{2}\oint_{\substack{\delta \mathcal{C}_x \times \delta \mathcal{C}_x}}A_{\mu}(y_1)A_{\nu}(y_2)dy_1^{\mu}dy_2^{\nu}+\cdots \right) e^{\oint_{\mathcal{C}}iA_{\mu}dy^{\mu}}\right]\right>
\end{aligned}
\eeq
To proceed, we use the Stokes theorem to the first term and write
\beq\label{eq:stokes}
\oint_{\delta \mathcal{C}_x}i A_{\mu} dy^{\mu}=\frac{i \left(\del_{\mu}A_{\nu}(x)-\del_{\nu}A_{\mu}(x)\right)}{2}\delta \sigma^{\mu \nu}\period
\eeq
For the second term, we split it into the symmetric and the antisymmetric parts
\beq
\begin{aligned}
&-{\rm P}\left(\frac{1}{2}\oint_{\substack{\delta \mathcal{C}_x \times \delta \mathcal{C}_x}}A_{\mu}(y_1)A_{\nu}(y_2)dy_1^{\mu}dy_2^{\nu}\right)=\\
&\underbrace{-\frac{1}{2}\left(\oint_{\delta \mathcal{C}_x}A_{\mu}(x)dx^{\mu}\right)^2}_{\tt Sym} \underbrace{-\frac{1}{2}\oint_{\substack{\delta \mathcal{C}_x \times \delta \mathcal{C}_x\\y_1>y_2}}[A_{\mu}(y_1),A_{\nu}(y_2)]dy_1^{\mu}dy_2^{\nu}}_{\tt Asym}\period
\end{aligned}
\eeq
Here $y_1>y_2$ means that the integral is path-ordered and $y_2$ is always behind $y_1$. 
Since $\delta \mathcal{C}_x$ is an infinitesimal contour around $x$, the antisymmetric part can be approximated as
\beq\label{eq:loopanti}
\begin{aligned}
{\tt Asym}&\simeq -\frac{[A_{\mu}(x),A_{\nu}(x)]}{2}\oint_{\substack{\delta \mathcal{C}_x \times \delta \mathcal{C}_x\\y_1>y_2}}dy_{1}^{\mu}dy_{2}^{\nu}\\
&=\frac{[A_{\mu}(y),A_{\nu}(y)]}{2}\oint_{\delta\mathcal{C}_x}y_2^{\mu}dy_2^{\nu}=\frac{[A_{\mu}(x),A_{\nu}(x)]}{2} \delta \sigma^{\mu\nu}\period
\end{aligned}
\eeq
The sum of \eqref{eq:stokes} and \eqref{eq:loopanti} gives $i F_{\mu\nu}(y)/2$ with $F_{\mu\nu}=\del_{\mu}A_{\nu}-\del_{\mu}A_{\nu}-i[A_{\mu},A_{\nu}]$.

On the other hand, the symmetric part at one loop reads (see figure \ref{fig:loopeqderivation})
\beq\label{eq:symtree}
\begin{aligned}
\left.\langle {\tt Sym}\rangle\right|_{\text{1-loop}}&=- \frac{N}{2}\oint_{\delta C_{x}}dy_1^{\mu}\oint_{\delta C_{x}}dy_2^{\nu}G_{\mu\nu}(y_1-y_2)\times  {\bf 1}
\end{aligned}
\eeq
Here ${\bf 1}$ is the identity matrix for the color factors and $G_{\mu\nu}(x_1-x_2)$ is a propagator of gauge fields without color factors. In the axial gauge ($A_{1}=0$), it reads\fn{Note that our normalization for the coupling constant in 2d YM is different from \cite{Kazakov:1980zi}: $(g_{\rm 2d}^{\text{\cite{Kazakov:1980zi}}})^2=(g^{\rm ours}_{\rm 2d})^2/2$.}
\beq
\begin{aligned}
&G_{\mu\nu}(x-y)=-\delta^{2}_{\mu}\delta^{2}_{\nu}\frac{g_{\rm 2d}^2}{4}\delta (x^{2}-y^{2})|x^{1}-y^{1}|\period
\end{aligned}
\eeq
Plugging this into \eqref{eq:symtree}, we get
\beq\label{eq:symtree2}
\left.\langle {\tt Sym}\rangle\right|_{\text{1-loop}}=-\frac{g_{\rm 2d}^2N}{4}|\delta \sigma|\period
\eeq
A couple of comments are in order: First, the result \eqref{eq:symtree} receives higher loop corrections. However since $g_{\rm 2d}$ has mass dimension $1$, it follows from dimensional analysis that the loop corrections come with higher powers of $\delta\sigma^{\mu\nu}$ and therefore can be neglected in the small area limit $\delta \sigma^{\mu\nu}\to 0$. Second, although \eqref{eq:symtree2} was computed in a specific gauge, the result is actually gauge-invariant. This is because the tree-level\fn{As stated above, the higher-loop corrections can be neglected in the small area limit.} gauge transformation
\beq
A_{\mu}(y) \quad \mapsto \quad A_{\mu}(y)+\del_{\mu}\alpha (y)\comma
\eeq 
only changes \eqref{eq:symtree} by a term that integrates to zero:
\beq
\delta {\tt Sym}\propto \oint_{\delta \mathcal{C}_x} dy_1^{\mu}\del_{\mu}\alpha (y_1)\oint_{\delta \mathcal{C}_x} dy_2^{\nu}\del_{\nu}\alpha (y_2)=0\period
\eeq
 Third, the result \eqref{eq:symtree}  is for the U$(N)$ gauge group and the answer will be different for other gauge groups. It would be interesting to derive the loop equation for general gauge groups.
 
 Now, combining the symmetric and antisymmetric parts, we arrive at the following expression for the first-order variation:
 \beq\label{eq:firstorderarea}
 \begin{aligned}
 \frac{\delta \langle \tilde{\mathcal{W}}_{\mathcal{C}}\rangle}{\delta \sigma^{\mu\nu}(x)}=\left<{\rm tr}{\rm P}\left[i F_{\mu\nu}(x)e^{\oint_{\mathcal{C}}iA_{\mu}dy^{\mu}}\right]\right>-\frac{g_{\rm 2d}^2 N}{4}n_{\mu\nu}(x)\langle \tilde{\mathcal{W}}_{\mathcal{C}}\rangle\period
 \end{aligned}
 \eeq
Let us make some clarification on the second term in \eqref{eq:firstorderarea}: First, it contains the orientation tensor $n_{\mu\nu}$ and changes discontinuously across the contour. This property plays a crucial role in the derivation below. Second, it may appear to contradict a familar statement that the small deformation of the Wilson loop is simply given by the insertion of $F_{\mu\nu}$. This apparent contradiction comes from the difference of the regularizations: Normally we insert $F_{\mu\nu}$ on top of the Wilson loop and regularize divergent diagrams which contract $F_{\mu\nu}$ and the loop by the principal value prescription. On the other hand, $F_{\mu\nu}$ in \eqref{eq:firstorderarea}  is inserted slightly above or below the loop, which gives a different regularization. This alternative regularization produces a different answer, but one can show by the explicit computation\fn{In the holomorphic gauge $A_{\bar{z}}=0$, the contraction of $F_{\mu\nu}$ and the straight-line loop reads (see e.g.~\cite{Bonini:2015fng})
\beq
\int_{-\infty}^{\infty} dy\langle i \ast F_{z\bar{z}}(x) i A_{z} (y)\rangle =-\frac{g_{\rm 2d}^2 N}{4\pi}\int_{-\infty}^{\infty} dy \frac{i}{1+y^2}\frac{1+xy}{x-y}\period
\eeq
In the principal value prescription, this integral vanishes while if we shift $x$ slightly above or below by adding $\pm i\epsilon$, it gives $\pm g_{\rm 2d}^2 N/(4\pi)$, which are precisely canceled by the second term in \eqref{eq:firstorderarea}.
} that the sum of the two terms in \eqref{eq:firstorderarea} reproduces the result in the principal value prescription.  We thus have
\beq
 \frac{\delta \langle \tilde{\mathcal{W}}_{\mathcal{C}}\rangle}{\delta \sigma^{\mu\nu}(x)}=\left.\left<{\rm tr}{\rm P}\left[i F_{\mu\nu}(x)e^{\oint_{\mathcal{C}}iA_{\mu}dy^{\mu}}\right]\right>\right|_{\text{principal value}}\comma
\eeq
 which is consistent with the familiar statement.
 
\begin{figure}[t]
\centering
\includegraphics[clip,height=2cm]{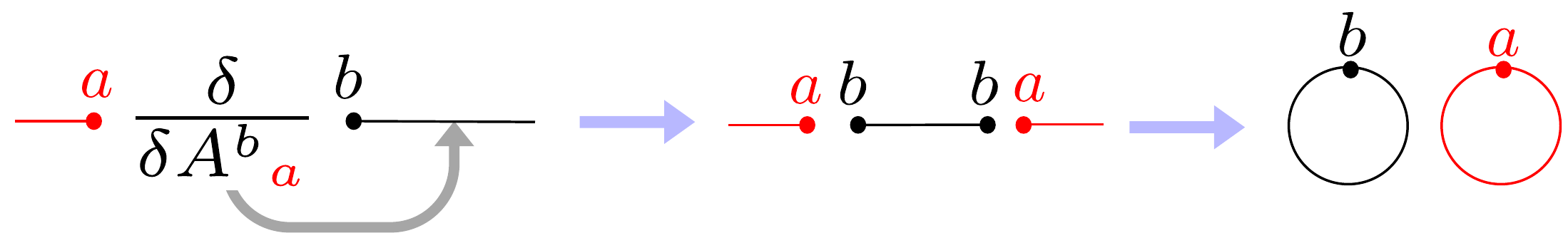}
\caption{The Ward identity on the Wilson loop. The insertion of $\nabla^{\mu}F_{\mu\nu}$ on the loop can  be replaced with the functional derivative $\delta/\delta A_{\nu}$. When it acts on other parts of the contour, it cuts the loop and reconnects it as shown in the figure.} \label{fig:Aonloop}
\end{figure}

 Now we derive the loop equation. First we differentiate  \eqref{eq:firstorderarea} and get
 \beq\label{eq:derloop1}
 \del^{\mu}\frac{\delta \langle \tilde{W}_{\mathcal{C}}\rangle}{\delta \sigma^{\mu\nu}(x)}=\left<{\rm tr}{\rm P}\left[i \nabla^{\mu}F_{\mu\nu}(x)e^{\oint_{\mathcal{C}}iA_{\mu}dy^{\mu}}\right]\right>-\frac{g_{\rm 2d}^2 N}{4}\del^{\mu}n_{\mu\nu}(x)\langle \tilde{\mathcal{W}}_{\mathcal{C}}\rangle\period
 \eeq
 To evaluate $\del^{\mu}n_{\mu\nu}$ that appears in the second term, it is useful to go back to the definition
 \beq\label{eq:looplaplacian}
 \del^{\mu}\frac{\delta \langle \tilde{W}_{\mathcal{C}}\rangle}{\delta \sigma^{\mu\nu}(x)}:= \lim_{\varepsilon\to 0}\frac{1}{\varepsilon}\left( \frac{\delta \langle \tilde{\mathcal{W}}_{\mathcal{C}}\rangle}{\delta \sigma^{\mu\nu}(x+\epsilon_{\mu}/2)}- \frac{\delta \langle \tilde{\mathcal{W}}_{\mathcal{C}}\rangle}{\delta \sigma^{\mu\nu}(x-\epsilon_{\mu}/2)}\right)\period
 \eeq
Here we take $x$ to be a point on the Wilson loop and $\epsilon_{\mu}$ to be a vector of length $\varepsilon$ in the $\mu$ direction; $ \left(\epsilon_{\mu}\right)^{\rho}:=\varepsilon\delta^{\rho}_{\mu}$.
 Since the points $x\pm \epsilon_{\mu}/2$ sit on different sides of the loop, we have $n_{\mu\nu}(x\pm \epsilon_{\mu}/2)=\pm 1$. Thus the derivative $\del^{\mu}n_{\mu\nu}$ picks up a contribution from a delta function
\beq\label{eq:deltanmu}
\del^{\mu}n_{\mu\nu}(x)=2\int_{I_{x}} dx^{\prime}_{\nu} \,\delta^{(2)}(x^{\prime}-x)\comma
\eeq 
where the contour $I_x$ is an infinitesimal one-dimensional segment which passes through $x$. Second we use the fact that $\nabla^{\mu}F_{\mu\nu}$ is proportional to the equation of motion\fn{$S_{\text{2d}}$ is the Euclidean action for 2d YM given by \eqref{eq:2dYMaction}.}
\beq
\nabla^{\mu}F_{\mu\nu}=-\frac{g_{\text{2d}}^2}{2}\frac{\delta S_{\text{2d}}}{\delta A^{\nu}}\period
\eeq 
We can then replace it inside the expectation value as
\beq
\Big< \left(\nabla^{\mu}F_{\mu\nu}(x)\right)^{a}{}_{b} \cdots \Big>\mapsto -\frac{g_{\text{2d}}^2}{2}\left< \frac{\delta }{\delta \left(A^{\nu}(x)\right)^{b}{}_{a}}\left(\cdots \right)\right>\period
\eeq
Here we wrote the gauge indices $a$ and $b$ explicitly for convenience. As a result we get (see also figure \ref{fig:Aonloop})
 \beq\label{eq:wardid}
 \begin{aligned}
 &\left<{\rm tr}{\rm P}\left[i \nabla^{\mu}F_{\mu\nu}(x)e^{\oint_{\mathcal{C}}iA_{\mu}dy^{\mu}}\right]\right> =-i\frac{g_{\rm 2d}^2}{2}\left<\frac{\delta}{\delta \left(A^{\nu}(x)\right)^{b}{}_{a}}\left[{\rm P}e^{\oint_{\mathcal{C}}i A_{\mu}dy^{\mu}}\right]^{b}{}_{a}\right>\\
 &=\frac{g_{\rm 2d}^2}{2}\oint_{\mathcal{C}} dx^{\prime}_{\nu}\,\,\delta^{(2)}(x^{\prime}-x) \left<{\rm tr}{\rm P}\left[e^{\oint_{x}^{x^{\prime}}i A_{\mu}dy^{\mu}}\right]{\rm tr}{\rm P}\left[e^{\oint_{x^{\prime}}^{x}i A_{\mu}dy^{\mu}}\right]\right>\period
\end{aligned}
 \eeq

 The integral \eqref{eq:wardid} is nonzero even for nonintersecting Wilson loops since it receives a contribution from a ``self contraction'', namely a contribution from a trivial coincidence point $x^{\prime}=x$ on the loop. It can be evaluated as
 \beq
 \begin{aligned}
\left(\text{self-contraction}\right)&=\frac{g_{\rm 2d}^2}{2}\int_{I_x}dx^{\prime}_{\nu}\delta^{(2)}(x^{\prime}-x)\left<{\rm tr}{\rm P}\left[e^{\oint_{\mathcal{C}}i A_{\mu}dy^{\mu}}\right]{\rm tr}\left[{\bf 1}\right]\right>\\
&=\frac{g_{\rm 2d}^2N}{2}\int_{I_x}dx^{\prime}_{\nu}\,\delta^{(2)}(x^{\prime}-x)\left<{\rm tr}{\rm P}\left[e^{\oint_{\mathcal{C}}i A_{\mu}dy^{\mu}}\right]\right>\period
\end{aligned}
 \eeq 
 As can be seen from \eqref{eq:deltanmu}, this turns out to cancel the second term in \eqref{eq:derloop1}. Thus, when the two terms in \eqref{eq:derloop1} are combined, we are left with contributions only from intersecting points and obtain the loop equation
 \beq\label{eq:loopeq2dYM}
 \del^{\mu}\frac{\delta \langle \tilde{W}_{\mathcal{C}}\rangle}{\delta \sigma^{\mu\nu}(x)}=\frac{g_{\rm 2d}^2}{2}{{\rm -}\!\!\!\!\!\!\int_{\mathcal{C}}} dx^{\prime}_{\nu}\,\,\delta^{(2)}(x^{\prime}-x) \left<{\rm tr}{\rm P}\left[e^{\oint_{x}^{x^{\prime}}i A_{\mu}dy^{\mu}}\right]{\rm tr}{\rm P}\left[e^{\oint_{x^{\prime}}^{x}i A_{\mu}dy^{\mu}}\right]\right>\comma
 \eeq
 where the symbol ${\rm -}\!\!\!\!\!\int_{\mathcal{C}}$ means that we remove an infinitesimal segment around $x$ when we perform the integral. For the actual application, it is useful to choose $\nu$ to be the direction of the Wilson loop near the intersection and $\mu$ to be the direction orthogonal to it. In addition, we integrate \eqref{eq:loopeq2dYM} along a small path in the $\mu$ direction. For the left hand side, this is basically equivalent to going back to the definition \eqref{eq:looplaplacian} and removing $1/\varepsilon$, which can be pictorially represented as\fn{In the last equality, we used $\delta \sigma^{\mu\nu}(x+\epsilon_{\mu}/2)=-\delta \sigma^{\mu\nu}(x-\epsilon_{\mu}/2)=|\delta \sigma|$.}
 \begin{align}
 &\hat{L}_{x}\langle \tilde{\mathcal{W}}_{\mathcal{C}}\rangle:=\lim_{\varepsilon\to 0}\frac{\delta \langle \tilde{\mathcal{W}}_{\mathcal{C}}\rangle}{\delta \sigma^{\mu\nu}(x+\epsilon_{\mu}/2)}-\frac{\delta \langle \tilde{\mathcal{W}}_{\mathcal{C}}\rangle}{\delta \sigma^{\mu\nu}(x-\epsilon_{\mu}/2)}\nn\\
 &=\frac{1}{\delta \sigma^{\mu\nu}(x+\frac{\epsilon_{\mu}}{2})}\left(\raisebox{-0.7cm}{\includegraphics[clip,height=1.5cm]{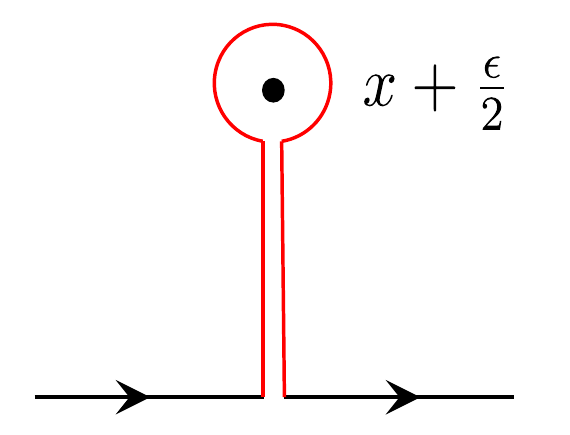}}-\raisebox{-0.7cm}{\includegraphics[clip,height=1.5cm]{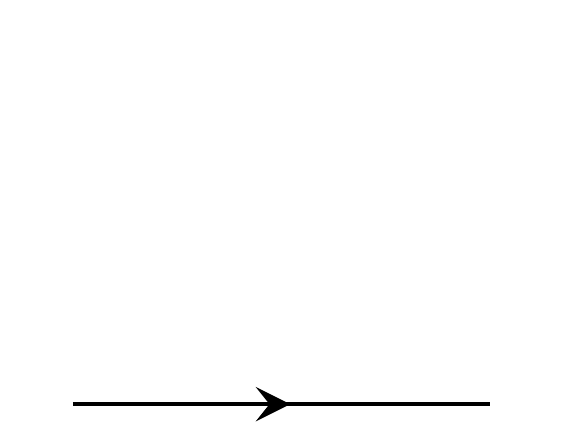}}\right)-\frac{1}{\delta \sigma^{\mu\nu}(x-\frac{\epsilon_{\mu}}{2})}\left(\raisebox{-0.7cm}{\includegraphics[clip,height=1.5cm]{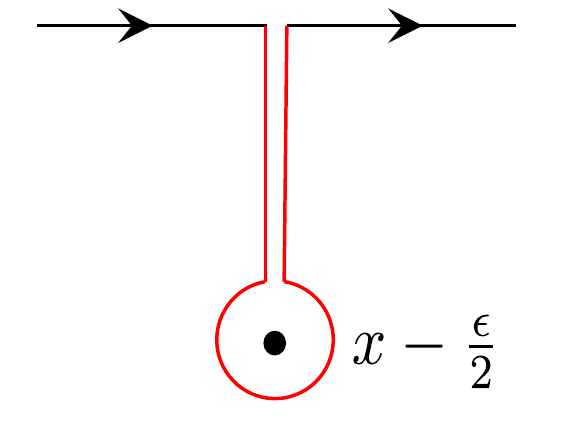}}-\raisebox{-0.7cm}{\includegraphics[clip,height=1.5cm]{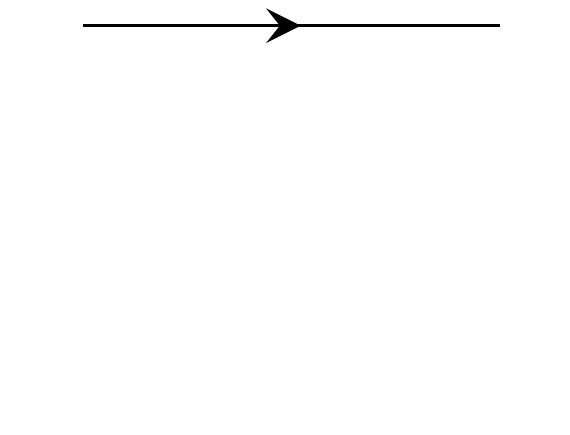}}\right)\nn\\
 &=\frac{1}{|\delta \sigma|}\left(\raisebox{-0.7cm}{\includegraphics[clip,height=1.5cm]{def1.pdf}}+\raisebox{-0.7cm}{\includegraphics[clip,height=1.5cm]{def3.pdf}}-2\raisebox{-0.7cm}{\includegraphics[clip,height=1.5cm]{def2.pdf}}\right)\period\label{eq:deflooplap}
 \end{align}
 We then get
 \beq\label{eq:loopeq2dYMfinal}
\hat{L}_x\langle \tilde{\mathcal{W}}_{\mathcal{C}}\rangle=\frac{g_{\rm 2d}^2}{2}\left<\tilde{\mathcal{W}}_{\mathcal{C}_1}\tilde{\mathcal{W}}_{\mathcal{C}_2}\right>\comma
 \eeq
 where $\mathcal{C}_{1,2}$ are the two contours obtained by reconnecting the loop $\mathcal{C}$ at the point $x$. Here we assumed that only two lines intersect at the point $x$. When more than two lines intersect, the right hand side of \eqref{eq:loopeq2dYMfinal} is replaced by a sum of all possible reconnections.

\begin{figure}[t]
\centering
\includegraphics[clip,height=2.5cm]{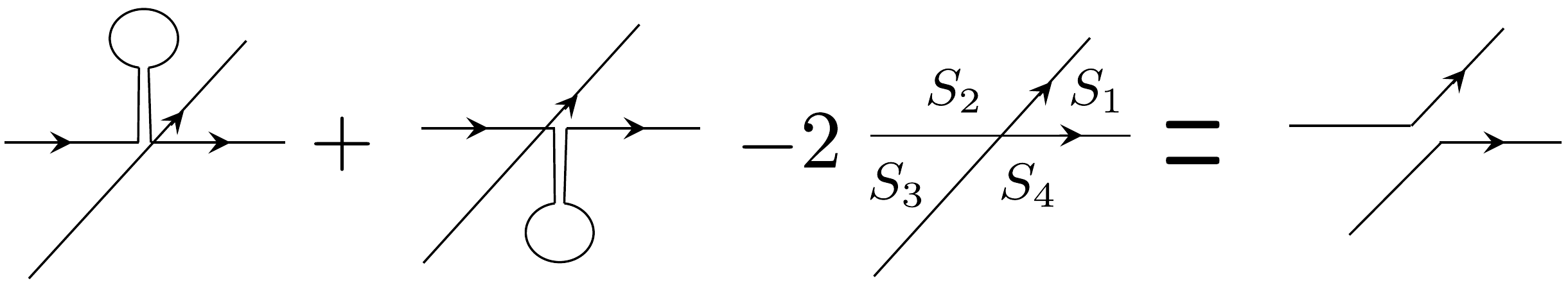}
\caption{The area derivatives and the loop laplacian $\hat{L}_x$. The left hand side is the definition of $\hat{L}_x$ which can be easily translated into area derivatives. (Adapted from \cite{Kazakov:1980zi}.)} \label{fig:areader}
\end{figure}

 Using the fact that the Wilson loops in 2d YM only depend on the area, we can convert \eqref{eq:loopeq2dYMfinal} to area derivatives (see figure \ref{fig:areader}),
 \beq\label{eq:intloop}
\hat{L}_x\langle \tilde{\mathcal{W}}_{\mathcal{C}}\rangle= \left(\del_{S_1}+\del_{S_3}-\del_{S_2}-\del_{S_4}\right)\langle \tilde{\mathcal{W}}_{\mathcal{C}}\rangle=\frac{g_{\rm 2d}^2}{2}\left<\tilde{\mathcal{W}}_{\mathcal{C}_1}\tilde{\mathcal{W}}_{\mathcal{C}_2}\right>\period
 \eeq
This formula can be applied straightforwardly when the loop is on $R^2$. For loops on $S^2$, not all the areas ($S_i$'s) are independent. In such a case, we go back to the pictorial definition of $\hat{L}_x$ given in \eqref{eq:deflooplap} and translate it into area derivatives as we see in the next subsection.

\subsection{Intersecting BPS loops in $\mathcal{N}=4$ SYM\label{subsec:integralintersect}}
We now apply the loop equation \eqref{eq:intloop} and compute the expectation value of intersecting $1/8$-BPS loops in $\mathcal{N}=4$ SYM.
\subsubsection{Figure eight loop\label{subsubsec:figureeight}} 
\begin{figure}[t]
\centering
\includegraphics[clip,height=2.5cm]{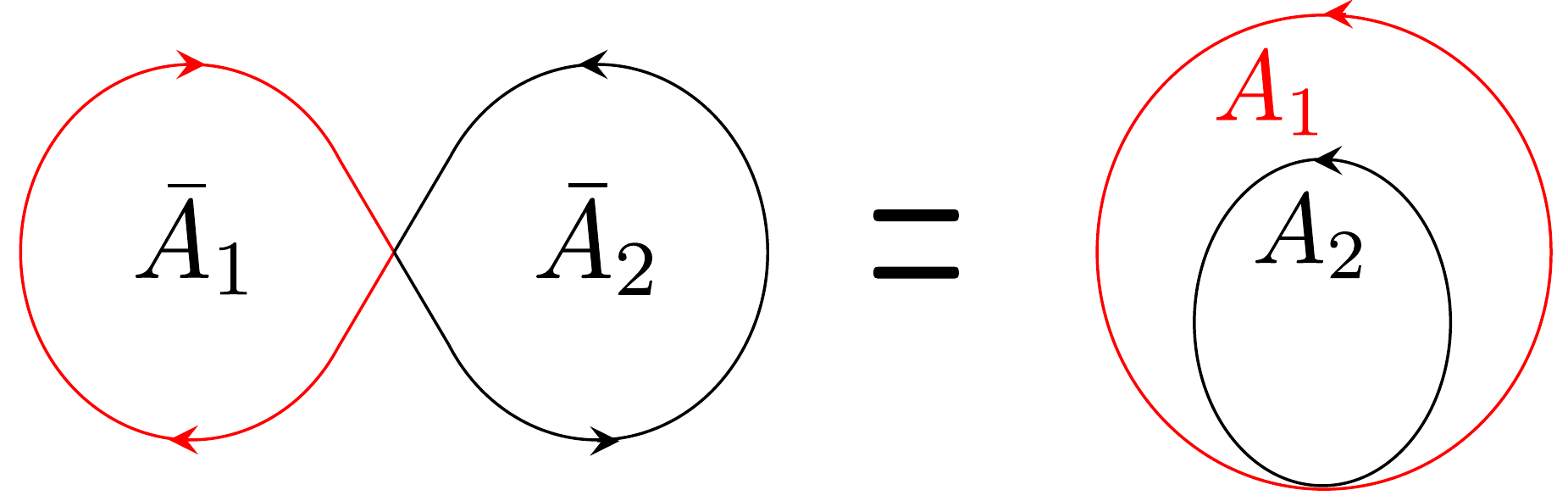}
\caption{The figure eight loop and the one-intersection loop. On $S^2$, one can continuously deform one to the other. In the right figure, the areas of the outer and inner circles are denoted by $A_{1,2}$ respectively. The areas are related by $A_1=4\pi-\bar{A}_1$ and $A_2=\bar{A}_2$.} \label{fig:eightandloop}
\end{figure}

The simplest intersecting loop is the ``figure eight'' loop.  Since the figure eight loop on $S^2$ is equivalent to a one-intersection loop (see figure \ref{fig:eightandloop}), we start with the latter, and the 
result for the figure-eight can be simply obtained by relating the areas as in 
figure \ref{fig:eightandloop}.

Under the action of $\hat{L}_x$, the one-intersection loop $\tilde{\mathcal{W}}_{A_1,A_2}$ reconnects to a product of two disconnected loops $\mathcal{W}_{1,2}$ with areas $A_{1,2}$. Using \eqref{eq:deflooplap}, this can be translated into area derivatives as follows\fn{To derive this, one simply needs to note that the first term \eqref{eq:deflooplap} decreases $A_2$ while the second term increases $A_1$ when applied to the one-intersection loop.}:
\beq
(\del_{A_1}-\del_{A_2})\langle \tilde{\mathcal{W}}_{A_1,A_2}\rangle=-\frac{4\pi g^2}{N} \langle\tilde{\mathcal{W}}_{1} \tilde{\mathcal{W}}_{2}\rangle\period
\eeq
Here we rewrote $g_{\rm 2d}$ in terms of the coupling in $\mathcal{N}=4$ SYM using \eqref{eq:2dto4d} and \eqref{eq:integrabilitycoupling}: $g_{\rm 2d}^2=-g_{\rm YM}^{2}/(2\pi)=-8\pi g^2/N$. Dividing both sides by $N$, we get the equation for the normalizaed expectation value
\beq
(\del_{A_1}-\del_{A_2})\langle \mathcal{W}_{A_1,A_2}\rangle=-4\pi g\langle\mathcal{W}_{1} \mathcal{W}_{2}\rangle\period
\eeq
This can be solved by replacing the right hand side with the integrals \eqref{eq:twocortwo}. The result reads
\beq\label{eq:singleintersectingresult}
\begin{aligned}
&\langle \mathcal{W}_{A_1,A_2}\rangle=\\
&\left<4\pi g^2 i\oint \frac{du_1}{8\pi^2 g^2}\frac{du_2}{8\pi^2 g^2} \bar{\Delta} (u_1,u_2) \frac{f_{A_1}(u_1)f_{A_2}(u_2)}{u_1-u_2}+\frac{1}{2}\oint \frac{du}{8\pi^2 g^2}f_{A_1}(u)f_{A_2}(u-i\epsilon)\right>_M\period
\end{aligned}
\eeq 
Alternatively, one can use \eqref{eq:twocorrfinal} to write
\beq\label{eq:contourseparatesingle}
\langle \mathcal{W}_{A_1,A_2}\rangle=\left<4\pi g^2 i\oint_{\mathcal{C}_1\prec\mathcal{C}_2} \frac{du_1}{8\pi^2 g^2}\frac{du_2}{8\pi^2 g^2} \bar{\Delta} (u_1,u_2) \frac{f_{A_1}(u_1)f_{A_2}(u_2)}{u_1-u_2}\right>_M\period
\eeq
Translating this to the figure eight loop using the relation (see figure \ref{fig:eightandloop}),
\beq\label{eq:arearel}
A_1=4\pi -\bar{A}_1\qquad A_2=\bar{A}_2\comma
\eeq 
 we obtain \eqref{eq:FigureEight} in the introduction. A simple consistency check of (\ref{eq:singleintersectingresult}) is to consider the special case $A_2=A_1$, which yields a doubly-wound fundamental Wilson loop. For $A_2=A_1$, the double-integral in (\ref{eq:singleintersectingresult}) vanishes due to antisymmetry of the integrand, and the single integral reproduces the expected result for the doubly-wound loop, which is given by (\ref{eq:singleresult}) with $g_{\rm YM}\rightarrow 2 g_{\rm YM}$.\footnote{For a $k$-wound fundamental loop, it is easy to see from the Gaussian matrix model (\ref{GaussMat}) that 
$$
\langle W_{\rm k-wound}\rangle = \langle W_{\rm single}\rangle|_{g_{\rm YM}\rightarrow k g_{\rm YM}}.
$$} 

\paragraph{Large N} In the large $N$ limit, one can evaluate the integral \eqref{eq:contourseparatesingle} explicitly. To do so, we first replace the expectation values of $f_A$'s with their large $N$ results,
\beq\label{eq:ftofrakf}
\begin{aligned}
&\left< \prod_{k}f_{A_k}(u_k)\right>_M=\left< \prod_{k}e^{iA_k(u_k-\frac{i\epsilon}{2})}\frac{\det \left(u_k-M-i\epsilon\right)}{\det \left(u_k-M\right)}\right>_M\overset{N\to \infty}{=}\prod_k \mathfrak{f}_{A_k}(u_k)\comma
\end{aligned}
\eeq
where $\mathfrak{f}_A$ is given by
\beq\label{eq:defoffrakfuseless}
\mathfrak{f}_{A}(u):=e^{iA u-4\pi i g^2 G(u)}\comma
\eeq
and $G(u)$ is the large $N$ resolvent
\beq
G(u):=\lim_{N\to \infty}\frac{1}{N}\left< {\rm tr}\left(\frac{1}{u-M}\right)\right>_M=\frac{1}{2g^2}\left(u-\sqrt{u^2-4g^2}\right)\period
\eeq
We then get
\beq\label{eq:largeNfigure8inte}
\begin{aligned}
&\langle \mathcal{W}_{A_1,A_2}\rangle\overset{N\to\infty}{=}4\pi g^2 i\oint_{\mathcal{C}_1\prec\mathcal{C}_2} \frac{du_1}{8\pi^2 g^2}\frac{du_2}{8\pi^2 g^2}   \frac{\mathfrak{f}_{A_1}(u_1)\mathfrak{f}_{A_2}(u_2)}{u_1-u_2}\period
\end{aligned}
\eeq
Here we replaced $\bar{\Delta}/(u_1-u_2)$ with its large $N$ limit, $1/(u_1-u_2)$.
This coincides with the expression obtained in \cite{Daul:1993xz} for the figure eight loop at large $N$ (after the redefinition of the areas \eqref{eq:arearel}). Below we evaluate this integral explicitly in terms of the Bessel functions.

To proceed, we introduce the Zhukowski variable\fn{As shown in \cite{Giombi:2018qox,Giombi:2018hsx}, the expressions obtained from localization often coincide with the ones from integrability when expressed in terms of $x$. This suggests that our integration variable $u$ is related to the rapidity of magnons, defined by $u_{\rm rapidity}=g (x+1/x)$, by $u_{\rm rapidity}=\sqrt{4g^2-u^2}$..}
\beq
u=-ig \left(x-\frac{1}{x}\right)\comma
\eeq
and express $\mathfrak{f}_{A}$ as
\beq\label{eq:defoffrakfusefull}
\mathfrak{f}_{A}(x)=e^{2g\pi (x+1/x)}e^{2g a (x-1/x)}\comma\qquad\qquad  a:=\frac{A-2\pi}{2}\period
\eeq
Substituting this into \eqref{eq:largeNfigure8inte}, we get
\beq\label{eq:forsaddlepoint}
\begin{aligned}
\langle \mathcal{W}_{A_1,A_2}\rangle\overset{N\to\infty}{=}-\frac{1}{4\pi g}\oint_{\mathcal{C}_1\prec\mathcal{C}_2} \frac{dx_1(x_1+x_1^{-1})}{2\pi i x_1}\frac{dx_2(x_2+x_2^{-1})}{2\pi i x_2}   \frac{\mathfrak{f}_{A_1}(x_1)\mathfrak{f}_{A_2}(x_2)}{(x_1-x_2)(1+1/x_1x_2)}\period
\end{aligned}
\eeq
We then expand $1/(x_1-x_2)(1+1/x_1x_2)$, close the contour $\mathcal{C}_1$ at the origin, and then close the contour $\mathcal{C}_2$ at infinity using the formula \cite{Gromov:2013qga,Giombi:2018qox}
\beq
\rho_a^{k}\,I_k(4\pi g_{a})=\int \frac{dx}{2\pi i x^{k+1}}e^{2\pi g (x+1/x)}e^{2g a(x-1/x)}\comma
\eeq
with
\beq
g_a :=g\sqrt{1-\frac{a^2}{\pi^2}}\comma\qquad\qquad  \rho_a :=\sqrt{\frac{\pi+a}{\pi-a}}\period
\eeq

As a result, we arrive at the following expression for the one-intersection loop,
\beq\label{eq:singleintersectionfinal}
\begin{aligned}
\!\langle \mathcal{W}_{A_1,A_2}\rangle\!\overset{N\to\infty}{=}\!\frac{\mathcal{I}_{0}^{a_1}\mathcal{I}_{1}^{a_2}}{2\pi g_{a_2}}+\sum_{k=1}^{\infty}\frac{\rho_{a_1}^{-k}\mathcal{I}_{k}^{a_1}}{4\pi g}\left[\left(\rho_{a_2}^{k+1}+\frac{(-1)^{k}}{\rho_{a_2}^{k+1}}\right)\mathcal{I}_{k+1}^{a_2}+\left(\rho_{a_2}^{k-1}+\frac{(-1)^{k}}{\rho_{a_2}^{k-1}}\right)\mathcal{I}_{k-1}^{a_2}\right],
\end{aligned}
\eeq
where we used a shorthand notation
\beq
\mathcal{I}_{k}^{a}:=I_{k}(4\pi g_a)\period
\eeq
Using the relation \eqref{eq:arearel}, one can translate this to the result for the figure-eight loop,
\begin{align}\label{eq:figureeightfinal}
\begin{aligned}
&\langle \mathcal{W}_{\text{figure-eight}}\rangle\overset{N\to\infty}{=}\\
&\frac{\mathcal{I}_{0}^{\bar{a}_1}\mathcal{I}_{1}^{\bar{a}_2}}{2\pi g_{\bar{a}_2}}+\sum_{k=1}^{\infty}\frac{\rho_{\bar{a}_1}^{k}\mathcal{I}_{k}^{\bar{a}_1}}{4\pi g}\left[\left(\rho_{\bar{a}_2}^{k+1}+\frac{(-1)^{k}}{\rho_{\bar{a}_2}^{k+1}}\right)\mathcal{I}_{k+1}^{\bar{a}_2}+\left(\rho_{\bar{a}_2}^{k-1}+\frac{(-1)^{k}}{\rho_{\bar{a}_2}^{k-1}}\right)\mathcal{I}_{k-1}^{\bar{a}_2}\right]\period
\end{aligned}
\end{align}

Two remarks are in order. First, the expression \eqref{eq:figureeightfinal} is not manifestly symmetric under $\bar{A}_1\leftrightarrow \bar{A}_2$. One can bring it into a symmetric form using the identities for the infinite sum of the Bessel functions. See Appendix \ref{ap:besselsum}. Second, the infinite sum in \eqref{eq:figureeightfinal} is convergent at finite coupling owing to the large $k$ behavior of $I_k(z)$
\beq
I_{k}(z)\overset{k\to\infty}{\sim}\frac{(z/2)^{k}}{k!}\comma\qquad\qquad z~\text{ fixed}\period
\eeq
 However, there is a subtlety in the strong-coupling expansion: The asymptotic expansion of the modified Bessel function reads
\beq
I_{k}(z)=\frac{e^{z}}{\sqrt{2\pi x}}\left(1+\frac{1-4k^2}{8 z}+\frac{9-40k^2 +16k^4}{128 x^2}+\cdots\right)\comma
\eeq
and it gives a polynomial in $k$ at each order. This leads to a divergence of the infinite sum when $\rho_{\bar{a}_1}>\rho_{\bar{a}_2}$ (i.e.~$\bar{A}_1>\bar{A}_2$). A more reliable approach at strong coupling is to perform the saddle-point analysis to the integral \eqref{eq:forsaddlepoint} as we see below.

\paragraph{Weak coupling expansion} Here we present the weak-coupling expansion (up to two loops) of the figure eight loop obtained from \eqref{eq:figureeightfinal}: 
\begin{align}\label{eq:figureeightweak}
\begin{aligned}
&\langle \mathcal{W}_{\text{figure-eight}}\rangle\overset{N\to\infty}{=}1+\frac{g^2}{2}\left[-(\bar{A}_1-\bar{A}_2)^2+4\pi(\bar{A}_1+\bar{A}_2)\right]\\
&\qquad +\frac{g^4}{12}\left[(\bar{A}_1-\bar{A}_2)^4-8\pi (\bar{A}_1^3+\bar{A}_2^{3})+16\pi^2 (\bar{A}_1^2+4 \bar{A}_1 \bar{A}_2 +\bar{A}_2^2)\right]+O(g^{6})\comma
\end{aligned}
\end{align}
As expected, the result is symmetric under the exchange $\bar{A}_{1}\leftrightarrow\bar{A}_2$ and reduces to the one for a single Wilson loop when $\bar{A}_2=0$. The result for the one-intersection loop can be obtained by the replacement $\bar{A}_1=4\pi -A_1$ and $\bar{A}_2=A_2$.

\paragraph{Strong coupling expansion} To perform the strong coupling expansion of the figure eight loop,  we do the saddle point analysis to the integral \eqref{eq:largeNfigure8inte} with the replacement \eqref{eq:arearel}. For definiteness, we assume\fn{Since the result \eqref{eq:figureeightfinal} is symmetric under the exchange of $\bar{A}_{1,2}$, the result for the other case $\bar{A}_1\geq \bar{A}_2$ can be obtained by a simple replacement $\bar{A}_{1}\leftrightarrow \bar{A}_2$.} $\bar{A}_1<\bar{A}_2$ below.

The saddle points are determined purely by the exponent of $\mathfrak{f}_{A}$'s and we find two saddle points for each variable
\beq
x_{1}=\pm \rho_{\bar{a}_1} \comma\qquad x_{2}=\pm 1/\rho_{\bar{a}_{2}}\period
\eeq
Owing to the geometrical constraint $\bar{A}_1+\bar{A}_2\leq 4\pi$ and the assumption $\bar{A}_{1}<\bar{A}_2$, we have $|\rho_{\bar{a}_1}|<|1/\rho_{\bar{a}_2}|$. Thus, the saddle points can be reached by deforming the original contours $\mathcal{C}_1\prec\mathcal{C}_2$ without making the contours pass through\fn{When the contours pass through each other, the saddle point approximation can potentially fail owing to the term $1/(x_1-x_2)(1+1/x_1x_2)$ in the integrand.} each other. Among those four saddle points, the ones with plus signs are always dominant. Expanding the integral around the saddle point using the coordinates $x_1=i t_1 +\rho_{\bar{a}_1}$ and $x_2=it_2+1/\rho_{\bar{a}_2}$, we get
\begin{align}
\langle \mathcal{W}_{\text{figure-eight}}\rangle&\overset{N\to\infty}{=}\underbrace{\frac{-e^{4\pi (g_{\bar{a}_1}+g_{\bar{a}_2})}}{4\pi g(\pi+\bar{a}_1)(\pi-\bar{a}_2)(\rho_{\bar{a}_1}-\frac{1}{\rho_{\bar{a}_2}})(1+\frac{\rho_{\bar{a}_1}}{\rho_{\bar{a}_2}})}}_{{\tt saddle}}\underbrace{\int dt_1 \,e^{-\frac{2\pi g_{\bar{a}_1}}{\rho_{\bar{a}_1}^2} t_1^2} \int dt_2 \,e^{-2\pi g_{\bar{a}_2}\rho_{\bar{a}_2}^2 t_2^2}}_{\text{\tt 1-loop}}\nn\\
&=-\frac{e^{4\pi (g_{\bar{a}_1}+g_{\bar{a}_2})}}{16\pi^2 (g_{\bar{a}_1}\bar{a}_2 +g_{\bar{a}_2} \bar{a}_1)\sqrt{g_{\bar{a}_1}g_{\bar{a}_2}}}+\cdots\period
\end{align}
Performing similar analysis to other saddle points, we get a sum
\beq\label{eq:strongcouplingfinal}
\begin{aligned}
&\langle \mathcal{W}_{\text{figure-eight}}\rangle\overset{N\to\infty}{=}\\
&\frac{-1}{16\pi^2 \sqrt{g_{\bar{a}_1}g_{\bar{a}_2}}}\left[\frac{e^{4\pi (g_{\bar{a}_1}+g_{\bar{a}_2})}-e^{-4\pi (g_{\bar{a}_1}+g_{\bar{a}_2})}}{g_{\bar{a}_1}\bar{a}_2+g_{\bar{a}_2}\bar{a}_1}-\frac{e^{4\pi (-g_{\bar{a}_1}+g_{\bar{a}_2})}-e^{4\pi (g_{\bar{a}_1}-g_{\bar{a}_2})}}{g_{\bar{a}_1}\bar{a}_2-g_{\bar{a}_2}\bar{a}_1}\right]\period
\end{aligned}
\eeq

\begin{figure}[t]
\centering
\begin{minipage}{0.45\hsize}
\centering
\includegraphics[clip,height=2.5cm]{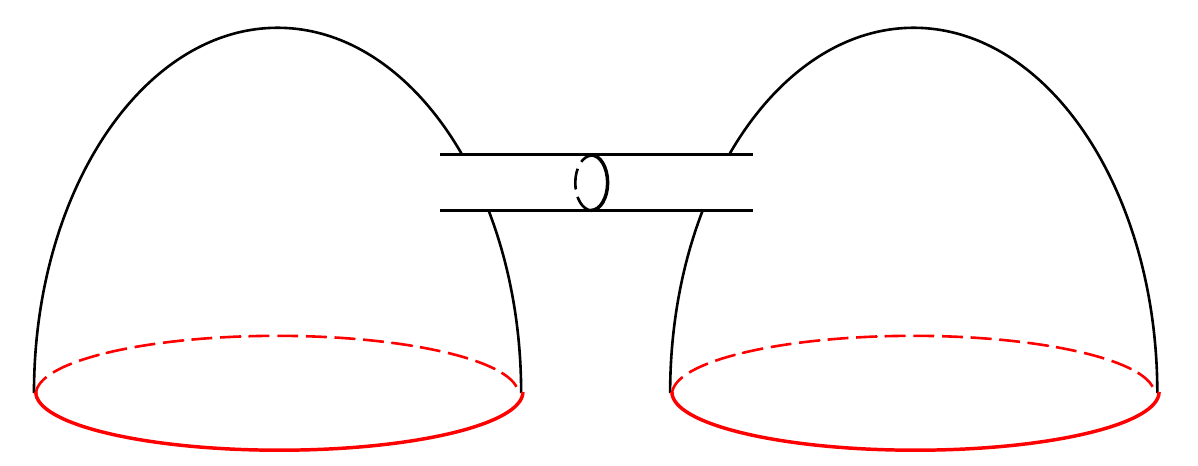} \\
$(a)$
\end{minipage}
\begin{minipage}{0.45\hsize}
\centering
\includegraphics[clip,height=2.5cm]{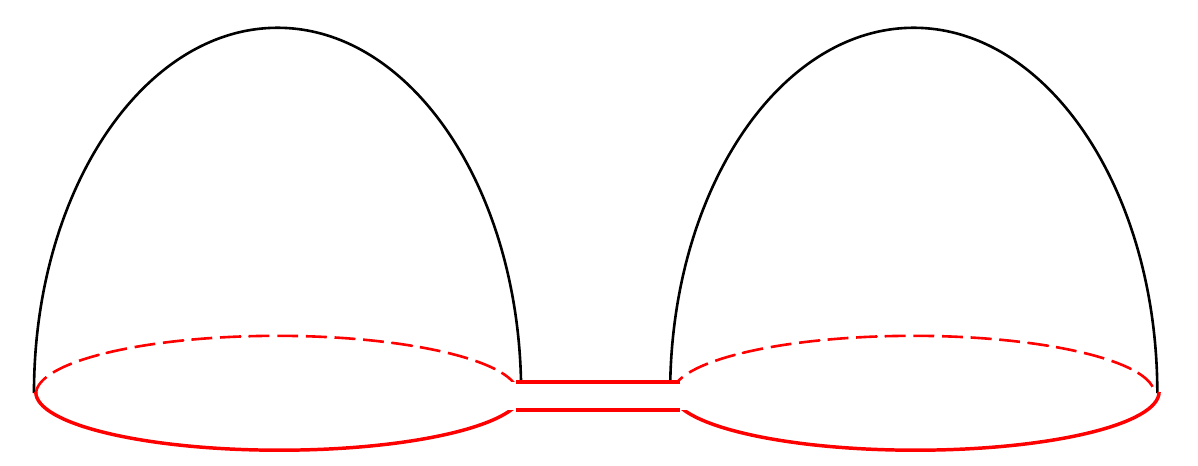} \\
$(b)$
\end{minipage}
\caption{$(a)$ The classical string configuration for the connected part of $\langle \mathcal{W}_{A_1}\mathcal{W}_{A_2}\rangle$. It consists of two disconnected surfaces joined together by an infinitesimal tube. $(b)$ The classical string configuration for the figure eight loop. In this case, two disconnected surfaces are connected by an infinitesimal strip. This explains the difference of the prefactors in \eqref{eq:connectedstrong} and \eqref{eq:connectedstrong2}.} \label{fig:saddles}
\end{figure}

To understand \eqref{eq:strongcouplingfinal} from the string worldsheet, it is useful to recall the connected correlator of two Wilson loops $\langle \mathcal{W}_{A_1}\mathcal{W}_{A_2}\rangle$ analyzed in \cite{Giombi:2009ms}. In that case, the BPS equation of the string worldsheet does not allow any connected surface and the only allowed surface is a degenerate cylinder made of two disks connected by a zero-area tube (see figure \ref{fig:saddles}). Physically the zero-area tube corresponds to a propagator of a graviton and gives a factor of $1/N^2$. Combined with a factor $g^2$ which comes from integrating over the endpoints of the propagator, we obtain the correct $g$-dependence at strong coupling,
\beq\label{eq:connectedstrong}
\langle \mathcal{W}_{A_1}\mathcal{W}_{A_2}\rangle\sim  \underbrace{\frac{e^{4\pi g_{a_1}}}{g^{3/2}}\frac{e^{4\pi g_{a_2}}}{g^{3/2}}}_{\text{disconnected disks}} \times \underbrace{\frac{1}{N^2}}_{\rm propagator}\times\underbrace{g^2}_{\rm endpoints}=\frac{e^{4\pi(g_{a_1}+g_{a_2})}}{gN^2}\period
\eeq

A similar argument seems to hold for \eqref{eq:strongcouplingfinal}: The idea is to consider a degenerate disk made of two disconnected worldsheet with disk topology (ending on the two loops of the figure-eight), joined together by a zero-area {\it strip} (see figure \ref{fig:saddles}). The zero-area strip can be viewed as a propagator of an open string and gives a factor of $1/g$. Combined with other factors, it gives
\beq\label{eq:connectedstrong2}
\langle \mathcal{W}_{\text{figure-eight}}\rangle\sim  \underbrace{\frac{e^{4\pi g_{\bar{a}_1}}}{g^{3/2}}\frac{e^{4\pi g_{\bar{a}_2}}}{g^{3/2}}}_{\text{disconnected disks}} \times \underbrace{\frac{1}{g}}_{\rm propagator}\times\underbrace{g^2}_{\rm endpoints}=\frac{e^{4\pi(g_{\bar{a}_1}+g_{\bar{a}_2})}}{g^2}\comma
\eeq
which is the correct $g$-dependence in \eqref{eq:strongcouplingfinal}. The other three saddle points can be interpreted similarly as contributions from stable/unstable disk solutions \cite{Drukker:2006ga,Giombi:2009ms} joined by a zero-area strip. It would be interesting to perform more detailed analysis and reproduce the full one-loop answer including the numerical coefficients. (See \cite{Forini:2017whz,Medina-Rincon:2018wjs} for recent progress on the one-loop computation on the worldsheet for a single Wilson loop.)

\subsubsection{Two-intersection loop} 
\begin{figure}[t]
\centering
\includegraphics[clip,height=4.5cm]{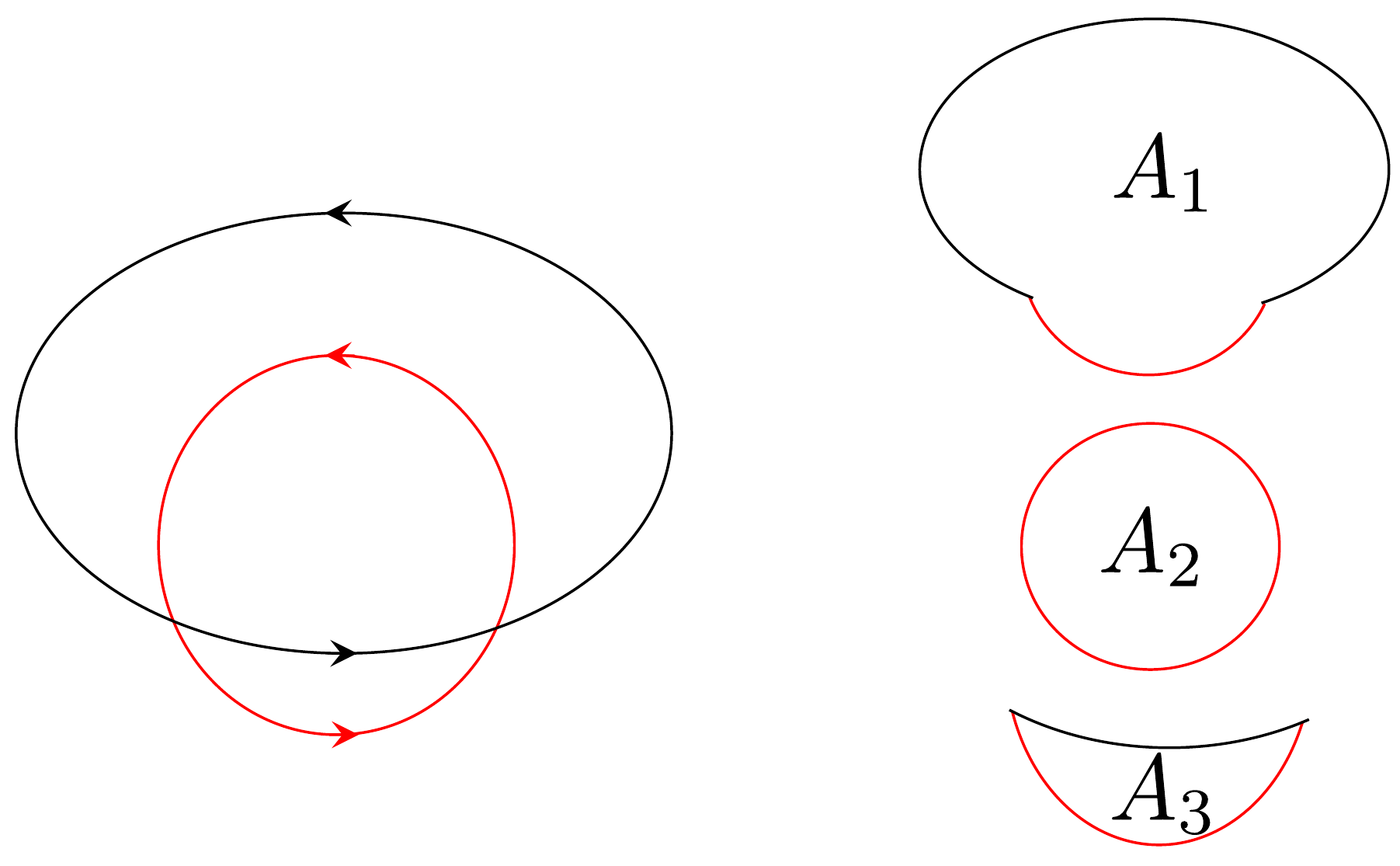}
\caption{The two intersection loop and the definition of the areas. $A_1$ is the area of the whole regions inside the loop while $A_2$ is the area inside the red circle. $A_3$ is the area inside a small subregion in the bottom bounded by the red and black curves.} \label{fig:twointersection}
\end{figure}

\begin{figure}[t]
\centering
\includegraphics[clip,height=3cm]{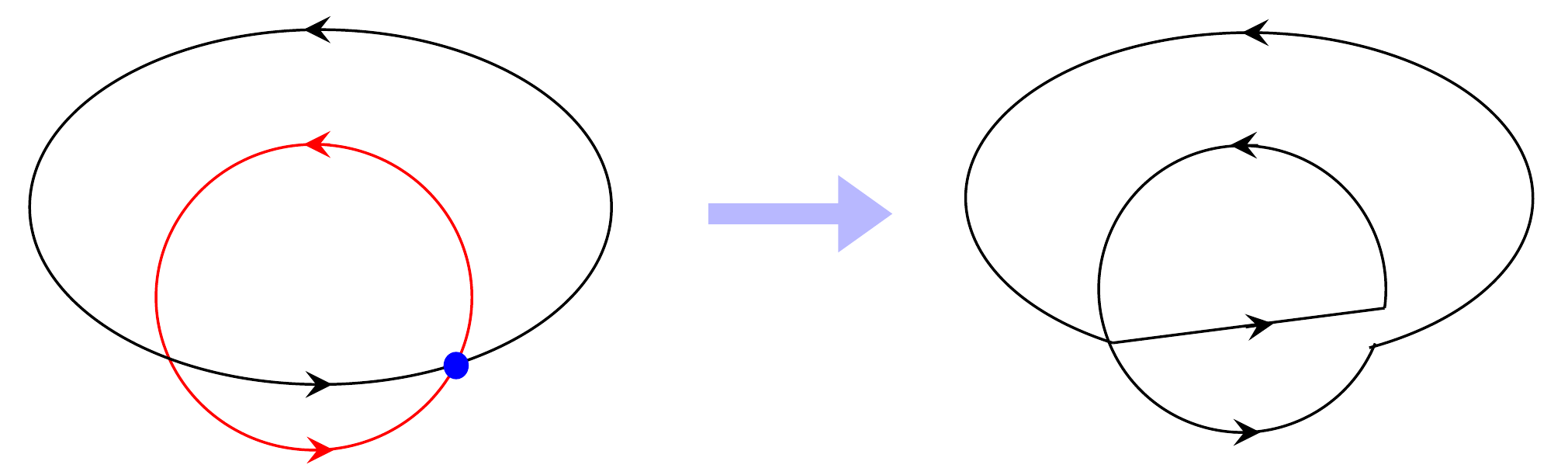}
\caption{The action of the loop equation on the two intersection loop. After using the loop equation to the point colored in blue, we obtain a one-intersection loop with areas $A_1$ and $A_2-A_3$.} \label{fig:actinglooplap}
\end{figure}

We now generalize the analysis to the two-intersection loop $\tilde{\mathcal{W}}_{A_1,A_2,A_3}$ depicted in figure \ref{fig:twointersection}. Applying the loop equation to the intersection shown in figure \ref{fig:actinglooplap}, we obtain
\beq
\left(\del_{A_1}+\del_{A_3}\right)\langle \tilde{\mathcal{W}}_{A_1,A_2,A_3}\rangle=-\frac{4\pi g^2}{N}\langle \tilde{\mathcal{W}}_{A_1,(A_{2}-A_{3})}\rangle\comma
\eeq
where $\tilde{\mathcal{W}}_{A_1,(A_{2}-A_{3})}$ is a one-intersection loop with areas $A_1$ and $A_2-A_3$. Solving this equation using \eqref{eq:contourseparatesingle} and normalizing it as\fn{The normalization is chosen such that the expectation value is $O(1)$ in the large $N$ limit.}
\beq
\mathcal{W}_{A_1,A_2,A_3}:=\frac{\tilde{\mathcal{W}}_{A_1,A_2,A_3}}{N^2}\comma
\eeq
we get
\beq
\begin{aligned}
\langle \mathcal{W}_{A_1,A_2,A_3}\rangle=&F(A_1-A_3,A_2)-\frac{(4\pi g^2)^2}{N^2}\left<\oint_{\mathcal{C}_1\prec\mathcal{C}_2}\frac{du_1}{8\pi^2 g^2}\frac{du_2}{8\pi^2 g^2}\bar{\Delta} \frac{f_{A_1}(u_1)f_{A_2-A_3}(u_2)}{(u_1-u_2)^2}\right>_M\comma\nn
\end{aligned}
\eeq
where $F(A_1-A_3,A_2)$ is the integration constant. To determine $F$, we consider the limit $A_3\to 0$, in which the two-intersection Wilson loop reduces to two disconnected Wilson loops with areas $A_{1}$ and $A_2$ respectively. We then get the constraint
\beq
\begin{aligned}
&\left.\langle \mathcal{W}_{A_1,A_2,A_3}\rangle\right|_{A_3=0}=\left<\oint_{\mathcal{C}_1\prec\mathcal{C}_2} \frac{du_1}{8\pi^2g^2}\frac{du_2}{8\pi^2 g^2}\bar{\Delta}f_{A_1}(u_1)f_{A_2}(u_2)\right>_M\comma
\end{aligned}
\eeq
which allows us to compute $F$. As a result, we get
\beq\label{eq:finiteNABC2}
\begin{aligned}
&\langle \mathcal{W}_{A_1,A_2,A_3}\rangle=\left<\oint_{\mathcal{C}_1\prec\mathcal{C}_2}\frac{du_1}{8\pi^2g^2}\frac{du_2}{8\pi^2g^2}\bar{\Delta}f_{A_1-A_3}(u_1)f_{A_2}(u_2)\right>_M\\
&+\left(\frac{4\pi g^2}{N}\right)^2\left<\oint_{\mathcal{C}_1\prec\mathcal{C}_2}\frac{du_1}{8\pi^2 g^2}\frac{du_2}{8\pi^2 g^2}\bar{\Delta} \frac{f_{A_1-A_3}(u_1)f_{A_2}(u_2)-f_{A_1}(u_1)f_{A_2-A_3}(u_2)}{(u_1-u_2)^2}\right>_M\period
\end{aligned}
\eeq

\paragraph{Large N} Let us next discuss the large $N$ expansion of \eqref{eq:finiteNABC2}. For later purposes, we compute it up to the first subleading order in $1/N$. 

 First we consider the first term of \eqref{eq:finiteNABC2} (to be denoted by $W_1$). It coincides with the integral representation of the two disconnected Wilson loops with areas $A_1-A_3$ and $A_2$ (see \eqref{eq:twocorrfinal}). Thus, using the result in the literature, we get the large $N$ expansion
\begin{align}\label{eq:twodisconnected}
\begin{aligned}
\!W_{1}\!\overset{N\to\infty}{=}\!&\frac{\mathcal{I}_1^{a_{13}}\mathcal{I}_1^{a_2}}{4\pi^2 g_{a_{13}}g_{a_2}}+\frac{\pi}{6N^2}\left[\frac{g_{a_2}^2\mathcal{I}_1^{a_{13}}\mathcal{I}_2^{a_2}}{g_{a_{13}}}+\frac{g_{a_{13}}^2\mathcal{I}_1^{a_{2}}\mathcal{I}_2^{a_{13}}}{g_{a_{2}}}\right]+\frac{1}{N^2}\sum_{k=1}^{\infty}k\left(\frac{\rho_{a_{2}}}{\rho_{a_{13}}}\right)^{k}\mathcal{I}_k^{a_{13}}\mathcal{I}_{k}^{a_2},
\end{aligned}
\end{align}
with $a_{ij}:=(A_i-A_j-2\pi)/2$. Here the first two terms come from the disconnected part, which is a product of the expectation values of a single loop\fn{The expectation value of a single loop up to $O(1/N^2)$ is given by \cite{Drukker:2000rr}
\beq
\langle \mathcal{W}_{A}\rangle=\frac{I_{1}(4\pi g_a)}{2\pi g_{a}} +\frac{\pi^2 g_{a}^2}{3 N^2}I_2 (4\pi g_a)+\cdots\period
\eeq
} \cite{Drukker:2000rr}, while the last term is the connected part computed in \cite{Giombi:2009ms}.

The second line of \eqref{eq:finiteNABC2} (to be denoted by $W_2$) can be evaluated in a manner similar to the figure eight loop: Namely we replace $f_{A}$ and $\bar{\Delta}/(u_1-u_2)^2$ with $\mathfrak{f}_{A}$ and $1/(u_1-u_2)^2$ respectively, rewrite it in terms of the Zhukowski variables and perform the integrals by closing the contour $\mathcal{C}_1$ at the origin and $\mathcal{C}_2$ at infinity. The result reads
\beq\nn
\begin{aligned}
W_2\overset{N\to\infty}{=}\frac{1}{N^2}\sum_{k=1}^{\infty} k\left[\frac{\mathcal{I}_k^{a_{1}}\mathcal{I}_k^{a_{23}}}{\rho_{a_1}^{k}}\left(\rho_{a_{23}}^{k}-\frac{(-1)^{k}}{\rho_{a_{23}}^{k}}\right)-\frac{\mathcal{I}_k^{a_{13}}\mathcal{I}_k^{a_{2}}}{\rho_{a_{13}}^{k}}\left(\rho_{a_{2}}^{k}-\frac{(-1)^{k}}{\rho_{a_{2}}^{k}}\right)\right]\period
\end{aligned}
\eeq 

Adding the two contributions, we get the large $N$ answer
\beq
\begin{aligned}\label{eq:largeNdoubleinter}
&\langle \mathcal{W}_{A_1,A_2,A_3}\rangle\overset{N\to \infty}{=}\frac{\mathcal{I}_1^{a_{13}}\mathcal{I}_1^{a_2}}{4\pi^2 g_{a_{13}}g_{a_2}}+\frac{\pi}{6N^2}\left(\frac{g_{a_2}^2\mathcal{I}_1^{a_{13}}\mathcal{I}_2^{a_2}}{g_{a_{13}}}+\frac{g_{a_{13}}^2\mathcal{I}_1^{a_{2}}\mathcal{I}_2^{a_{13}}}{g_{a_{2}}}\right)\\
&+\frac{1}{N^2}\sum_{k=1}^{\infty} k\left[\frac{\mathcal{I}_k^{a_{1}}\mathcal{I}_k^{a_{23}}}{\rho_{a_1}^{k}}\left(\rho_{a_{23}}^{k}-\frac{(-1)^{k}}{\rho_{a_{23}}^{k}}\right)+(-1)^{k}\frac{\mathcal{I}_k^{a_{13}}\mathcal{I}_k^{a_{2}}}{(\rho_{a_{13}}\rho_{a_2})^{k}}\right]\period
\end{aligned}
\eeq
\paragraph{Weak coupling expansion} From \eqref{eq:largeNdoubleinter}, the weak coupling expansion can be obtained straightforwardly by expanding each term in powers of $g^2$. The result up to one loop reads
\begin{align}
&\langle \mathcal{W}_{A_1,A_2,A_3}\rangle\overset{N\to \infty}{=}\\
&1+g^2\left(\frac{4\pi (A_1+A_2-A_3)-A_1^2-A_2^2-A_3^2+2A_1A_3}{2}+\frac{4\pi (A_2-A_3)-A_2(A_1-A_3)}{N^2}\right)\period\nn
\end{align}
One can check that the result reproduces the one for the two disconnected loops \cite{Drukker:2000rr,Giombi:2009ms} in the limit $A_3\to 0$.
\paragraph{Strong coupling expansion} We now evaluate the strong-coupling limit of \eqref{eq:finiteNABC2}. The first term, $W_1$, is easy to evaluate since it coincides with the correlator of disconnected Wilson loops with areas $A_1-A_3$ and $A_{2}$. Using the result in the literature \cite{Drukker:2000rr,Giombi:2009ms}, we get
\beq
\begin{aligned}
&W_1\overset{N\to\infty}{=}W_{1,{\rm disc}}+W_{1,{\rm conn}}\comma\\
&W_{1,{\rm disc}}=\frac{e^{4\pi (g_{a_{13}}+g_{a_2})}}{8\pi^2 (g_{a_{13}}g_{a_2})^{3/2}}\left(1+\frac{2(\pi g_{a_{13}})^{3}}{3N^2}\right)\left(1+\frac{2(\pi g_{a_{2}})^{3}}{3N^2}\right)\comma\\
&W_{1,{\rm conn}}=\frac{e^{4\pi (g_{a_{13}}+g_{a_2})}}{8\pi^2N^2\sqrt{g_{a_{13}}g_{a_2}}}\frac{\rho_{a_{13}}\rho_{a_2}}{(\rho_{a_{13}}-\rho_{a_2})^2}\comma
\end{aligned}
\eeq
where we kept only the leading exponential. $W_{1,{\rm disc}}$ and $W_{1,{\rm conn}}$ are the contributions from the disconnected part and the connected part respectively. The second term in \eqref{eq:finiteNABC2}, $W_2$, can be evaluated by the saddle point analysis. The result reads
\beq\label{eq:twodifferentexp}
W_2\overset{N\to\infty}{=}\frac{1}{8N^2}\left(\frac{\sqrt{g_{a_{1}}g_{a_{23}}}e^{4\pi (g_{a_1}+g_{a_{23}})}}{(g_{a_{1}}a_{23}+g_{a_{23}}a_{1})^2}-\frac{\sqrt{g_{a_{13}}g_{a_{2}}}e^{4\pi (g_{a_{13}}+g_{a_{2}})}}{(g_{a_{13}}a_{2}+g_{a_{2}}a_{13})^2}\right)\period
\eeq

Let us discuss the worldsheet interpretation. $W_{1,{\rm disc}}$ is simply a product of the contributions from two disconnected surfaces whose worldsheet interpretation is already discussed in \cite{Drukker:2000rr}. The rest ($W_{1,{\rm disc}}$ and $W_{2}$) scales as
$
W_{1.{\rm disc}}, W_2 \sim e^{4\pi g}/(g N^2)
$,
which coincides with \eqref{eq:connectedstrong}. This shows that the relevant worldsheet configurations are again two disconnected surfaces connected by a zero-area tube. The only complication here is that there are two different disconnected surfaces ending on the Wilson loop, corresponding to two different exponentials in \eqref{eq:twodifferentexp}. The first one ends on the closed loops with areas $A_1$ and $A_2-A_3$ while the second one ends on the closed loops with areas $A_1-A_3$ and $A_2$ (see also figure \ref{fig:twointersection}). However owing to the geometrical constraint $A_1>A_2>A_3$, we have $g_{a_{13}}+g_{a_{2}}\geq g_{a_{1}}+g_{a_{23}}$. Thus the leading strong coupling answer is always given by $\sim e^{4\pi (g_{a_{13}}+g_{a_{2}})}$.
\section{Cross Anomalous Dimension at Small Angle\label{sec:cross}}
\begin{figure}[t]
\centering
\begin{minipage}{0.45\hsize}
\centering
\includegraphics[clip,height=3.5cm]{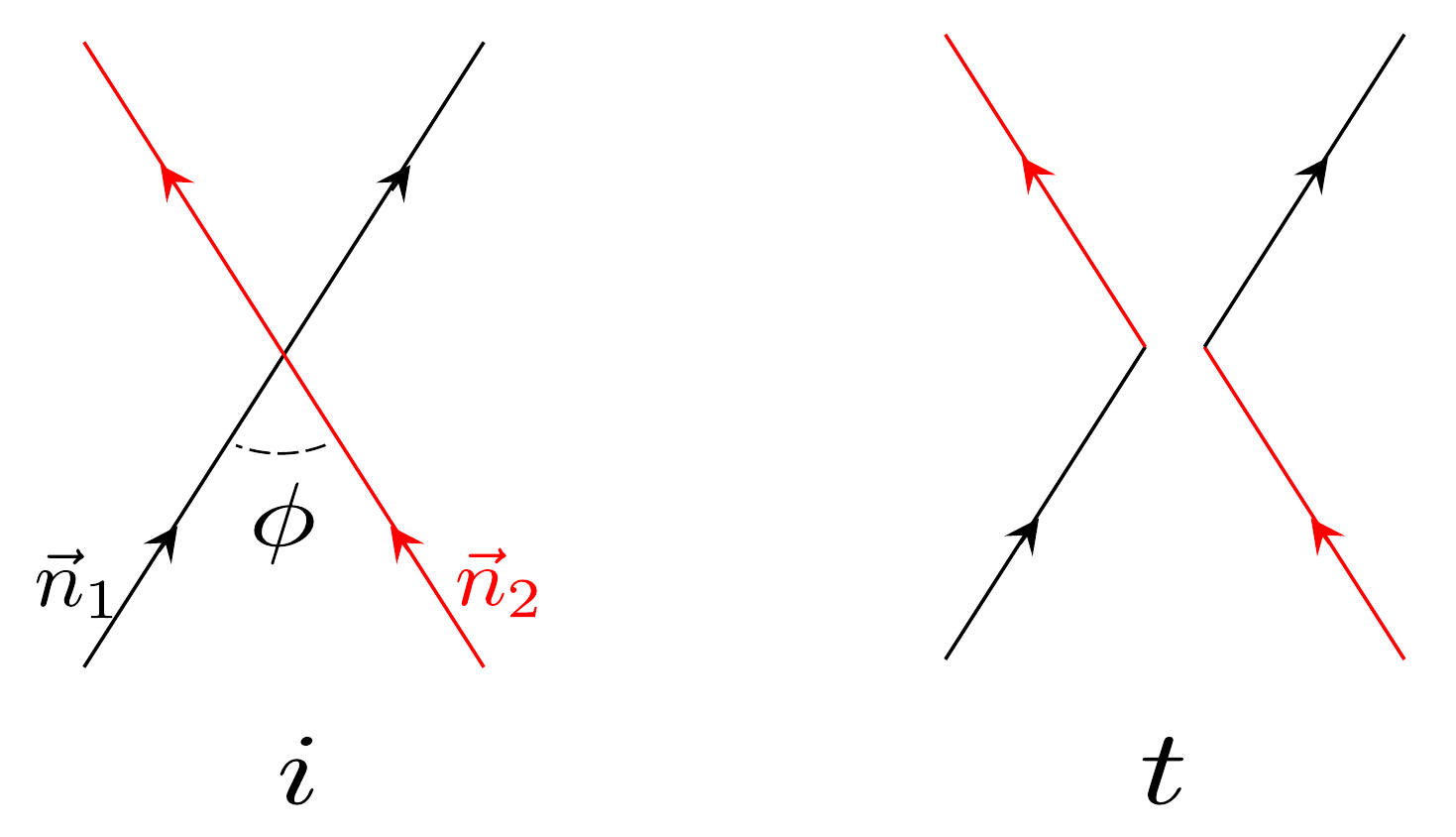} \\
$(a)$
\end{minipage}
\hspace{40pt}
\begin{minipage}{0.45\hsize}
\centering
\includegraphics[clip,height=3.5cm]{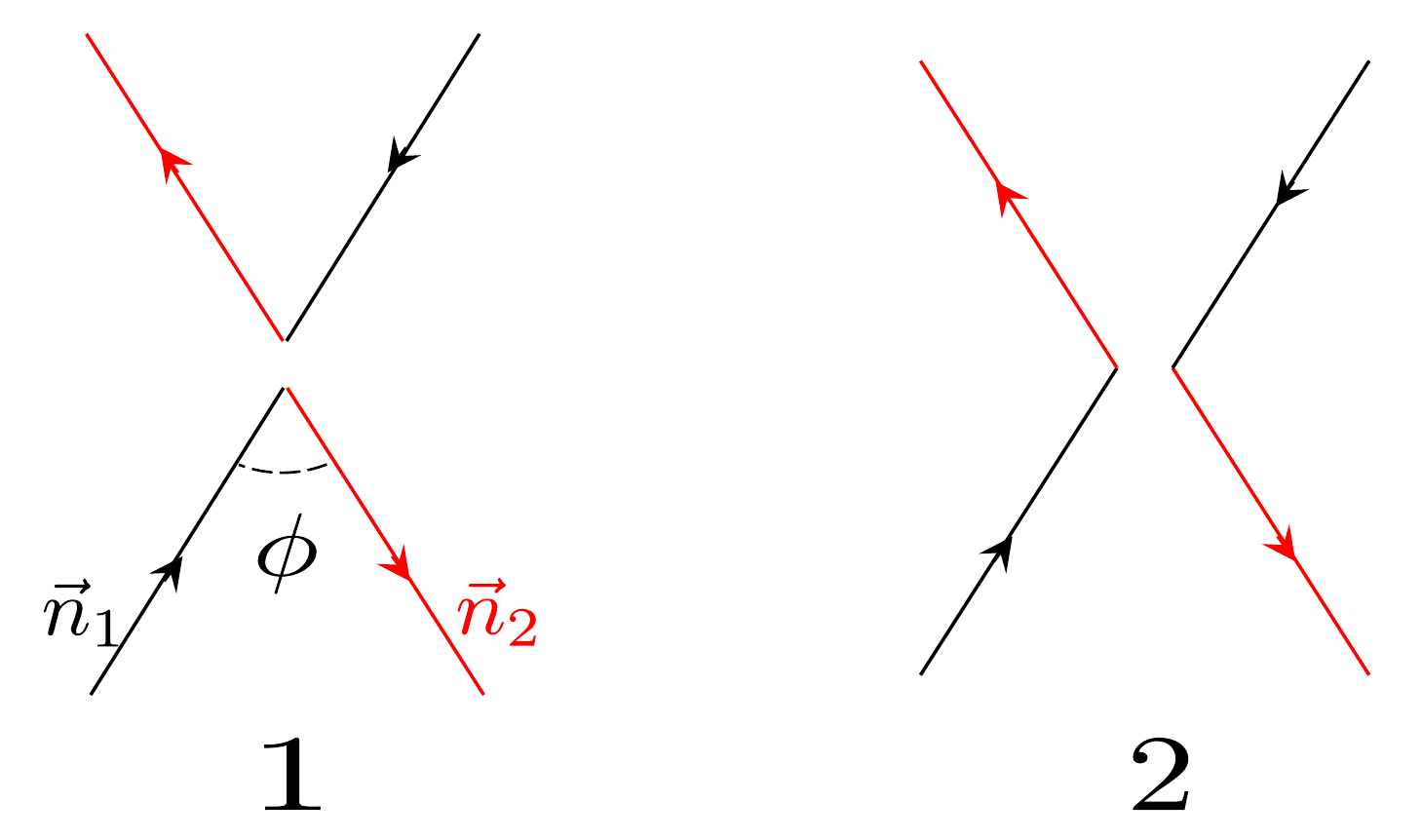} \\
$(b)$
\end{minipage}
\caption{$(a)$ Intersecting lines (denoted by $i$) and ``touching'' lines (denoted by $t$). They are characterized by a geometrical angle $\phi$ and mix under the renormalization. In $\mathcal{N}=4$ SYM, we can consider the generalization of these lines which couple also to scalars. In that case, the black lines couple to $\vec{n}_1\cdot \vec{\Phi}$ and the red lines couple to $\vec{n}_2\cdot \vec{\Phi}$. This introduces an additional angle $\cos\theta =\vec{n}_1\cdot \vec{n}_2$.  $(b)$ Two touching configurations ($1$ and $2$) whose cross anomalous dimension will be computed in Appendix \ref{ap:cross}.} \label{fig:cross}
\end{figure}

We now apply the results in the previous sections to the computation of the {\it cross anomalous dimension}. The cross anomalous dimension is a quantity which governs the UV divergence associated to an intersection of two Wilson lines. In some respects, it is similar to the cusp anomalous dimension, which governs the UV divergence associated to a cusp of the Wilson line. However, one notable difference is that the cross anomalous dimension is a $2\times 2$ matrix since the intersecting Wilson lines mix with the ``touching'' Wilson lines (depicted in figure \ref{fig:cross}-$(a)$) under the renormalization group flow. 

Note that there is another cross anomalous dimension which governs the mixing of two different touching Wilson lines depicted in figure \ref{fig:cross}-$(b)$. As we show in Appendix \ref{ap:cross}, our formalism can be applied to this quantity as well.

\subsection{Cross anomalous dimension in $\mathcal{N}=4$ SYM}
\paragraph{Definition and the relation to amplitudes}
The cross anomalous dimension matrix $\Gamma_{\rm cross}$ determines the renormalization group (RG) property of the Wilson lines with an intersection  \cite{Korchemsky:1993hr,Korchemskaya:1994qp}:
\beq\label{eq:RGcross}
\left(\mu\frac{\del}{\del \mu}+\beta (g)\frac{\del}{\del g}\right)\mathcal{W}^{\rm R}_{A}+\left(\Gamma_{\rm cross}\right)_{A}{}^{B}\mathcal{W}^{\rm R}_B=0\qquad (A,B=i,t)\comma
\eeq
Here $\mu$ is the RG scale and $\mathcal{W}_{i}$ and $\mathcal{W}_{t}$ denote the intersecting and the touching configurations respectively. The superscript {\rm R} signifies the fact that the Wilson lines are renormalized and are related to the {\it bare} Wilson lines $\mathcal{W}_A$ by the multiplicative renormalziation,
\beq
\mathcal{W}_{A}^{\rm R}=\left(Z(\mu,\epsilon)\right)_{A}{}^{B}\mathcal{W}_B \period
\eeq
with $\epsilon$ being the UV cut-off. As we discuss in more detail later, in conformal field theories one can compute the cross anomalous dimension more directly from the expectation value of the bare Wilson lines by reading off the coefficient of $\log \epsilon$, $\langle \mathcal{W}\rangle \sim e^{\Gamma_{\rm cross} \log \epsilon}$.

The cross anomalous dimension is a function of the angle $\phi$ between the two intersecting lines. When the angle is analytically continued as $\phi\to i\varphi$, it gives the so-called {\it soft anomalous dimension}. The soft anomalous dimension controls  the IR divergence of the scattering amplitude of two massive quarks in the Regge kinematics, $s,m^2\gg -t \gg \Lambda_{\rm QCD}^2$, and describes how the soft gluons transfer the color degrees of freedom of the quarks. In that context, $\varphi$ is the boost angle between the two quarks defined by 
\beq
\cosh \varphi:= -\frac{p_1\cdot p_2}{\sqrt{p_1^2 p_2^2}}\comma
\eeq
with $p_{1,2}$ being the four-momenta of the quarks. See \cite{Korchemsky:1993hr,Korchemskaya:1994qp} for more details.

For the application to QCD, the limit $\varphi\to \infty$ is of particular interest since it describes the high energy scattering of light partons. On the Wilson line side, this corresponds to an intersection of two light-like lines, see for instance \cite{Becher:2009cu,Gardi:2009qi,Becher:2009qa}. The limit was studied also in $\mathcal{N}=4$ SYM: In \cite{Dixon:2016epj}, the self-crossing lightlike loop was analyzed up to nine loops. The analysis was pushed further in \cite{Caron-Huot:2019vjl} in which the anomalous dimensions relevant for the limit were determined exactly in the large $N$ limit using the pentagon OPE decomposition \cite{Basso:2013vsa}. In this paper, we focus on different limits, namely $\phi\sim\varphi \sim 0$ and the near BPS limit, and compute the anomalous dimension exactly at finite $N$.
\paragraph{Generalization in $\mathcal{N}=4$ SYM} In $\mathcal{N}=4$ SYM, one can consider a generalization of the cross anomalous dimension $\Gamma_{\rm cross}(\phi,\theta)$ which depends on another angle $\theta$. To do so, we consider an intersection of the supersymmetric Wilson lines,
\beq\label{eq:locallysusy}
\mathcal{W}\sim {\rm tr} {\rm P}\exp \left(\int (i A_{\mu}\dot{x}^{\mu}+ \vec{n}\cdot \vec{\Phi}|\dot{x}|)d\tau\right)\comma
\eeq
where the {\it R-symmetry polarization} $\vec{n}$ is a six-component unit vector which dictates the coupling to the scalars $\vec{\Phi}=(\Phi_1,\ldots, \Phi_6)$. The intersection of such lines can be characterized by the geometric angle $\phi$ and the {\it R-symmetry angle} $\theta$ which is defined by
\beq
\cos \theta:= \vec{n}_1\cdot \vec{n}_2\comma
\eeq
where $\vec{n}_{1,2}$ are the R-symmetry polarizations of the two lines at the intersection point.

The supersymmetric Wilson line \eqref{eq:locallysusy} naturally arises from the worldline action of the $W$-boson in the Coulomb branch of $\mathcal{N}=4$ SYM. To be concrete, let us consider the symmetry breaking phase ${\rm U}(N+2) \to {\rm U}(1)\times {\rm U}(1)\times {\rm U}(N)$ dictated by the scalar expectation value
\beq
\langle \vec{\Phi}\rangle ={\rm diag}\left(m_1 \vec{n}_1\comma m_2\vec{n}_2, \underbrace{0,\ldots,0}_{N}\right)\period
\eeq
In this phase, we have two kinds of $W$-bosons, one coming from the $(1,k)$ (or $(k,1)$) component and the other coming from the $(2,k)$ (or $(k,2)$) component of the gauge field. In the limit $m_{1,2}\gg 1$, they can be treated as classical probe particles and their coupling to the unbroken U$(N)$ degrees of freedom is given precisely by \eqref{eq:locallysusy}. Thus $\Gamma_{\rm cross}(\phi,\theta)$ gives a natural generalization of the cross anomalous dimension in QCD and it characterizes the IR divergence of the scattering of massive $W$-bosons after the analytic continuation $\phi\to i \varphi$.

When $\phi= \theta$, the whole configuration becomes BPS and the anomalous dimension $\Gamma_{\rm cross}(\phi,\theta)$ vanishes. Expanding $\Gamma_{\rm cross}(\phi,\theta)$ around this limit, we obtain\fn{The relation between $\Gamma_{\rm cross}$ and $\hat{\gamma}_{\rm cross}$ parallels the relation between the cusp anomalous dimension and the Bremsstralung function \cite{Correa:2012at}.}
\beq\label{eq:firstder}
\Gamma_{\rm cross}(\phi,\theta)=(\phi-\theta)\gamma_{\rm cross}(\theta) +O((\phi-\theta)^2)\period
\eeq 
In what follows, we compute the first coefficient $\gamma_{\rm cross}(\theta)$ exactly as a function of $\lambda$ and $N$. 
\subsection{Two-point function of intersections\label{subsec:2ptint}}
Let us explain in more detail how to extract the cross anomalous dimension from the expectation values of the bare Wilson lines. The key idea is to regard them as the two-point functions of intersections.

\begin{figure}[t]
\centering
\includegraphics[clip,height=6.5cm]{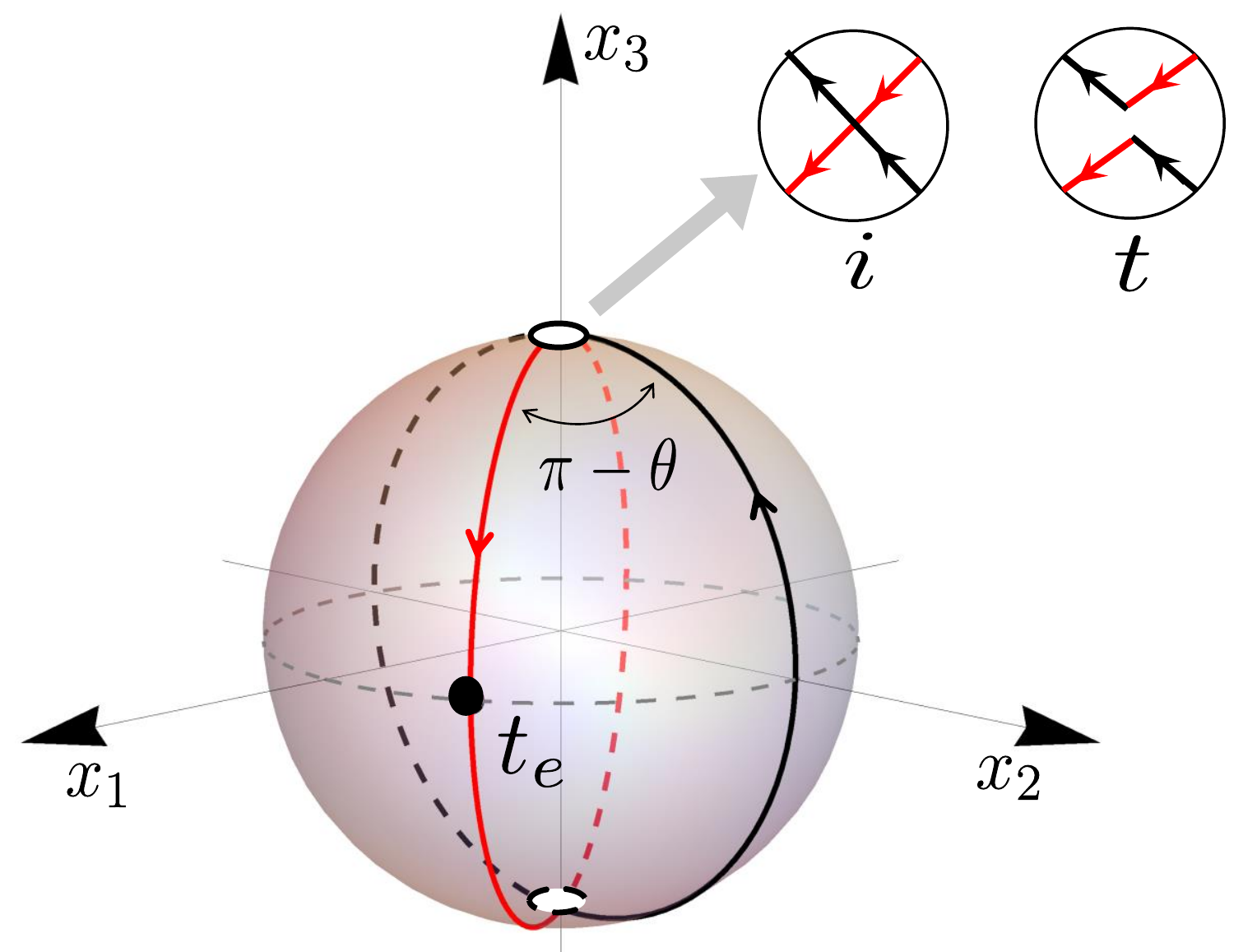}
\caption{Intersecting lines mapped onto $S^2$. The black and the red lines denote $\mathcal{W}_1$ and $\mathcal{W}_2$ respectively. On $S^2$, the configuration contains two intersections, one at the north pole and the other at the south pole. At each intersection (denoted by white circles), one can make either of the two choices, $i$ and $t$. This leads to the $2\times 2$ matrix structure of the cross anomalous dimension.} \label{fig:twopntsphere}
\end{figure}

To be concrete, consider the intersection of the following two Wilson lines on $R^{4}$,
\beq\label{eq:twolineR2}
\mathcal{W}_{1,2}={\rm P}\exp \int^{\infty}_{-\infty}d\tau \left(i A\cdot \dot{x}_{1,2}+\vec{\Phi}\cdot \vec{n}_{1,2}|\dot{x}_{1,2}|\right) \comma
\eeq
with
\beq
\begin{aligned}
&\dot{x}_1 =(1,0,0,0)\comma \qquad && \dot{x}_2 =(\cos\phi,\sin\phi,0,0)\comma\\
&n_1 =(1,0,0,0,0,0)\comma \qquad && n_2 =(\cos\theta,\sin\theta,0,0,0,0)\period
\end{aligned}
\eeq
In addition to the obvious intersection at the origin, the lines intersect also at infinity. This is easier to see if one maps the configuration to $S^2$ using the conformal transformation
\beq\label{eq:R2toS2}
X_1=\frac{2x_1}{1+x_1^2 +x_2^2}\comma\quad X_2=\frac{-2x_1}{1+x_1^2+x_2^2}\comma\quad X_3=\frac{1-x_1^2-x_2^{2}}{1+x_1^{2}+x_2^{2}}\period
\eeq 
Here $X$'s are the embedding coordinates of $S^2$ while $x$'s are the coordinates on $R^2$ inside $R^4$. The two intersections are mapped to the north and the south poles of $S^2$, see figure \ref{fig:twopntsphere}.

To extract the cross anomalous dimension, one has to consider the operator mixing: For each intersection, we can either let the lines intersect (to be denoted by $i$) or resolve the intersection and make the lines touching (to be denoted by $t$). Since there are two intersections, we have in total four different configurations of the Wilson lines which we denote by $\mathcal{W}_{ii}$, $\mathcal{W}_{it}$, $\mathcal{W}_{ti}$ and $\mathcal{W}_{tt}$. They can be regarded as two-point functions of ``operators'', labeled by $i$ and $t$, sitting at the intersections.

These four choices of two-point functions can be naturally organized into a $2\times 2$ matrix
\beq
\mathbb{W}:=\pmatrix{cc}{\langle\mathcal{W}_{ii}\rangle&\langle\mathcal{W}_{it}\rangle\\\langle\mathcal{W}_{ti}\rangle&\langle\mathcal{W}_{tt}\rangle}=\pmatrix{cc}{\langle i |\exp\left(D \log \frac{\epsilon_{\rm UV}}{r_{\rm IR}}\right)|i\rangle&\langle i |\exp\left(D \log \frac{\epsilon_{\rm UV}}{r_{\rm IR}}\right)|t\rangle\\\langle t |\exp\left(D \log \frac{\epsilon_{\rm UV}}{r_{\rm IR}}\right)|i\rangle&\langle t |\exp\left(D \log \frac{\epsilon_{\rm UV}}{r_{\rm IR}}\right)|t\rangle}\period
\eeq
Here $\langle i,t |$ and $|i,t\rangle$ denote the intersections at the origin and infinity respectively and $D$ is the dilatation operator. $\epsilon_{\rm UV}$ and $r_{\rm IR}$ are the UV and the IR cutoffs. $\mathbb{W}$ can be expressed alternatively in terms of the cross anomalous dimension $\Gamma_{\rm cross}$ and  the overlap $\eta$ as
\beq
\mathbb{W}=e^{\Gamma_{\rm cross} \log (\epsilon_{\rm UV}/r_{\rm IR})}\cdot \eta\comma
\eeq
where $\Gamma_{\rm cross}$ and $\eta$ are $2\times 2$ matrices defined by 
\beq
\begin{aligned}
\pmatrix{cc}{\langle i|\comma&\langle t|}\Gamma_{\rm cross}:=\pmatrix{cc}{ \langle i|D\comma& \langle t|D}\comma\qquad \quad \eta := \pmatrix{cc}{\langle i |i\rangle&\langle i |t\rangle\\\langle t |i\rangle&\langle t |t\rangle}\period
\end{aligned}
\eeq

In the near BPS limit $\theta\sim\phi$, both $\eta$ and $\Gamma_{\rm cross}$ can be expanded in powers of $(\phi-\theta)$,
\beq
\eta=\eta_0 +(\phi-\theta)\eta_1+\cdots \comma\qquad \Gamma_{\rm cross}=(\phi-\theta)\gamma_{\rm cross}+\cdots\period
\eeq
We then obtain the following expansion of the two-point function matrix $\mathbb{W}$:
\beq
\begin{aligned}
&\mathbb{W}=\mathbb{W}_0+(\phi-\theta)\mathbb{W}_1+\cdots\comma\\
&\mathbb{W}_0=\eta_0 \comma\qquad \mathbb{W}_1=\eta_1 +(\gamma_{\rm cross}\cdot \eta_0)\log (\epsilon_{\rm UV}/r_{\rm IR})\period
\end{aligned}
\eeq
Thus $\gamma_{\rm cross}$ is given by the coefficient of the logarithm in $\mathbb{W}_1$, multiplied by $(\mathbb{W}_0)^{-1}$, 
\beq\label{eq:gammacrossgg}
\gamma_{\rm cross}=\left(\left.\mathbb{W}_1\right|_{\log \frac{\epsilon_{\rm UV}}{r_{\rm IR}}}\right) \cdot (\mathbb{W}_0)^{-1}\comma
\eeq 
where $\mathbb{W}_0$ is nothing but the expectation value in the BPS limit:
\beq\label{eq:defWbb0}
\mathbb{W}_0(\theta)=\left.\pmatrix{cc}{\langle\mathcal{W}_{ii}\rangle&\langle\mathcal{W}_{it}\rangle\\\langle\mathcal{W}_{ti}\rangle&\langle\mathcal{W}_{tt}\rangle}\right|_{\phi=\theta}\period
\eeq
\subsection{Cross anomalous dimension from localization}
We now relate $\gamma_{\rm cross}$ given in \eqref{eq:gammacrossgg} to the localization computation. This can be done by following the arguments in \cite{Correa:2012at,Bonini:2015fng,Giombi:2018qox}, which we briefly review below\fn{See section 4.2 of \cite{Giombi:2018qox} for more detailed explanation.}.

The first step is to start with the BPS limit $\theta=\phi$ of \eqref{eq:twolineR2} and deform $\theta$ slightly. This amounts to inserting a scalar on the second Wilson line
\beq
\delta\mathcal{W}_2 =\delta\theta \times  \int_{-\infty}^{\infty}d\tau \, {\rm P}\left[ \Phi^{\prime}(\tau)\mathcal{W}_2\right]\comma
\eeq
with
\beq
\Phi^{\prime}(\tau)=\left.-\sin \theta \Phi_1 +\cos \theta\Phi_2\right|_{x^{\mu}=(\tau\cos \theta ,\tau\sin\theta ,0,0,0,0)} \period
\eeq
Now, using the invariance under the dilatation around the origin, we can determine the position dependence of the scalar insertion as
\beq
\langle \Phi^{\prime}(\tau) \cdots \rangle =\frac{1}{|\tau|}\langle \Phi^{\prime}(\tau=1) \cdots\rangle\comma
\eeq
where we denoted all the other parts (the Wilson lines $\mathcal{W}_{1,2}$ and possible resolutions of the intersections etc.) by $\cdots$. From this, one can see that the integral of $\tau$ produces the logarithmic divergence and its coefficient is given by the expectation value of the Wilson loops with the $\Phi^{\prime}$ insertion. More explicitly, we have the relation
\beq\label{eq:tryingtorewrite}
\begin{aligned}
\left.\mathbb{W}_1\right|_{\log \frac{\epsilon_{\rm UV}}{r_{\rm IR}}}=2\left.\pmatrix{cc}{\langle\mathcal{W}^{\prime}_{ii}\rangle&\langle\mathcal{W}^{\prime}_{it}\rangle\\\langle\mathcal{W}^{\prime}_{ti}\rangle&\langle\mathcal{W}^{\prime}_{tt}\rangle}\right|_{\phi=\theta}\comma
\end{aligned}
\eeq
where $\mathcal{W}^{\prime}_{AB}$ is the Wilson line $\mathcal{W}_{AB}$ with $\Phi^{\prime}$ inserted at $(\cos\theta,\sin\theta,0,0,0,0)$. The factor of $2$ comes from summing up contributions from $\int^{\infty}_0d\tau $ and $\int ^{0}_{-\infty}d\tau$.

The second step is to map the whole configuration to $S^2$ using \eqref{eq:R2toS2}. We then get two great circles whose contour are given by
\beq
(x_{1},x_{2},x_{3})=\begin{cases}(0,\sin t,-\cos t)\qquad &(\mathcal{W}_1)\\(-\sin\theta\sin t,\cos \theta\sin t,-\cos t)\qquad &(\mathcal{W}_2)\end{cases}\comma
\eeq 
with $t\in (-\pi,\pi)$. They couple to $\Phi^{1}$ and $\cos \theta \Phi^{1}+\sin\theta\Phi^{2}$ respectively and satisfy the $1/8$ BPS condition \eqref{eq:eightBPScond}. 
Under this map, the insertion $\Phi^{\prime}(\tau=1)$ is mapped to the insertion at a point $t_e$ where $\mathcal{W}_2$ intersects with the equator of $S^2$ (see figure \ref{fig:twopntsphere}). 

The third step is to replace $\Phi^{\prime} $ at $t_e$ with $\Phi^{\prime}-i \Phi_4=:-\tilde{\Phi}$. This replacement does not affect the expectation values since the correlator with $\Phi_4$ vanishes owing to the charge conservation. We then use the fact that the insertion of $\tilde{\Phi}$ corresponds to the insertion of the field strength in 2d YM \cite{Pestun:2009nn,Giombi:2009ds}. Therefore its expectation value can be computed by taking the area derivative. In our setup there are two regions with area $2(\pi-\theta)$, and changing $\theta$ by $\delta \theta$ leads to a total change (decrease) of the area by $4\delta\theta$. We thus have the relation
\beq
\left.\pmatrix{cc}{\langle\mathcal{W}^{\prime}_{ii}\rangle&\langle\mathcal{W}^{\prime}_{it}\rangle\\\langle\mathcal{W}^{\prime}_{ti}\rangle&\langle\mathcal{W}^{\prime}_{tt}\rangle}\right|_{\phi=\theta}=\frac{1}{4}\del_{\theta}\mathbb{W}_0(\theta)\comma
\eeq
with $\mathbb{W}_0$ is the expectation value in the BPS limit \eqref{eq:defWbb0}. Combined with \eqref{eq:tryingtorewrite}, it gives
\beq
\left.\mathbb{W}_1\right|_{\log \frac{\epsilon_{\rm UV}}{r_{\rm IR}}}=\frac{1}{2}\del_{\theta}\mathbb{W}_0(\theta)\period
\eeq
Using \eqref{eq:gammacrossgg}, we get the formula relating $\gamma_{\rm cross}$ to the BPS Wilson loops
\beq\label{eq:nearBPSrelation}
\gamma_{\rm cross}=\frac{1}{2}\left(\del_{\theta}\mathbb{W}_0\right) \cdot\left(\mathbb{W}_0\right)^{-1}\period 
\eeq

 Before we proceed, let us comment on the normalization. In what follows, we normalize $\mathbb{W}_0$ by dividing the path-ordered exponentials by a common factor $N^2$. This is a natural normalization for discussing the renormalization of {\it open} intersecting lines (see \cite{Munkler:2018cvu,Korchemsky:1993hr,Korchemskaya:1994qp}) and makes $\Gamma_{\rm cross}$ more symmetric. However it does not coincide with the normalization used in some of the literature in which they discuss the renormalization of closed loops. We will later translate the final result to such a normalization.

The last step is to compute $\mathbb{W}_0$. Let us first consider $\langle \mathcal{W}_{tt}\rangle$. Since both intersections are resolved, it coincides with a correlator of two disconnected loops. Thus, setting $A_1=2(\pi+\theta)$ and $A_2=2(\pi-\theta)$ in \eqref{eq:twocorrfinal}, we get
\beq\label{eq:WttfiniteN}
\left.\langle \mathcal{W}_{tt}\rangle\right|_{\phi=\theta}=\left<\oint_{\mathcal{C}_1\prec \mathcal{C}_2}\frac{du_1}{8\pi^2 g^2}\frac{du_2}{8\pi^2 g^2}\bar{\Delta}(u_1,u_2)f_{2(\pi+\theta)}(u_1)f_{2(\pi-\theta)}(u_2)\right>_M\period
\eeq
Using the result in the literature, we can compute its large $N$ limit as\fn{The result \eqref{eq:WttlargeN} can be obtained from \eqref{eq:twodisconnected} by setting $A_1-A_3=2(\pi+\theta)$ and $A_2=2(\pi-\theta)$.}
\beq\label{eq:WttlargeN}
\left.\langle \mathcal{W}_{tt}\rangle\right|_{\phi=\theta}\overset{N\to\infty}{=}\left(\frac{\mathcal{I}_1^{\theta}}{2\pi g_{\theta}}\right)^2+\frac{\pi g_{\theta}\mathcal{I}_1^{\theta}\mathcal{I}_2^{\theta}}{3N^2}+\frac{1}{N^2}\sum_{k=1}^{\infty}\frac{k(\mathcal{I}_k^{\theta})^2}{(\rho_{\theta})^{2k}}\period
\eeq 
Second, $\langle \mathcal{W}_{ti}\rangle(=\langle \mathcal{W}_{it}\rangle)$ is a single-intersection loop with areas $A_1=2(\pi+\theta)$ and $A_2=2(\pi-\theta)$ (or equivalently, a figure-eight loop with areas $\bar{A}_1=\bar{A}_2=2(\pi-\theta)$). We thus get from \eqref{eq:contourseparatesingle}
\beq\label{eq:WtifiniteN}
\left.\langle \mathcal{W}_{ti}\rangle\right|_{\phi=\theta}=\frac{4\pi g^2 i}{N}\left<\oint_{\mathcal{C}_1\prec\mathcal{C}_2} \frac{du_1}{8\pi^2 g^2}\frac{du_2}{8\pi^2 g^2} \bar{\Delta} (u_1,u_2) \frac{f_{2(\pi+\theta)}(u_1)f_{2(\pi-\theta)}(u_2)}{u_1-u_2}\right>_M\period
\eeq
Here the extra factor $1/N$ comes from the normalization that we adopted. The large $N$ limit can be computed from \eqref{eq:singleintersectionfinal} as
\begin{align}\label{eq:WtilargeN}
&\left.\langle \mathcal{W}_{ti}\rangle\right|_{\phi=\theta}\overset{N\to\infty}{=}\frac{1}{2\pi N}\left(\frac{\mathcal{I}_{0}^{\theta}\mathcal{I}_{1}^{\theta}}{g_{\theta}}+\sum_{k=1}^{\infty}\frac{\mathcal{I}_{k}^{\theta}\mathcal{I}_{k+1}^{\theta}}{g}\left(\rho_{\theta}^{-2k-1}+(-1)^{k}\frac{\rho_{\theta}-\rho_{\theta}^{-1}}{2}\right)\right)\period
\end{align}
Finally, $\langle \mathcal{W}_{ii}\rangle$ is a two-intersection loop which we computed in \eqref{eq:finiteNABC2}. Setting $A_1=2(\pi+\theta)$, $A_2=2\pi$ and $A_3=2\theta$, we get
\beq\label{eq:WiifiniteN}
\begin{aligned}
&\left.\langle \mathcal{W}_{ii}\rangle\right|_{\phi=\theta}=\left<\oint_{\mathcal{C}_1\prec\mathcal{C}_2}\frac{du_1}{8\pi^2g^2}\frac{du_2}{8\pi^2g^2}\bar{\Delta}f_{2\pi}(u_1)f_{2\pi}(u_2)\right>_M\\
&+\frac{(4\pi g^2)^2}{N^2}\left<\oint_{\mathcal{C}_1\prec\mathcal{C}_2}\frac{du_1}{8\pi^2 g^2}\frac{du_2}{8\pi^2 g^2}\bar{\Delta} \frac{f_{2\pi}(u_1)f_{2\pi}(u_2)-f_{2(\pi+\theta)}(u_1)f_{2(\pi-\theta)}(u_2)}{(u_1-u_2)^2}\right>_M\period
\end{aligned}
\eeq
The large $N$ limit can be computed from \eqref{eq:largeNdoubleinter}, and the result reads
\beq\label{eq:WiilargeN}
\begin{aligned}
&\left.\langle \mathcal{W}_{ii}\rangle\right|_{\phi=\theta}\overset{N\to \infty}{=}\left(\frac{\mathcal{I}_1^{0}}{2\pi g}\right)^2+\frac{g\pi \mathcal{I}_1^{0}\mathcal{I}_2^{0}}{3N^2}+\sum_{k=1}^{\infty} \frac{k\left[\left(\rho_{\theta}^{-2k}-(-1)^{k}\right)(\mathcal{I}_k^{\theta})^2+(-1)^{k}(\mathcal{I}_k^{0})^2\right]}{N^2}\period
\end{aligned}
\eeq
Note that these expectation values satisfy the following relation (at finite $N$)
\beq\label{eq:loopintheta}
\del_{\theta}\left.\langle \mathcal{W}_{ii}\rangle\right|_{\phi=\theta}=-\frac{8\pi g^2}{N}\left.\langle \mathcal{W}_{ti}\rangle\right|_{\phi=\theta}\comma\qquad \del_{\theta}\left.\langle \mathcal{W}_{ti}\rangle\right|_{\phi=\theta}=-\frac{8\pi g^2}{N}\left.\langle \mathcal{W}_{tt}\rangle\right|_{\phi=\theta}\period
\eeq
They are simply the loop equations written in terms of $\theta$-derivatives, but one can also verify them directly  from \eqref{eq:WttfiniteN}, \eqref{eq:WtifiniteN} and \eqref{eq:WiifiniteN}.

These expressions, together with the relation \eqref{eq:nearBPSrelation}, give the exact cross anomalous dimension in the near BPS limit of the U$(N)$ theory. Using the relations \eqref{eq:loopintheta}, we get
\beq
\begin{aligned}
\gamma_{\rm cross}&=\left.\frac{1}{\det \mathbb{W}_0}\pmatrix{cc}{0&-\frac{4\pi g^2}{N}\det \mathbb{W}_0\\-\frac{\langle \mathcal{W}_{ti}\rangle\del_{\theta}\langle \mathcal{W}_{tt}\rangle}{2}-\frac{4\pi g^2\langle\mathcal{W}_{tt} \rangle^2}{N}&\frac{\langle \mathcal{W}_{ii}\rangle\del_{\theta}\langle \mathcal{W}_{tt}\rangle}{2}+\frac{4\pi g^2\langle\mathcal{W}_{ti} \rangle\langle \mathcal{W}_{tt}\rangle}{N}}\right|_{\phi=\theta}\period
\end{aligned}
\eeq
Note in particular that the entries in the first row are $0$ and $-4\pi g^2/N$ at all orders in $\lambda$ and $N$.
Epanding the result at large $N$, we get
\beq
\begin{aligned}
\gamma_{\rm cross}&\overset{N\to\infty}{=}\pmatrix{cc}{0&-\frac{4\pi g^2}{N}\\-\frac{4\pi g^2 h_1 +h_0h_2}{N}&h_0+\frac{4\pi g^2h_2+\frac{h_0 (h_2)^2}{h_1}-h_0 h_3+h_4}{N^2}}\comma
\end{aligned}
\eeq
with
\beq
\begin{aligned}
&h_0 =\frac{4\pi \theta g_{\theta}}{\theta^2-\pi^2}\frac{\mathcal{I}_2^{\theta}}{\mathcal{I}_1^{\theta}}\comma\qquad h_1=\left(\frac{g\mathcal{I}_1^{\theta}}{g_{\theta} \mathcal{I}_1^{0}}\right)^2\comma\\
& h_2=2\pi\left(\frac{g}{\mathcal{I}_1^{0}}\right)^2\left(\frac{\mathcal{I}_{0}^{\theta}\mathcal{I}_{1}^{\theta}}{g_{\theta}}+\sum_{k=1}^{\infty}\frac{\mathcal{I}_{k}^{\theta}\mathcal{I}_{k+1}^{\theta}}{g}\left(\rho_{\theta}^{-2k-1}+(-1)^{k}\frac{\rho_{\theta}-\rho_{\theta}^{-1}}{2}\right)\right)\comma\\
&h_3=\left(\frac{2\pi g_{\theta}}{\mathcal{I}_1^{\theta}}\right)^2\left(\frac{g\pi \mathcal{I}_1^{0}\mathcal{I}_2^{0}}{3}+\sum_{k=1}^{\infty} k\left[\left(\rho_{\theta}^{-2k}-(-1)^{k}\right)(\mathcal{I}_k^{\theta})^2+(-1)^{k}(\mathcal{I}_k^{0})^2\right]\right)\comma\\
&h_4=\left(\frac{2\pi g_{\theta}}{\mathcal{I}_1^{\theta}}\right)^2\sum_{k=1}^{\infty} \frac{k}{2}\del_{\theta}\left[\left(\rho_{\theta}^{-2k}-(-1)^{k}\right)(\mathcal{I}_k^{\theta})^2\right]\period
\end{aligned}
\eeq
Here $h_0$ is related to the Bremsstrahlung function $B(\lambda)$ in \cite{Correa:2012at} by
\beq
h_0 =\frac{4\pi^2\theta}{\theta^2-\pi^2}B(16\pi^2 g_{\theta}^2)\period
\eeq
The eigenvalues ($\gamma_{\pm}$) of $\gamma_{\rm cross}$ are given (up to $O(1/N^2)$) by
\beq
\begin{aligned}
&\gamma_{+}=h_0+\frac{1}{N^2}\left(\frac{h_0(h_2)^2}{h_1}+h_0 h_3+h_4+8\pi g^2 h_2+\frac{16\pi^2g^4 h_1}{h_0}\right)\comma\\
&\gamma_{-}=-\frac{1}{N^2}\left(4\pi g^2 h_2+\frac{16\pi^2g^4 h_1}{h_0}\right)\period
\end{aligned}
\eeq
\subsection{U(1) factor and weak- and strong-coupling expansions\label{subsec:U1}}
We now expand our results and compare them with the perturbative data.  However, since the results in the literature are for the ${\rm SU}(N)$ gauge group, we first need to strip off the ${\rm U}(1)$ factor from our results, which are for the ${\rm U}(N)$ gauge group. This can be done by computing the expectation values in the ${\rm U}(1)$ theory since the Wilson loop in the ${\rm U}(N)$ theory factorizes as
\beq
\mathcal{W}_{{\rm U}(N)}=\mathcal{W}_{{\rm U}(1)}\mathcal{W}_{{\rm SU}(N)}\period
\eeq

Since the ${\rm U}(1)$ theory is free, the computation is rather straightforward. In addition, as the gauge group is Abelian, the path-ordering is unnecessary and all the four entries of $\mathbb{W}$ become identical. The result can be read off from the  Appendix A of \cite{Giombi:2012ep}, or from the two-matrix model in \cite{Giombi:2009ms} specialized to $U(1)$. For the two disconnected loops with areas defined as in figure \ref{fig:twoloop}, this gives
\beq
\left.\langle\mathcal{W}_{1}\mathcal{W}_{2} \rangle\right|_{{\rm U}(1)}=\exp\left(\frac{g^2}{N^2}\frac{4\pi (4\pi -A_1+A_2)-(4\pi-A_1-A_2)^2}{2}\right)\period
\eeq
Setting $A_1=2(\pi+\theta)$ and $A_2=2(\pi-\theta)$, we get
\beq
\begin{aligned}
\left.\langle \mathcal{W}_{tt}\rangle\right|_{{\rm U}(1)}&=\left.\langle \mathcal{W}_{ti}\rangle\right|_{{\rm U}(1)}=\left.\langle \mathcal{W}_{it}\rangle\right|_{{\rm U}(1)}=\left.\langle \mathcal{W}_{ii}\rangle\right|_{{\rm U}(1)}=e^{\frac{8\pi^2g^2}{N^2}\left(1-\frac{\theta}{\pi}\right)}\left(=:w_{{\rm U}(1)}\right)\period
\end{aligned}
\eeq
We then obtain
\beq\label{eq:SUNresult}
\gamma_{\rm cross}^{{\rm SU}(N)}=\gamma_{\rm cross}-\frac{\del_{\theta}w_{{\rm U}(1)}}{2w_{{\rm U}(1)}}{\bf 1}=\gamma_{\rm cross}+\frac{4\pi g^2}{N^2}{\bf 1}\comma
\eeq
where $\gamma_{\rm cross}$ is the result for the U$(N)$ theory \eqref{eq:nearBPSrelation} and ${\bf 1}$ is the $2\times 2$ identity matrix.
\paragraph{Weak coupling} Expanding \eqref{eq:nearBPSrelation} and \eqref{eq:SUNresult} at weak coupling, we get the following result for $\gamma_{\rm cross}$ up to three loops:
\beq
\begin{aligned}
&\gamma_{\rm cross}^{\rm SU(N)}=g^2 \gamma_{(1)} +g^4\gamma_{(2)}+g^{6}\gamma_{(3)}+\cdots\comma\\
&\gamma_{(1)}=\pmatrix{cc}{\frac{4\pi}{N^2}&-\frac{4\pi}{N}\\\frac{4(\theta-\pi)}{N}&-4\theta+\frac{4\pi}{N^2}}\comma\quad
\gamma_{(2)}=\pmatrix{cc}{0&0\\\frac{8\theta(\theta-\pi)(\theta-5\pi)}{3N}&\frac{8\theta (\pi^2-\theta^2)}{3}+\frac{16\pi\theta (\theta-\pi)}{N^2}}\comma\\
&\gamma_{(3)}=\pmatrix{cc}{0&0\\\frac{8(\theta-\pi)\theta (5\pi^3+\pi^2\theta-\pi\theta^2+\theta^3)}{3N}&-\frac{8\theta(\pi^2-\theta^2)^2}{3}+\frac{16\pi\theta(\theta-\pi)(\pi^2-9\pi\theta+5\theta^2)}{3N^2}}\period
\end{aligned}
\eeq
\paragraph{Closed-loop normalization} As mentioned below \eqref{eq:nearBPSrelation}, the normalization we used is suited for the renomalization of open intersecting lines. To translate our result into the normalization commonly used for the renormalization of closed loops, we simply need to perform the following conjugation as explained in \cite{Munkler:2018cvu}\fn{In \cite{Munkler:2018cvu}, $\Gamma_{\rm cross}$ and $\Gamma_{\rm cross}^{\rm closed}$ were denoted as $\widehat{\Gamma}_{\rm cross}$ and $\Gamma_{\rm cross}$ respectively.}
\beq
\Gamma_{\rm cross}^{\rm closed}=S\Gamma_{\rm cross}S^{-1}\comma\qquad \gamma_{\rm cross}^{\rm closed}=S\gamma_{\rm cross}S^{-1}\comma
\eeq  
with $S:={\rm diag}(\sqrt{N},1/\sqrt{N})$.
We then get
\begin{align}
&\gamma_{\rm cross}^{\text{SU(N),closed}}=g^2 \gamma^{\rm closed}_{(1)} +g^4\gamma^{\rm closed}_{(2)}+g^{6}\gamma^{\rm closed}_{(3)}+\cdots\comma\\
&\gamma^{\rm closed}_{(1)}=\pmatrix{cc}{\frac{4\pi}{N^2}&-4\pi\\\frac{4(\theta-\pi)}{N^2}&-4\theta+\frac{4\pi}{N^2}}\comma\,\,\,\gamma^{\rm closed}_{(2)}=\pmatrix{cc}{0&0\\\frac{8\theta(\theta-\pi)(\theta-5\pi)}{3N^2}&\frac{8\theta (\pi^2-\theta^2)}{3}+\frac{16\pi\theta (\theta-\pi)}{N^2}}\comma\nn\\
&\gamma_{(3)}^{\rm closed}=\pmatrix{cc}{0&0\\\frac{8(\theta-\pi)\theta (5\pi^3+\pi^2\theta-\pi\theta^2+\theta^3)}{3N^2}&-\frac{8\theta(\pi^2-\theta^2)^2}{3}+\frac{16\pi\theta(\theta-\pi)(\pi^2-9\pi\theta+5\theta^2)}{3N^2}}\period\nn
\end{align}
$\gamma_{(1,2)}^{\rm closed}$ are in perfect agreement\fn{\cite{Munkler:2018cvu} uses a slightly non-standard convention in which the cross anomalous dimension is defined with an extra minus sign. (This can be seen by comparing \eqref{eq:RGcross} of our paper and (2.3) of \cite{Munkler:2018cvu}). Thus, to compare with our results, we need to consider $-\Gamma_{\rm cross}$ in \cite{Munkler:2018cvu}.} with the near-BPS limit of the two-loop results in \cite{Munkler:2018cvu}. It would be interesting to perform a direct three-loop computation and reproduce $\gamma_{(3)}^{\rm closed}$.
\paragraph{Strong coupling} The strong coupling limit of $\gamma_{\rm cross}$ can be computed from the results for intersecting loops in section \ref{subsec:integralintersect}. The result in the planar limit reads
\beq
\begin{aligned}
&\mathbb{W}_0 \sim \pmatrix{cc}{(c_0+\frac{c_1}{N^2})e^{8\pi g}+\frac{c_2 (\theta)e^{8\pi g_{\theta}}}{N^2}&\frac{c_3(\theta)e^{8\pi g_{\theta}}}{N} \\\frac{c_3(\theta)e^{8\pi g_{\theta}}}{N}&c_4(\theta)e^{8\pi g_{\theta}}}\comma\\
&\mathbb{W}_0^{-1} \sim \pmatrix{cc}{\frac{e^{-8\pi g}}{c_0}&-\frac{c_3(\theta)e^{-8\pi g}}{c_0 c_4(\theta)N} \\-\frac{c_3(\theta)e^{-8\pi g}}{c_0 c_4(\theta)N}&\frac{e^{-8\pi g_{\theta}}}{c_4(\theta)}}\comma
\end{aligned}
\eeq
where $c_0$ and $c_1$ are $\theta$-independent while $c_2$-$c_4$ are $\theta$-dependent prefactors, among which $c_3$ and $c_4$ are relevant for the analysis:
\beq
c_3(\theta)=-\frac{1}{32\pi^2(g_{\theta}^2\theta)}\left(1+O(g^{-1})\right)\comma\qquad c_4(\theta)=\frac{1}{8\pi^2(g_{\theta})^3}\left(1+O(g^{-1})\right)\period
\eeq
 We then get the following result in the planar limit:
\beq
\gamma_{\rm cross}^{\text{SU(N),closed}}=4\pi \del_{\theta}g_{\theta}\pmatrix{cc}{0&\frac{c_3}{c_4}\\0&1}+\frac{1}{2}\pmatrix{cc}{0&\frac{\del_{\theta}c_3}{c_4}\\0&\frac{\del_{\theta}c_4}{c_4}}+O(N^{-2})\period
\eeq
The leading strong-coupling answer for the lower-diagonal component $-4\pi \del_{\theta}g_{\theta}$ reproduces the prediction made in \cite{Munkler:2018cvu} using the classical worldsheet. By contrast, the leading strong-coupling answer for the upper-right component is given by
\beq\label{eq:mismatch}
4\pi \del_{\theta}g_{\theta}\frac{c_3}{c_4} \sim \frac{g^2}{\pi}\comma
\eeq
and does not match with the one in \cite{Munkler:2018cvu}, which predicts the same answer as the lower-diagonal component, $4\pi \del_{\theta}g_{\theta}$. This however does not immediately imply contradiction: As is well-known, the individual matrix elements of the anomalous dimension depend on the choice of the basis of operators. It is likely that the choice we made here for supersymmetric localization is different from the choice implicitly made in the analysis of the classical worldsheet. To avoid such ambiguities, we should compare the eigenvalues of the anomalous dimension matrix, which in fact agree with the ones in \cite{Munkler:2018cvu}. It would be important to understand this point further and also perform a comparison at the nonplanar level. 
\section{Conclusion\label{sec:conclusion}}
In this paper, we computed the expectation values of intersecting $1/8$ BPS Wilson loops in $\mathcal{N}=4$ SYM at finite $\lambda$ and $N$ using supersymmetric localization and the loop equation. The results are given by a coupled system of the Gaussian matrix model and multiple contour integrals, which in the planar limit give an infinite sum of products of modified Bessel functions. Applying the formalism to near-BPS limits of the cross anomalous dimension, we reproduced the perturbative data in \cite{Munkler:2018cvu}.

The main message of this paper is that the loop equation provides a powerful computational tool in $\mathcal{N}=4$ SYM when combined with localization.\fn{See \cite{Drukker:1999gy,Polyakov:2000ti,Polyakov:2000jg} for previous attempts to analyze the loop equation in $\mathcal{N}=4$ SYM and gauge/string duality.} It would be interesting to explore the connection with other nonperturbative techniques such as integrability and the conformal bootstrap\fn{See \cite{Anderson:2016rcw,Lin:2020mme} for interesting proposals on (different) bootstrap approaches to the loop equation.}: The intersecting lightlike Wilson lines were studied in the planar limit \cite{Caron-Huot:2019vjl} from integrability by the pentagon OPE \cite{Basso:2013vsa}. However the relation to the loop equation was not explored. Studying them through the lens of the loop equation may lead to stronger results, or at least would lead to a deeper understanding of the pentagon OPE. 

The $1/2$ BPS Wilson loop in $\mathcal{N}=4$ SYM is an example of a conformal defect \cite{Billo:2016cpy,Giombi:2017cqn,Kiryu:2018phb}. The insertion of $F_{\mu\nu}$---which plays the central role in the derivation of the loop equation---is a displacement operator, which is present in any conformal defect. Rephrasing the loop equation in the language of the defect CFT may allow us to use it as a dynamical input\fn{Here we have in mind the loop equation for the {\it non-intersecting line}, for which the right hand side of the loop equation vanishes. } for the conformal bootstrap. This would be perhaps useful for the nonsupersymmetric Wilson line discussed in \cite{Beccaria:2017rbe,Beccaria:2019dws,Correa:2018fgz}. Of course, in the absence of supersymmetric localization, one would need to study in this case the loop equation in the 4d gauge theory. Another direction is to analyze intersecting conformal defects in general CFTs. For the case of two intersecting 1d defects, one should be able to interpret them as a conformal two-point function of intersections as discussed in section \ref{subsec:2ptint}.  

Regarding the cross anomalous dimension, the simplest next step would be to generalize our computation to multiple lines intersecting at a point. This would shed light on the structure of the soft anomalous dimension of multileg amplitudes, studied for instance in \cite{Almelid:2015jia}. Of course, our analysis only applies to a small angle (or near-BPS) limit but it might be possible to combine it with the bootstrap approach in \cite{Almelid:2017qju} and constrain the full answer.

Another interesting direction is to perform the computation in different setups. For instance, the exact Bremsstrahlung function in $\mathcal{N}=2$ SCFT was studied in \cite{Fiol:2015spa,Gomez:2018usu,Bianchi:2019dlw}. Generalizing it to the small angle limit of the cross anomalous dimension is an important problem. It would also be interesting to study the ladder limit of the Wilson loop in $\mathcal{N}=4$ SYM, in which the R-symmetry angle $\theta$ is sent to $i\infty$ while the combination $\hat{\lambda}:=\lambda e^{-i\theta}$  is held fixed. This limit selects the ladder diagrams which can be resummed analytically \cite{Correa:2012nk,Kim:2017sju,Cavaglia:2018lxi,Correa:2018lyl,Correa:2018pfn,McGovern:2019sdd}. Last but not least, it is important to further study the cross anomalous dimension in perturbation theory. In particular, it would be desirable to generalize the result for the supersymmetric Wilson lines in \cite{Munkler:2018cvu} to nonsupersymmetric Wilson lines. \vspace{-10pt}
\subsection*{Acknowledgement}
We thank Jiaqi Jiang for collaboration on a related topic, and Hagen M\"{u}nkler for discussions and comments on the draft. SK thanks Lance Dixon, Grigory Korchemsky, Enrico Herrmann, and Ian Moult for discussions on the cross anomalous dimension. We thank CERN for hospitality during completion of this work. The work of SG is supported in part by the US  NSF under Grants No.~PHY-1620542 and PHY-1914860.  The work of SK is supported by DOE grant number DE-SC0009988.
\appendix
\section{Infinite Sum of Modified Bessel Functions\label{ap:besselsum}}
In this appendix, we derive identities for the infinite sum of modified Bessel functions and apply it to \eqref{eq:figureeightfinal} to rewrite it in a more symmetric form. 
The starting point is the integral representation
\beq
I_{n}(z)=\oint\frac{dx}{2\pi i x}\frac{e^{\frac{z}{2}(x+1/x)}}{x^{n}}\period
\eeq
We then rescale the integration variable $x \to \alpha x$ to get
\beq
I_{n}(z)=\frac{1}{\alpha^{n}}\oint\frac{dx}{2\pi i x}\frac{e^{\frac{z}{2}(\alpha x+1/(\alpha x))}}{x^{n}}\period
\eeq
We now factorize the exponential into two pieces
\beq
\exp\left[\frac{z}{2}\left(\alpha x+\frac{1}{\alpha x}\right)\right]=\exp \left[\frac{z_1}{2}\left(x+\frac{1}{x}\right)\right]\times \exp\left[\frac{z_2}{2}\left(\beta x+\frac{1}{\beta x}\right)\right]\comma
\eeq
with
\beq
z =\sqrt{z_1^{2}+z_2^{2}+\left(\beta +\frac{1}{\beta}\right)z_1 z_2}\comma\qquad \alpha=\sqrt{\frac{z_1+z_2\beta}{z_1+z_2\beta^{-1}}}\period
\eeq
and use the generating function representation for each exponential:
\beq
e^{\frac{y}{2}(x+1/x)}=\sum_{k=-\infty}^{\infty}I_k(y) y^{k}\period
\eeq
After performing the integral of $x$, we get
\beq
I_{n}(z)=\frac{1}{\alpha^{n}}\sum_{k=-\infty}^{\infty}\beta^{k}I_{n-k}(z_1)I_{k}(z_2)\period
\eeq
Specifying $n$ to be $1$ and using $I_{-k}=I_k$, we get the identity
\beq\label{eq:identityofinfintiesum}
I_{1}(z)=\frac{1}{\alpha}\sum_{k=0}^{\infty}\beta^{k+1}I_{k}(z_1)I_{k+1}(z_2)+\beta^{-k}I_{k+1}(z_1)I_{k}(z_2)\period
\eeq

Now applying this identity \eqref{eq:identityofinfintiesum}, we can exchange $\bar{A}_{1,2}$ in \eqref{eq:figureeightfinal}:
\begin{align}
\begin{aligned}
&\langle \mathcal{W}_{\text{figure-eight}}\rangle\overset{N\to\infty}{=}\\
&\frac{\mathcal{I}_{0}^{\bar{a}_2}\mathcal{I}_{1}^{\bar{a}_1}}{2\pi g_{\bar{a}_1}}+\sum_{k=1}^{\infty}\frac{\rho_{\bar{a}_2}^{k}\mathcal{I}_{k}^{\bar{a}_2}}{4\pi g}\left[\left(\rho_{\bar{a}_1}^{k+1}+\frac{(-1)^{k}}{\rho_{\bar{a}_1}^{k+1}}\right)\mathcal{I}_{k+1}^{\bar{a}_1}+\left(\rho_{\bar{a}_1}^{k-1}+\frac{(-1)^{k}}{\rho_{\bar{a}_1}^{k-1}}\right)\mathcal{I}_{k-1}^{\bar{a}_1}\right]\period
\end{aligned}
\end{align}
It is also possible to make it manifestly symmetric under $\bar{A}_1\leftrightarrow \bar{A}_2$:
 \begin{align}
&\langle \mathcal{W}_{\text{figure-eight}}\rangle\overset{N\to\infty}{=}\frac{I_{1}^{\frac{\bar{a}_1-\bar{a}_2}{2}}(2g)}{4\pi g}+\frac{1}{2\pi g}\sum_{k=1}^{\infty}\left(I_{k+1}^{\bar{a}_1}(g)I_{k}^{\bar{a}_2}(g)+I_{k+1}^{\bar{a}_2}(g)I_{k}^{\bar{a}_1}(g)\right)\comma
\end{align}
Here $I^{\theta}(g)$ is the modified Bessel function introduced in \cite{Gromov:2013qga},
\beq\label{eq:defbessel0}
\begin{aligned}
I_k^{\theta}(g)&:=\frac{I_k (4\pi g_{\theta})}{2} \left[\left(\frac{\pi+\theta}{\pi-\theta}\right)^{\frac{k}{2}}-(-1)^{k}\left(\frac{\pi-\theta}{\pi+\theta}\right)^{\frac{k}{2}}\right]\period
\end{aligned}
\eeq
\section{Cross Anomalous Dimension of Two Touching Lines\label{ap:cross}}
In this appendix, we compute the cross anomalous dimension of two touching Wilson lines. The basic strategy is the same as in section \ref{sec:cross}: We map it to a sphere, view it as two-point functions of intersections and differentiate it with respect to the angle $\theta$. 

In this case, the analogue of  \eqref{eq:defWbb0} is given by
\beq
\bar{\mathbb{W}}_0=\left.\pmatrix{cc}{\langle \mathcal{W}_{11}\rangle&\langle \mathcal{W}_{12}\rangle\\\langle\mathcal{W}_{21}\rangle&\langle \mathcal{W}_{22}\rangle}\right|_{\phi=\theta}\comma
\eeq
where $\mathcal{W}_{ij}$ denotes a Wilson loop whose intersections at the north and the south poles are resolved into configurations $i$ and $j$ in figure \ref{fig:cross}-$(b)$. The formula \eqref{eq:nearBPSrelation}  applies also to this case and the computation boils down to computing the BPS loops $\langle \mathcal{W}_{ij}\rangle|_{\phi=\theta}$. The main difference from section \ref{sec:cross} is that all the relevant loops are non-intersecting and one can simply use the results in the literature.

Let us first consider $\langle \mathcal{W}_{11}\rangle$. It corresponds to two oppositely-oriented Wilson loops with areas $A_1=4\pi-2\theta$ and $A_2=2\theta$. Using the result in \cite{Giombi:2009ms}, we find\fn{In this appendix, we focus on the large $N$ limit for simplicity. The result at finite $N$ can be obtained using matrix models in \cite{Giombi:2012ep}.}
\beq\label{eq:W1221}
\left.\langle \mathcal{W}_{11}\rangle\right|_{\phi=\theta}=\left(\frac{\mathcal{I}_1^{\pi-\theta}}{2\pi g_{\pi-\theta}}\right)^2+\frac{\pi g_{\pi-\theta}\mathcal{I}_1^{\pi-\theta}\mathcal{I}_2^{\pi-\theta}}{3N^2}+\frac{1}{N^2}\sum_{k=1}^{\infty}\frac{k(-1)^{k}(\mathcal{I}_k^{\pi-\theta})^2}{(\rho_{\pi-\theta})^{2k}}+O(1/N^4)\period
\eeq
On the other hand, $ \mathcal{W}_{12}$ and $\mathcal{W}_{21}$ are single Wilson loops with areas $4\theta$ and $4(\pi-\theta)$ respectively. They have the same expectation values given by\fn{As in section \ref{sec:cross}, here we normalized the Wilson loops by dividing by a common factor $N^2$. This is the origin of the extra factor of $1/N$ in \eqref{eq:W1221}.}
\beq
\left.\langle \mathcal{W}_{12}\rangle\right|_{\phi=\theta}=\left.\langle \mathcal{W}_{21}\rangle\right|_{\phi=\theta}=\frac{1}{N}\left(\frac{\mathcal{I}_1^{2\theta-\pi}}{2\pi g_{2\theta-\pi}}+\frac{\pi^2g_{2\theta-\pi}^2\mathcal{I}_2^{2\theta-\pi}}{3N^2}\right)+O(1/N^4)\period
\eeq
Finally $\mathcal{W}_{22}$ corresponds to two oppositely-oriented loops with areas $A_1=2(\pi+\theta)$ and $A_2=2(\pi-\theta)$. We then have
\beq
\begin{aligned}
&\left.\langle \mathcal{W}_{22}\rangle\right|_{\phi=\theta}=\left(\frac{\mathcal{I}_1^{\theta}}{2\pi g_{\theta}}\right)^2+\frac{\pi g_{\theta}\mathcal{I}_1^{\theta}\mathcal{I}_2^{\theta}}{3N^2}+\frac{1}{N^2}\sum_{k=1}^{\infty}\frac{k(-1)^{k}(\mathcal{I}_k^{\theta})^2}{(\rho_{\theta})^{2k}}+O(1/N^4)\period
\end{aligned}
\eeq

From $\bar{\mathbb{W}}_0$, the near-BPS limit of the cross anomalous dimension $\bar{\Gamma}_{\rm cross}=(\phi-\theta)\bar{\gamma}_{\rm cross} +O((\phi-\theta)^2)$ can be computed by $\bar{\gamma}_{\rm cross}=\frac{1}{2}(\del_{\theta}\bar{\mathbb{W}}_0)\cdot (\bar{\mathbb{W}}_0)^{-1}$. The result reads
\begin{align}
&\bar{\gamma}_{\rm cross}=\\
&\pmatrix{cc}{-\bar{h}_0^{-}+\frac{1}{N^2}\left(\bar{h}_0^{-}\bar{h}_1^{-}-\frac{\bar{h}_2^{-}}{2}-(\bar{h}_0^{+}+\bar{h}_0^{-})\bar{h}_3^{-}\bar{h}_3\right)&\frac{(\bar{h}_0^{+}+\bar{h}_0^{-})\bar{h}_3}{N}\\\frac{(\bar{h}_0^{+}-\bar{h}_0)\bar{h}_3^{-}}{N}&\bar{h}_0+\frac{1}{N^2}\left(-\bar{h}_0\bar{h}_1+\frac{\bar{h}_2}{2}-(\bar{h}_0^{+}-\bar{h}_0)\bar{h}_3^{-}\bar{h}_3\right)}\comma\nn
\end{align}
with $\bar{h}_k:= \bar{h}_k (\theta)$, $\bar{h}_{k}^{-}:=\bar{h}_k (\pi-\theta)$ and $\bar{h}_k^{+}:=\bar{h}_k(2\theta-\pi)$ and
\beq
\begin{aligned}
&\bar{h}_0(\theta)=\frac{4\pi \theta g_{\theta}}{\theta^2-\pi^2}\frac{\mathcal{I}_2^{\theta}}{\mathcal{I}_1^{\theta}}\comma\qquad \bar{h}_1(\theta)=\left(\frac{\mathcal{I}_1^{\theta}}{2\pi g_{\theta}}\right)^2\left[\frac{\pi g_{\theta}\mathcal{I}_1^{\theta}\mathcal{I}_2^{\theta}}{3}+\sum_{k=1}^{\infty}\frac{k(-1)^{k}(\mathcal{I}_k^{\theta})^2}{(\rho_{\theta})^{2k}}\right]\comma\\
&\bar{h}_2 (\theta)=\frac{2\pi g_{\theta}^2\mathcal{I}_1^{2\theta-\pi}}{(\mathcal{I}_1^{\theta})^2}\comma\qquad \bar{h}_3 (\theta)=\left(\frac{\mathcal{I}_1^{\theta}}{2\pi g_{\theta}}\right)^2\del_{\theta}\left[\frac{\pi g_{\theta}\mathcal{I}_1^{\theta}\mathcal{I}_2^{\theta}}{3}+\sum_{k=1}^{\infty}\frac{k(-1)^{k}(\mathcal{I}_k^{\theta})^2}{(\rho_{\theta})^{2k}}\right]\period
\end{aligned}
\eeq
The eigenvalue ($\bar{\gamma}_{\pm}$) are given by
\beq
\begin{aligned}
\bar{\gamma}_{+}&=\bar{h}_0 +\frac{(\bar{h}_0+\bar{h}_0^{-})(-2\bar{h}_0\bar{h}_1+\bar{h}_2^{-})+2 (\bar{h}_0-\bar{h}_0^{-})^2\bar{h}_3\bar{h}_3^{-}}{2N^2 (\bar{h}_0+\bar{h}_0^{-})}\comma\\
\bar{\gamma}_{-}&=-\bar{h}_0^{-} +\frac{(\bar{h}_0+\bar{h}_0^{-})(2\bar{h}_0^{-}\bar{h}_1^{-}-\bar{h}_2^{-})-2 (\bar{h}_0^{-}+\bar{h}_0^{+})^2\bar{h}_3\bar{h}_3^{-}}{2N^2 (\bar{h}_0+\bar{h}_0^{-})}\period
\end{aligned}
\eeq

To perform a comparison with the results in \cite{Munkler:2018cvu}, one needs to consider the ${\rm SU}(N)$ theory by stripping off the U$(1)$ factor
\beq
 \bar{\gamma}_{\rm cross}^{{{\rm SU}(N)}}=\bar{\gamma}_{\rm cross}-\frac{\del_{\theta}\bar{w}_{{\rm U}(1)}}{2\bar{w}_{{\rm U}(1)}}{\bf 1}\period
\eeq
Here $\bar{w}_{{\rm U}(1)}$ is the ${\rm U}(1)$ factor which in this case is the expectation value of the Wilson loop with area $4\theta$ (see \cite{Giombi:2012ep}),
\beq
\bar{w}_{{\rm U}(1)}=\exp\left(\frac{8g^2 \theta (\pi-\theta)}{N^2}\right)\period
\eeq
Converting the result to the closed-loop normalization by $\bar{\gamma}_{\rm cross}^{\rm closed}=S\bar{\gamma}_{\rm cross}S^{-1}$ with $S={\rm diag}(\sqrt{N},1/\sqrt{N})$, and expanding the result at weak coupling, we get
\begin{align}
&\bar{\gamma}_{\rm cross}^{\text{SU(N),closed}}=g^2 \bar{\gamma}^{\rm closed}_{(1)}+g^{4}\bar{\gamma}^{\rm closed}_{(2)}+g^{6}\bar{\gamma}^{\rm closed}_{(3)}+\cdots\comma\\
&\bar{\gamma}^{\rm closed}_{(1)}=-4\pmatrix{cc}{(\theta-\pi)+\frac{\pi-2\theta}{N^2}&\theta\\\frac{\theta-\pi}{N^2}&\theta+\frac{\pi-2\theta}{N^2}}\comma\nn\\
&\bar{\gamma}^{\rm closed}_{(2)}=-\frac{8}{3}\pmatrix{cc}{\theta(\theta-\pi)(\theta-2\pi)+\frac{2\theta(\pi^2-\theta^2)}{N^2}&\theta (\theta-\pi)(\theta+4\pi)\\\frac{\theta(\theta-\pi)(\theta-5\pi)}{N^2}&(\theta^2-\pi^2)-\frac{2\theta(\theta-\pi)(\theta-2\pi)}{N^2}}\comma\nn\\
&\bar{\gamma}^{\rm closed}_{(3)}=-\frac{8}{3}\pmatrix{cc}{(\theta-\pi)(\theta-2\pi)^2 \theta^2+\frac{2\theta(\theta-\pi)(3\pi^3-3\pi^2\theta+5\pi\theta^2-\theta^3)}{N^2}&\theta (\theta-\pi)(\theta^3+2\pi\theta^2+2\pi^2\theta-6\pi^3)\\\frac{\theta (\theta-\pi) (\pi^3+9\pi^2\theta-5\pi\theta^2+\theta^3)}{N^2}&\theta(\pi^2-\theta^2)^2-\frac{2\theta(\theta-\pi)(4\pi^3-4\pi^2\theta+2\pi\theta^2+\theta^3)}{N^2}}\period\nn
\end{align}
The results up to two loops reproduce the perturbative computation in \cite{Munkler:2018cvu}.

The strong coupling expansion can be obtained from the results in section \ref{subsec:integralintersect}. In the planar limit, we have
\beq\nn
\begin{aligned}
&\bar{\mathbb{W}}_0 \sim \pmatrix{cc}{\bar{c}_1e^{8\pi g_{\pi-\theta}}&\frac{\bar{c}_2 e^{4\pi g_{2\theta-\pi}}}{N} \\\frac{\bar{c}_2 e^{4\pi g_{2\theta-\pi}}}{N}&\bar{c}_3e^{8\pi g_{\theta}}}\comma\quad
\bar{\mathbb{W}}_0^{-1} \sim \pmatrix{cc}{\frac{e^{-8\pi g_{\pi-\theta}}}{\bar{c}_1}&-\frac{\bar{c}_2e^{4\pi(g_{2\theta-\pi}-2g_{\theta}-2g_{\pi-\theta})}}{\bar{c}_1 \bar{c}_3N} \\-\frac{\bar{c}_2e^{4\pi(g_{2\theta-\pi}-2g_{\theta}-2g_{\pi-\theta})}}{\bar{c}_1 \bar{c}_3N}&\frac{e^{-8\pi g_{\theta}}}{\bar{c}_3}}\comma
\end{aligned}
\eeq
where the prefactors all depend on $\theta$ and are given (at strong coupling) by
\beq
\bar{c}_1=\frac{1}{8\pi^2 (g_{\pi-\theta})^3}\comma\qquad \bar{c}_2=\frac{1}{\pi (2g_{2\theta-\pi})^{3/2}}\comma\qquad\bar{c}_3=\frac{1}{8\pi^2 g_{\theta}^3}\period
\eeq
Then the leading answer at strong coupling is given by
\beq
\gamma_{\rm cross}^{\text{SU(N),closed}}=\pmatrix{cc}{4\pi \del_{\theta}g_{\pi-\theta}&\frac{2\pi\bar{c}_2}{\bar{c}_3}(\del_{\theta}g_{2\theta-\pi}-2\del_{\theta}g_{\pi-\theta})e^{4\pi (g_{2\theta-\pi}-2g_{\theta})}\\0&4\pi \del_{\theta}g_{\theta}}+\cdots \period
\eeq
As in section \ref{subsec:U1}, the diagonal components reproduce the results in \cite{Munkler:2018cvu} while the upper off-diagonal component does not match with the one in \cite{Munkler:2018cvu}. However the off-diagonal component is exponentially suppressed at strong coupling since $g_{2\theta-\pi}<2g_{2\theta}$. Thus the eigenvalues of the matrix do coincide with the ones in \cite{Munkler:2018cvu}. (See the discussion below \eqref{eq:mismatch}).
\bibliographystyle{JHEP}
\bibliography{LoopRef}
\end{document}